\documentclass{SciPost}

\hypersetup{
    colorlinks,
    linkcolor={red!50!black},
    citecolor={blue!50!black},
    urlcolor={blue!80!black}
}

\usepackage[bitstream-charter]{mathdesign}
\usepackage{bm}
\usepackage[vcentermath]{youngtab}
\DeclareSymbolFont{usualmathcal}{OMS}{cmsy}{m}{n}
\DeclareSymbolFontAlphabet{\mathcal}{usualmathcal}

\fancypagestyle{SPstyle}{
\fancyhf{}
\lhead{\colorbox{scipostblue}{\bf \color{white} ~SciPost Physics }}
\rhead{{\bf \color{scipostdeepblue} ~Submission }}

\fancyfoot[C]{\textbf{\thepage}}
}
\newcommand{\diff}{d}

\newcommand{\im}{i}
\newcommand{\tr}{\operatorname{Tr}}
\newcommand{\diag}{\operatorname{diag}}

\newcommand{\U}{\mathrm{U}}
\newcommand{\SU}{\mathrm{SU}}   
\newcommand{\PSU}{\mathrm{PSU}}
\newcommand{\SO}{\mathrm{SO}}
\newcommand{\Spin}{\mathrm{Spin}}
\newcommand{\Sp}{\mathrm{Sp}}
\newcommand{\USp}{\mathrm{USp}}
\newcommand{\calS}{\mathcal{S}}
\newcommand{\rme}{\mathrm{e}}
\newcommand{\bolN}{\boldsymbol{N}}

\newsavebox{\dotdot}
\savebox{\dotdot}[3mm]{\shortstack{\circle{2.0}\\ \\ \circle{2.0}}}
\newcommand{\nord}[1]{\usebox{\dotdot} \! #1 \usebox{\dotdot}} 
\begin{document}
\pagestyle{SPstyle}
\begin{center}{\Large \textbf{\color{scipostdeepblue}{
Infrared properties of two-dimensional $\mathrm{SU}(N)/\mathrm{H}$ nonlinear $\sigma$ models at nonzero $\theta$ angles\\
}}}\end{center}
\begin{center}\textbf{
Philippe Lecheminant\textsuperscript{1},
Yuya Tanizaki\textsuperscript{2$\star$} and
Keisuke Totsuka\textsuperscript{2}
}\end{center}

\begin{center}
{\bf 1} Laboratoire de Physique Th\'eorique et Mod\'elisation, CNRS, CY Cergy Paris Universit\'e, 95302 Cergy-Pontoise Cedex, France
\\
{\bf 2} Yukawa Institute for Theoretical Physics, Kyoto University, Kitashirakawa Oiwakecho, Sakyo-ku, Kyoto 606-8502, Japan
\\[\baselineskip]
$\star$ \href{mailto:yuya.tanizaki@yukawa.kyoto-u.ac.jp}{\small yuya.tanizaki@yukawa.kyoto-u.ac.jp} 
\end{center}

\section*{\color{scipostdeepblue}{Abstract}}
\textbf{\boldmath{%
A general strategy is proposed to explore the low-energy properties of two-dimensional nonlinear $\sigma$ models with $\theta$ terms. 
We demonstrate its application to nonlinear $\sigma$ models with the target space $\SU(N)/\mathrm{H}$, 
which include $\mathbb{C}P^{N-1}$, complex Grassmannian manifolds as well as the flag $\SU(N)/\text{U(1)}^{N-1}$ 
and $\SU(N)/\text{SO($N$)}$ manifolds. By analyzing the symmetry and its anomaly content, we realize these nonlinear $\sigma$ models by considering specific deformations of the $\SU(N)$$_1$ conformal field theory. For the flag-manifold $\SU(N)/\text{U}(1)^{N-1}$ 
and $\SU(N)/\text{SO($N$)}$ models, those deformations are shown to correspond to the marginal current-current operator with the specific sign which leads to a massless renormalization group flow to the $\SU(N)$$_1$ fixed point.  In contrast, a massive regime with a two-fold ground-state degeneracy is found for the  $\mathbb{C}P^{N-1}$ ($N >2$) and the Grassmannian nonlinear $\sigma$ models at $\theta=\pi$.
}}

\vspace{\baselineskip}



\vspace{10pt}
\noindent\rule{\textwidth}{1pt}
\tableofcontents
\noindent\rule{\textwidth}{1pt}
\vspace{10pt}

\section{Introduction}
\label{sec:intro}
Two-dimensional nonlinear $\sigma$ models play an important role in various areas of physics, ranging from high-energy physics to condensed matter. In high-energy physics, these models are much simpler field theories than (3+1)-dimensional non-Abelian gauge theories but still share similar properties, such as asymptotic freedom, nonperturbative generation of the mass gap~\cite{Polyakov-75}.
Both in high-energy and condensed-matter physics, nonlinear $\sigma$ models describe the long-wavelength and long-time dynamics of Nambu-Goldstone modes in various continuous symmetry-breaking phases of matter,
and a paradigmatic example in condensed matter is the $\mathbb{C}P^1(\,\cong S^2)$ nonlinear $\sigma$ model [also commonly called the $\mathrm{O}(3)$ nonlinear $\sigma$ model] that describes the long-wavelength spin-wave excitations of an antiferromagnet \cite{Sachdev-book, Tsvelik-book-95, Fradkin-book-13, Altland-S-book-10}.
More general nonlinear $\sigma$ models with target space defined by a symmetric space $\mathrm{G}/\mathrm{H}$ govern the low-energy properties of Nambu-Goldstone modes relevant to the physics of Anderson localization whose different symmetry classes are encoded in different target manifolds \cite{Evers-M-08}, and to the monitored dynamics of free fermions \cite{Fava-P-S-B-N23, Poboiko-P-G-M-23, Chao-Ming-S-B-L-23, Fava-P-B-N-24,Tiutiakina-L-G-DN-DL-24}.

Nonlinear $\sigma$ models have target spaces with nontrivial topologies, and their classical actions may include topological terms depending on the spacetime dimension, which give purely imaginary contributions to the Euclidean action and cause interferences among different field configurations in the functional integral \cite{Tsvelik-book-95, Fradkin-book-13, Altland-S-book-10}. The striking consequence of the topological interference has been first uncovered for the two-dimensional $\mathbb{C}P^{1}$ nonlinear $\sigma$ model \cite{Haldane-PLA-83, Haldane-PRL-83}, which enjoys the $\theta$ angle and describes the semiclassical properties of the spin-$S$ SO($3$) Heisenberg chain when $\theta=2\pi S$.
Two different universality classes emerge depending on whether the on-site representation of the SO($3$) group is linear (for integer spins) or projective (for half-integer spins); when $S$ is half-integer (respectively integer) the $\theta$ term causes destructive (respectively constructive) interferences in the partition function leading to the emergence of a gapless (respectively Haldane gap) behavior.

Since this Haldane conjecture,  the effect of topological terms in nonlinear $\sigma$ models has played a central role in
our understanding of phases of quantum matter and quantum critical phenomena. For instance, the tenfold-way classification of non-interacting topological insulators and topological superconductors can be derived from the study of nonlinear $\sigma$ models on symmetric spaces in the presence of ${\mathbb{Z}}_{2}$ and Wess-Zumino-Novikov-Witten (WZNW)  topological terms \cite{Ryu-S-F-L-10, Chiu-T-S-R-16}. Also, the classification of bosonic symmetry-protected topological (SPT) phases in all spatial dimensions $d$ can be obtained using semiclassical O($d+2$) nonlinear $\sigma$  models with a $\theta$ term \cite{Bi-R-S-X-15}.  Topological terms also induce exotic quantum phases transition such as the integer quantum Hall plateau phase transitions which may be governed by the gapless behavior of the Grassmannian $ \frac{\text{U($2k$)}}{\text{U($k$)} \times \text{U($k$)}}$ nonlinear $\sigma$ model with a  $\theta= \pi$ term when $k\rightarrow 0$ \cite{Levine-L-P-83, Pruisken-84}.  A second striking example is the competition between two conventional symmetry-breaking orders which can be understood from the analysis of a nonlinear $\sigma$ model with a WZNW term \cite{Tanaka-H-06, Senthil-F-06}.  The latter leads to an exotic phase transition, dubbed deconfined quantum criticality, which goes beyond the Landau-Ginzburg paradigm
\cite{Senthil-V-B-S-F-04, Senthil-23}.

Although nonlinear $\sigma$ models with a topological term have many applications, it is notoriously difficult to correctly identify their low-energy properties. In the simplest situation, the infrared (IR) limit of the two-dimensional $\mathbb{C}P^{1}$ nonlinear $\sigma$ model with $\theta=\pi$  has been studied over the years with a variety of approaches.
When $\theta=0$, it is known from the classic result of Polyakov \cite{Polyakov-75} and the exact Bethe ansatz solution \cite{Wiegmann-85} that a mass gap is generated dynamically. At $\theta=\pi$, the field theory enjoys a mixed anomaly between SO(3)  and
${\mathbb{Z}}_2$ (time-reversal symmetry for instance) symmetries \cite{Gaiotto-K-K-S-17, Metlitski-T-18}, which constrains the IR limit to be nontrivial with either a critical behavior or a ground-state degeneracy via the 't~Hooft anomaly-matching condition~\cite{tHooft-anomaly-80} or the Lieb-Schultz-Mattis theorem~\cite{Lieb-S-M-61}.  From the Haldane mapping \cite{Haldane-PLA-83, Haldane-PRL-83} of the spin-$S$ Heisenberg chain and the fact that spin-1/2 Heisenberg chain has a gapless ground state, the two-dimensional $\mathbb{C}P^{1}$ nonlinear $\sigma$ model with $\theta=\pi$ is expected to be massless and belong to the SU(2)$_1$ universality class \cite{Affleck-Haldane}.  Using non-Abelian bosonization of the spin-$S$ Heisenberg chain, Affleck and Haldane \cite{Affleck-Haldane} gave a strong argument in favor of this critical behavior based on an important observation that the two-dimensional $\mathbb{C}P^{1}$ nonlinear $\sigma$ model with the $\theta=2 \pi S$ topological term is obtained when an appropriate perturbation is added to the two-dimensional SU(2)$_{2S}$ WZNW model \cite{Witten-84, Knizhnik-Z-84}.
In the $S=1/2$ case, there is no symmetry-allowed relevant perturbation in the SU(2)$_{1}$ WZNW conformal field theory (CFT) except the marginal current-current interaction, which strongly suggests that the $\mathbb{C}P^{1}$ nonlinear $\sigma$ model at $\theta=\pi $ flows to the SU(2)$_1$ CFT with central charge $c=1$.
Shankar and Read \cite{Shankar-R-90} arrived at a similar conclusion by considering a lattice regularisation of the $\mathbb{C}P^{1}$ nonlinear $\sigma$ model with the $\theta=\pi $ topological term, which, in the strong-coupling limit, is mapped onto the spin-1/2 Heisenberg chain.  Zamolodchikov and Zamolodchikov \cite{Zamolodchikov-Z-92} found the exact $S$-matrix corresponding to an integrable massless renormalization group (RG) flow between the 
ultraviolet (UV) $\mathbb{C}P^{1}$ nonlinear $\sigma$ model at $\theta=\pi $ and the critical SU(2)$_1$ WZNW model in the far IR limit. This criticality has been confirmed numerically over the years employing Monte Carlo simulations and tensor network techniques \cite{Bietenholz-P-W-95, Azcoiti-D-F-G-12, Alles-G-P-14, Caspar-S-22, Tang-X-W-T-21, Araz-S-S-23}. The functional RG approach has also been applied to this problem to describe the nonperturbative RG flow of the model as a function of $\theta$ \cite{Fukushima-S-T-22, Haruna-S-Y-24}. Very recently,  an interesting nonperturbative renormalization approach has been developed to determine the RG flow of the two-dimensional $\mathbb{C}P^{1}$   nonlinear $\sigma$ model at $\theta=\pi$  to the SU(2)$_1$ fixed point \cite{Zirnbauer-24, McRoberts-H-G-24}.

In this paper, we  investigate the IR properties of two-dimensional nonlinear $\sigma$ models by extending the idea by Affleck and Haldane \cite{Affleck-Haldane, Affleck-88}. 
To demonstrate how it works, we concretely study nonlinear $\sigma$ models with topological $\theta$ terms, whose target spaces are of the form $\SU(N)$/H [with H being a subgroup of $\SU(N)$].
More specifically, our examples include the type-AI symmetric space $\SU(N)/\text{SO($N$)}$ [$\text{H}=\text{SO($N$)}$], the complex Grassmannian $\text{Gr}(N,k)= \text{U($N$)}/\text{U($k$)} {\times} \text{U($N-k$)}$  at $\theta=\pi$,
as well as the flag-manifold $\SU(N)/\text{U(1)}^{N-1}$ nonlinear
$\sigma$ model with multiple $\theta$ terms [$\theta_a = 2\pi a/N$  ($a = 1, \ldots, N-1$)] \cite{Affleck-B-W-22}. 

These nonlinear $\sigma$ models with the target spaces $\SU(N)/\mathrm{H}$ describe the long-distance physics of the $\SU(N)$ Heisenberg spin chains if the subgroups $\mathrm{H}$ are chosen properly for given on-site $\SU(N)$ representations. For instance, the two-dimensional Gr($N$, $k$) nonlinear $\sigma$ model at $\theta= \pi n_{\text{c}}$ ($n_{\text{c}}\in \mathbb{Z}$) term \eqref{ComplexGrassmannfields} is directly related to the semiclassical limit of  $\SU(N)$ Heisenberg spin chain with alternating $\SU(N)$ irreps described by a Young diagram with $k$ rows and $n_{\text{c}}$ column on even sites and its conjugate on odd sites \cite{Affleck-85, Read-S-NP-89, Wang-W-W-19}.
As a special case, the two-dimensional $\mathbb{C}P^{N-1}=\text{Gr}(N,1)$ nonlinear $\sigma$  model with $\theta=\pi p$ describes the low-energy properties of a $p$-leg $\SU(N)$ Heisenberg spin ladder with alternating Heisenberg spin chains in the $\bolN$-$\bar{\bolN}$ representations of the $\SU(N)$ group \cite{Beard-P-R-W-05,Laflamme-E-al-16,Laflamme-E-al-16,Evans-G-H-W-18,Nguyen-M-T-U-23}. Finally, the flag-manifold $\SU(N)/\U(1)^{N-1}$ nonlinear $\sigma$ model has recently been discovered as describing the continuum limit of antiferromagnetic Heisenberg $\SU(N)$ spin chain in various representations \cite{Bykov-12, Bykov-13, Lajko-W-M-A-17, Wamer-L-M-A-20, Wamer-A-20, Affleck-B-W-22}.

In many cases, the presence of the $\theta$ term produces mixed global anomalies, and the IR properties are severely constrained by the anomaly-matching condition. For instance, the $\SU(N)/\SO(N)$ nonlinear $\sigma$ model at $\theta=\pi$ has been conjectured to enjoy a massless integrable RG flow 
with emergent $\SU(N)_1$ criticality with central charge $c=N-1$ in the far IR limit \cite{Fendley-01, Fendley-JHEP-01, Marino-M-R-23}. Similarly, the flag-manifold $\SU(N)/\mathrm{U}(1)^{N-1}$ nonlinear $\sigma$ model with the $(N-1)$ topological angles $\theta_a = 2\pi a/N$  ($a= 1, \ldots, N-1$) is also expected to be massless exhibiting the same $\SU(N)$$_1$ behavior \cite{Lajko-W-M-A-17,Tanizaki-S-18,Ohmori-S-S-19,Wamer-L-M-A-20}. In contrast, the presence of the $\theta=\pi$ topological term in the Grassmannian Gr$(2N, N)$ nonlinear $\sigma$ model does not lead to a critical behavior but has nontrivial effects resulting in the emergence of a massive phase with two-fold ground-state degeneracy \cite{Affleck-88}. A similar result is also expected for $\mathbb{C}P^{N-1}$ nonlinear $\sigma$  model with $\theta=\pi$  with a fully gapped phase that breaks spontaneously
the charge conjugation symmetry \cite{Seiberg-84, Beard-P-R-W-05}.   

We propose here a systematic method, based on the pioneer work of Affleck and Haldane \cite{Affleck-Haldane}, that enables us to confirm all the conjectures and results derived from these previous works. The starting point of our approach is the two-dimensional  $\SU(N)_1$ WZNW model with the Euclidean action \cite{Witten-84,Knizhnik-Z-84}:
\begin{equation}
{\cal S}_{{\rm WZNW}_1} = \frac{1}{8\pi} \int_{M_2} d^2 x \; {\rm Tr} \; (\partial_{\mu} G^{\dagger} \partial_{\mu} G)
+ \frac{i }{12\pi}   \int_{M_3} d^3 y \; \epsilon^{\alpha \beta \gamma}
{\rm Tr} \; (G^{\dagger} \partial_{\alpha} G \, G^{\dagger} \partial_{\beta} G \, G^{\dagger} \partial_{\gamma} G) \; ,
\label{WZW}
\end{equation}
where a summation over repeated indices is implied in the following, $G$ is an $\SU(N)$ matrix-valued field (the WZNW field), and $M_3$ is a three-dimensional manifold whose boundary is the two-dimension Euclidean space: $\partial M_3 = M_2$.  The WZNW model (\ref{WZW}) is massless with its IR properties described by the $\SU(N)_1$  CFT
\cite{Knizhnik-Z-84}. This conformal symmetry is usually lost when generic perturbations are added to the WZNW action. We consider here a class of  deformed $\SU(N)$$_1$ WZNW models of the following general form:
\begin{equation}
{\cal S} =  {\cal S}_{{\rm WZNW}_1} +  \lambda \int_{M_2} d^2 x  \; \mathcal{V} (G, G^{\dagger}) \; ,
\label{WZWgenint}
\end{equation}
where $\mathcal{V} (G, G^{\dagger})$ is a potential compatible with the global symmetry of the system. 
We apply the Affleck-Haldane strategy \cite{Affleck-Haldane,Affleck-88,Tanizaki-S-18,Ohmori-S-S-19}
by tailoring the potential ${\cal V} (G, G^{\dagger})$ in such a way that we obtain the desired nonlinear $\sigma$ model with the target space $\SU(N)/\mathrm{H}$ in the limit $\lambda \rightarrow \infty$, where the $\theta$ terms are produced from the WZNW term in Eq. (\ref{WZW}). 
We then consider the small $\lambda$ perturbation, and study its RG flow in the vicinity of the WZNW fixed point. 
For models on the flag-manifold $\SU(N)/\text{U}(1)^{N-1}$ and $\SU(N)/\text{SO($N$)}$, the deformation ${\cal V} (G, G^{\dagger})$ enjoys a  $({\mathbb{Z}}_N)_L$ discrete symmetry $G \rightarrow e^{i 2 \pi/N}G$ which forbids the generation of any relevant fields of the $\SU(N)_1$  CFT under the RG flow. The potential is then at most marginal, the lowest dimensional operator compatible with the symmetries of $\mathcal{V} (G,G^{\dagger})$ being the $\SU(N)_1$ current-current interaction with scaling dimension $2$. 
After replacing the potential term by this marginal operator, the action (\ref{WZWgenint}) becomes the $\SU(N)$ Gross-Neveu model (also called chiral Gross-Neveu model), which is an integrable field theory solved by the Bethe ansatz \cite{Andrei-L-79,Andrei-L-80} or by using the factorized S-matrix approach \cite{Berg-W-78,Koberle-K-S-79}. The non-perturbative IR spectrum of the latter model is known  and strongly depends on the sign of the coupling of the current-current interaction. The crucial point in concluding on the IR properties of the model boils down to the determination of the sign of its coefficient.  To this end, we exploit the fact that the $\SU(N)$$_1$  CFT with central charge $c=N-1$ admits a free-field representation in terms of $N-1$ bosons \cite{DiFrancesco-M-S-book} and compute various operator product expansions (OPEs) to explicitly evaluate the point-splitting regularization of ${\cal V} (G, G^{\dagger})$ to fix {\em  the specific sign} of the marginal current-current interaction. We then find the negative sign of the coupling constants for the models on the flag-manifold $\SU(N)/\text{U}(1)^{N-1}$ and $\SU(N)/\text{SO($N$)}$, 
which leads to the massless RG flow logarithmically approaching to the $\SU(N)_1$ fixed point. In contrast, for the  
$\mathbb{C}P^{N-1}$ ($N >2$) and Grassmannian  nonlinear $\sigma$ models at $\theta=\pi$, we obtain the massive phase with a two-fold ground-state degeneracy.

The rest of the paper is structured as follows.
In Sec.~\ref{sec:demo_SU2}, we review the approach of Affleck-Haldane that connects the $\mathbb{C}P^{1}$ nonlinear $\sigma$  model  to the SU(2)$_1$ WZNW model, and we also explain the general idea of our strategy. 
The analysis to the  flag-manifold $\SU(N)$/U(1)$^{N-1}$ nonlinear $\sigma$  model with $\theta$ terms is presented in Sec.~\ref{sec:flagsigmamodel}, demonstrating  the emergence of an $\SU(N)$$_1$ massless behavior in the IR limit. A similar approach is applied to the $\SU(N)$/SO($N$) nonlinear $\sigma$ model at $\theta=\pi$ in Sec.~\ref{sec:SU/SOsigmamodel} to confirm the existence of the $\SU(N)$$_1$ massless flow proposed in Refs. \cite{Fendley-01, Fendley-JHEP-01}. We also discuss the IR properties of the $\SU(N)$/USp($N$) nonlinear $\sigma$ model when $N$ is even. In Sec.~\ref{sec:Grassmanniansigmamodel},  we investigate the Grassmannian  U($2k$)/U($k$)$\times$ U($k$) and  $\mathbb{C}P^{N-1}$ nonlinear $\sigma$ models at $\theta=\pi$. Finally, our concluding remarks are presented in Sec.~\ref{sec:conclusion} together with several technical Appendices.

\section{Demonstration for the strategy with the \texorpdfstring{$\SU(2)_1$}{SU(2) level-1} WZNW model}
\label{sec:demo_SU2}
In this section, we discuss the massless RG flow from the $\mathbb{C}P^1$ nonlinear $\sigma$ model at $\theta=\pi$ to the $\SU(2)_1$ WZNW CFT. 
We first give a review of the seminal work by Affleck and Haldane~\cite{Affleck-Haldane} that connects the $\SU(2)_1$ WZNW model with the $\mathbb{C}P^1$ nonlinear $\sigma$ model at $\theta=\pi$. We will present this discussion in a manner that allows for a direct generalization to other examples discussed later of this paper.  
The CFT analysis is then used to describe the RG flow as a current-current perturbation around the WZNW fixed point, providing a reasonable minimal scenario for the phase diagram. 

\subsection{Double-trace deformation of the \texorpdfstring{$\SU(2)_1$}{SU(2) level-1} WZNW model}
Let us consider the $\SU(2)_1$ WZNW model with the double-trace potential, 
\begin{equation}
    \calS=\calS_{{\rm WZNW}_1}+\lambda \int \diff^2 x\, |\tr G(x)|^2  \; ,   
    \label{eqn:double-trace-perturbed-WZW}
\end{equation}
where $G$ is the $\SU(2)$-valued scalar field and $\calS_{{\rm WZNW}_1}$ is the $\SU(2)$ level-$1$ WZNW action~\eqref{WZW}. 
At $\lambda=0$, this theory has the global symmetry:
\begin{equation}
    \frac{\SU(2)_L\times \SU(2)_R}{\mathbb{Z}_2}, 
\end{equation}
where $(\Omega_L,\Omega_R)\in \SU(2)_L\times \SU(2)_R$ acts as $G\mapsto \Omega_L G  \Omega_R^\dagger$ and the $\mathbb{Z}_2$ quotient is taken because $(-\bm{1}_2,-\bm{1}_2)\in \SU(2)_L\times \SU(2)_R$ acts trivially. This symmetry group has the 't~Hooft anomaly, which requires that the system is gapless, and the model at $\lambda=0$ actually describes the $\SU(2)_1$ CFT. 
At nonzero $\lambda$, the global symmetry is explicitly broken to the subgroup 
\begin{equation}
    \frac{\SU(2)_V}{\mathbb{Z}_2}\times (\mathbb{Z}_2)_L\subset \frac{\SU(2)_L\times \SU(2)_R}{\mathbb{Z}_2}, 
    \label{eq:SU2_symmetry_doubletrace}
\end{equation}
where $\SU(2)_V$ is the diagonal subgroup of $\SU(2)_L\times \SU(2)_R$ and $V\in \SU(2)_V$ acts as $G\mapsto VGV^\dagger$. The center of $\SU(2)_L$, denoted by $(\mathbb{Z}_2)_L$,  describes the discrete chiral symmetry, $G\mapsto -G$. 

This symmetry structure clarifies the importance of studying this double-trace perturbation in the context of the antiferromagnetic Heisenberg spin chain.  
The discrete chiral symmetry $(\mathbb{Z}_2)_L$ appears as the low-energy realization of the one-step lattice translation symmetry, and thus the symmetry~\eqref{eq:SU2_symmetry_doubletrace} has the legitimate origin in the antiferromagnetic Heisenberg spin chain \cite{Affleck-Haldane}. 
Moreover, there is the $\mathbb{Z}_2$ mixed 't~Hooft anomaly between PSU$(2)_V = \SU(2)_V/\mathbb{Z}_2$ and $(\mathbb{Z}_2)_L$, and thus the low-energy physics has to be nontrivial to satisfy the anomaly-matching constraint, which is the field-theoretic version of the LSM theorem \cite{Furuya-O-17}.
 In this case, the possible scenarios to satisfy the anomaly matching are either (i) the system remains gapless while respecting the symmetry \eqref{eq:SU2_symmetry_doubletrace}, or (ii) the ground states are two-fold degenerate by breaking the $(\mathbb{Z}_2)_L$ symmetry spontaneously.   
We will see below that both scenarios of the ground states are realized within the above effective Lagrangian by properly choosing the sign of $\lambda$. 

Now, let us assume that $|\lambda|\gg 1$ so that the dynamics of $G$ is restricted to the classical moduli space of the double-trace potential.\footnote{To be precise, $\lambda$ is the dimensionful quantity, and we cannot talk about whether it is small or large unless we have other scale. However, the WZNW model is conformal, and we need to combine it with other perturbation to discuss its magnitude. For this purpose, we secretly deform the kinetic term of the WZNW action away from the conformal fixed point $\frac{1}{8\pi}$, and assume its marginal perturbation is large enough so that $\lambda$ becomes relevant in the UV limit.}  
We first consider the simpler case, $\lambda \to -\infty$, where the potential is minimized when the trace is maximized: $|\tr G |=2$,   
and hence the two classical vacua are given by 
\begin{align}
    G=\pm \bm{1}_2 \; . 
\end{align}
These vacua are related to each other by the discrete chiral symmetry $(\mathbb{Z}_2)_L$, and thus we find the two-fold degenerate gapped vacua that breaks the $(\mathbb{Z}_2)_L$ symmetry spontaneously when $\lambda \to -\infty$. 

The case $\lambda \to +\infty$ is more nontrivial and corresponds to the analysis by Affleck and Haldane~\cite{Affleck-Haldane}. Our presentation here basically follows Ref.~\cite{Tanizaki-S-18} so that its $\SU(N)$ generalization 
can be obtained. 
To minimize the double-trace potential with $\lambda>0$, we require that $\tr G=0$, and this classical moduli can be parametrized as 
\begin{align}
    G(x)= U(x)\, \mathrm{diag}(\rme^{\pi \im/2}, \rme^{-\pi \im/2})\, U(x)^\dagger,   
    \label{SU2wzwfieldmanifold}
\end{align}
where $U(x)$ is ``locally'' an $\SU(2)$-valued field. We note that this decomposition has the local $\U(1)$ gauge redundancy, $U(x)\sim U(x) \rme^{\im \alpha(x) \sigma_3}$, which implies that the physical target space is $\SU(2)/\U(1)\simeq \mathbb{C}P^1$. 
This becomes more evident if we represent the $\SU(2)$ matrix $U(x)$ as\footnote{%
To make contact with the original $S^{2}$ formulation in Ref.~\cite{Affleck-Haldane}, we note that $G=U (i \sigma_{3}) U^{\dagger}$ can be written 
as $G= i \vec{\Omega} {\cdot} \vec{\sigma}$ with the unit vector $\Omega_{i} = \vec{z}^{\dagger} \sigma_{i} \vec{z}$.}
\begin{align}
    U(x) = \begin{pmatrix}
        z_1(x) & -z_2^*(x) \\
        z_2(x) & z_1^*(x)
    \end{pmatrix}
\end{align}
with the constraint $|z_1|^2+|z_2|^2=1$, and then the $\U(1)$ gauge transformation acts on $\vec{z}=(z_1,z_2)^T$ as $\vec{z}(x)\mapsto \rme^{\im \alpha(x)}\vec{z}(x)$, so the equivalence class $[\vec{z}\,]$ parametrizes $\mathbb{C}P^1$. 
Let us summarize the action of the global symmetry in this description:
The $\SU(2)_V$ transformation acts as $\vec{z}\mapsto V \vec{z}$, and the faithful global symmetry is $\SU(2)_V/\mathbb{Z}_2$ because $-\bm{1}_2$ can be absorbed into the $\U(1)$ gauge redundancy. 
The $(\mathbb{Z}_2)_L$ transformation should acts as $G\to -G$, and this can be achieved by $U\mapsto U (\im \sigma_2)$, which is equivalent to the charge conjugation, $\vec{z}\mapsto \vec{z}^*$, in the $\mathbb{C}P^1$ description up to the $\SU(2)_V/\mathbb{Z}_2$ transformation. 

By substituting $G=U (\im \sigma_3) U^\dagger$ into the WZNW action (\ref{WZW}), we can derive the effective Lagrangian of the $\mathbb{C}P^1$ nonlinear $\sigma$ model. 
The computation of the kinetic term gives 
\begin{align}
    \tr(\partial_\mu G^\dagger \partial_\mu G) \propto |(\partial_\mu + \im a_\mu)\vec{z}|^2
\end{align}
with the auxiliary $\U(1)$ gauge field $a=\im \vec{z}^\dagger . \diff \vec{z}$. 
The computation of the WZNW term $\Gamma[G]$ in Eq.~\eqref{WZW} is tricky since we first have to extend $G(x)$ to the three-dimensional manifold $M_3$ with $\partial M_3=M_2$. 
The convenient choice given in Ref.~\cite{Tanizaki-S-18} is to take $M_3=(M_2\times [0,1])/(M_2\times \{1\})$, and we extend $G(x)=U(x) \rme^{\im (\pi/2)\sigma_3} U(x)^\dagger$ as 
\begin{align}
    G(x,x_3)=U(x)\, \rme^{\frac{\im}{2} \theta(x_3)\sigma_3} U(x)^\dagger, 
\end{align}
by setting $\theta(0)=\pi$ and $\theta(1)=0$. Since $G(x,0)=G(x)$ and $G(x,1)=U(x) U(x)^\dagger =\bm{1}$, this extension satisfies the requirement. Then, the straightforward calculation gives 
\begin{equation}
    \Gamma [G(x,x_3)]=\im \frac{\theta(0)}{2\pi}\int_{M_2} \diff a \; ,
\end{equation}
and we obtain the $\mathbb{C}P^1$ nonlinear $\sigma$ model at $\theta=\theta(0)=\pi$ for $\lambda \to +\infty$. 
Instead of performing this direct calculation, one can also deduce the value of the topological angle $\theta=\pi$  from general principles. The charge conjugation symmetry $(\mathbb{Z}_2)_L$ is present only at $\theta=0$ and $\theta=\pi$, and the mixed anomaly between $\SU(2)_V/\mathbb{Z}_2$ and $(\mathbb{Z}_2)_L$ can be reproduced only at $\theta=\pi$. Therefore, $\theta=\pi$ should be selected just by symmetry and consistency. 

\subsection{Relating the double-trace deformation and the current-current perturbation}
\label{sec:demo_SU2_OPE}
Therefore, the classical analysis with the double-trace potential $|\tr G|^{2}$ suggests that 
\begin{itemize}
    \item $\lambda \ll 0$: Two-fold degenerate gapped vacua that spontaneously breaks the center symmetry $(\mathbb{Z}_2)_L$ (or $G \to -G$). 
    \item $\lambda \gg 0$: The physics is described by the $\mathbb{C}P^1$ nonlinear $\sigma$ model at $\theta=\pi$. 
\end{itemize}
We would like to compare this with the 
SU(2)$_1$  current-current ${\vec J}_L \cdot {\vec J}_R$ perturbation of  the $\SU(2)_1$ WZNW model. We note that the latter is the lowest-dimensional scalar operator that respects the $[\SU(2)_V/\mathbb{Z}_2]\times (\mathbb{Z}_2)_L$ symmetry since the strongly relevant
SU(2)$_1$ WZNW primary field $\tr G$ is odd under $(\mathbb{Z}_2)_L$ and cannot thus appear \cite{Affleck-Haldane}: 
\begin{equation}
    \calS'=\calS_{\mathrm{WZNW}_1} - \lambda' \int \diff^2 x\, {\vec J}_L \cdot {\vec J}_R(x). 
\label{SU(2)-current-current}
\end{equation}
The latter model is the SU(2) Gross-Neveu model which can be solved exactly by means of the Bethe ansatz  approach \cite{Andrei-L-79}. It has a gapless (respectively fully gapped) spectrum when $ \lambda'  >0$ (respectively $ \lambda'  <0$). The spectrum of the elementary excitations for $ \lambda'  >0$ consists of spin-1/2 massless particles called the spinons. The exact massless scattering amplitudes for the right and left moving spinons have been given in
Ref.~\cite{Zamolodchikov-Z-92}. In the $ \lambda'  >0$ case, the spinons become massive excitations whose non-perturbative mass gap as function of  $ \lambda'$  is known from the exact solution of the model. The $\beta$-RG function to all orders in a certain minimal prescription, valid for all $\lambda'$, has been proposed in Refs. \cite{Kutasov-89,Gerganov-L-M-01}. Here, we simply need the one-loop RG equation to review the sign dependence of  $ \lambda'$ on the IR physics of the model:

\begin{align}
    \frac{\diff \lambda'}{\diff \ln \mu}= \beta \lambda'^2,
\end{align}
with some positive numerical coefficient $\beta$. This can be solved as 
\begin{align}
    \lambda'(\mu)= \frac{1}{1/\lambda'(a^{-1})+\beta \ln (1/\mu a)}. 
\end{align}
If $\lambda'<0$ at the UV cutoff $\mu=a^{-1}$, it is marginally relevant and $\lambda'$ diverges at some low-energy scale within the one-loop RG flow, which indicates the generation of the mass gap. 
On the other hand, if $\lambda'>0$ at the UV cutoff, it is marginally irrelevant and the system goes back to the WZNW CFT logarithmically. 
This observation strongly suggests that the double-trace deformation \eqref{eqn:double-trace-perturbed-WZW} and the current-current term
\eqref{SU(2)-current-current} are related: 
\begin{align}
    |\tr G|^2\leftrightarrow - {\vec J}_L \cdot {\vec J}_R \; . 
    \label{eq:DTandJlJr}
\end{align}
Since both operators share the same symmetry structure, it is quite natural to expect the existence of such a correspondence. 
According to the above argument, however, we have to establish this including its sign. 

To justify this correspondence explicitly, 
we regularize the original double-trace deformation at the short-distance scale as 
\begin{equation}
     |\tr G(x)|^2 \Rightarrow \int \diff^2 r\, F_\varepsilon(|r|) \tr[G(x+r/2)] \tr[G^\dagger(x-r/2)] \;  , 
    \label{eqn:double-trace-point-split}
\end{equation}
where $F_{\varepsilon}(|r|)$ is a positive function sufficiently localized in the region $|r|\lesssim \varepsilon$. 
When we are interested in the physics at the length scale much larger than $\varepsilon$, this should give identical results as the double-trace deformation. 
After this UV regularization, the operator product on the right-hand side can be evaluated using the short-distance OPE of the primary operator $\tr G$ of the $\SU(2)_1$ WZNW CFT.   
Noting that the $\SU(2)_1$ WZNW model is equivalent to the free massless compact boson at the self-dual radius, 
one can explicitly compute this OPE using the free-field representation (see Appendix~\ref{sec:AppendixB}):
\begin{equation}
    \tr[G(x+r/2)] \tr[G^\dagger(x-r/2)]\sim \frac{1}{|r|}- 4 \pi^2 |r| {\vec J}_L \cdot {\vec J}_R +\cdots \;  . 
    \label{eqn:double-trace-normal-ordered-SU2}
\end{equation}
When this is substituted in the above UV regularized form \eqref{eqn:double-trace-point-split}, we obtain the current-current operator with the negative coefficient as the leading nontrivial contribution, which justifies the identification \eqref{eq:DTandJlJr} {\em including the sign}. 

Thanks to this correspondence, it is quite natural to conclude that the $\mathbb{C}P^1$ nonlinear $\sigma$ model at $\theta=\pi$ is gapless and logarithmically approaches the $\SU(2)_1$ WZNW conformal fixed point in the deep IR regime.  More rigorous statements have been obtained  
in Refs.~\cite{Fateev-Z-Zn-91,Zamolodchikov-Z-92} based on the integrability of the model \eqref{SU(2)-current-current} which corresponds to the integrable massless RG flow of the $\mathbb{C}P^1$ nonlinear $\sigma$ model at $\theta=\pi$, and more recently in Ref.~\cite{Zirnbauer-24} using an interesting approach. 

\subsection{Proposal of the general strategy to investigate the RG flow}
\label{sec:proposal}
\begin{figure}[t]
\begin{center}
\includegraphics[scale=0.6]{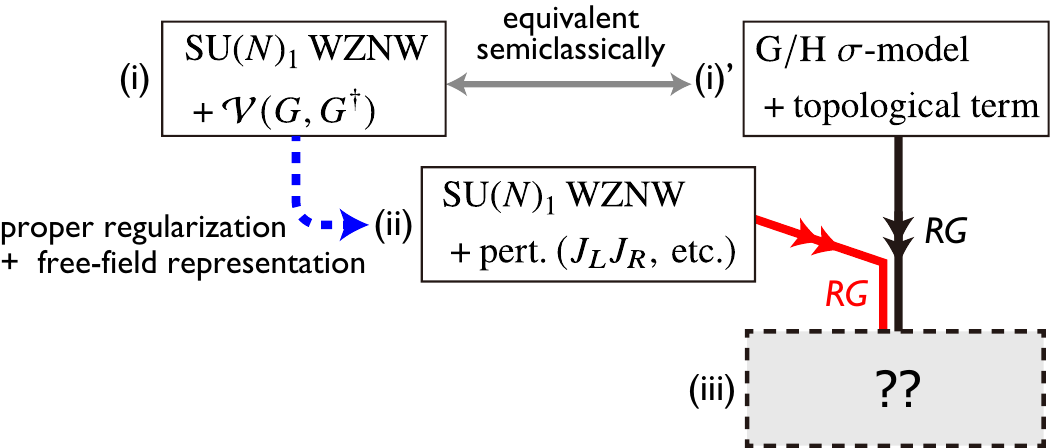}
\end{center}
\caption{Illustration of the strategy.  We start from the WZNW model (the principal chiral model with the WZNW term, precisely) 
with a potential term $\mathcal{V}(G,G^{\dagger})$ [model-(i)] which is designed to reproduce the desired nonlinear $\sigma$ model [(i)']  
in the semiclassical limit. 
We then properly regularize the potential (blue dashed line) and investigate what happens to the resulting deformed WZNW CFT [(ii)] by RG 
(red arrow) or by exploiting the integrability if the deformation leads to an integrable WZNW CFT.
\label{fig:strategy}}
\end{figure}

Here, let us extract the essence of the above RG analysis with the point-splitting regularization. 
Assume that we have established a connection between a certain WZNW CFT and the nonlinear $\sigma$ model of our interest using Affleck-Haldane's method 
(see Fig.~\ref{fig:strategy}): Introducing the potential term $\mathcal{V}(G(x))$, the nonlinear $\sigma$ model is obtained from the CFT by restricting the motion of $G(x)$ into the classical minima of $\mathcal{V}(G(x))$. 
As only the classical minima of $\mathcal{V} (G(x))$ is important in the Affleck-Haldane argument, it would be reasonable to assume that $\mathcal{V} (G(x))$ can be written in the form of $\mathcal{V} (G(x))=\sum_O |O(x)|^2$, where the sum runs over certain primary operators related to $G$ and its polynomials.  Several examples illustrating this form will be given in the following sections of this paper.
The primary operator $\tr G$ of the CFT itself can appear in $\mathcal{V} (G)$, which is indeed the case for the Grassmannian cases 
(see Sec.~\ref{sec:Grassmanniansigmamodel}). 
In such cases, the relevance of the perturbation can be determined directly by examining  the scaling dimension of $G$.  

A particularly interesting situation arises when all the primary fields are forbidden by certain discrete symmetries (see Secs.~\ref{sec:flagsigmamodel} and 
\ref{sec:SU/SOsigmamodel}), as in the  
$\mathbb{C}P^1$ case described above where the discrete $\mathbb{Z}_2$ chiral symmetry $G\mapsto -G$ excludes the only relevant operator $\tr G$. Under these symmetries, the potentials are at most marginal, the lowest dimensional operator compatible with the symmetries of $\mathcal{V} (G)$ being the current-current interaction with scaling dimension $2$:
\begin{equation}
    \calS'=\calS_{\mathrm{WZNW}_1} - \lambda' \int \diff^2 x\, \sum_{A=1}^{N^2-1} J^{A}_L J^{A}_R (x),
\label{SU(N)-current-currentaction}
\end{equation}
where we have introduced the chiral SU($N$)$_1$ current $J^{A}_{L,R}$. Action (\ref{SU(N)-current-currentaction}) 
is the SU($N$) Gross-Neveu model  which is an integrable field theory  \cite{Andrei-L-80}. 
The $\beta$-RG function of this field theory to all orders in a certain minimal prescription has been proposed in Refs. \cite{Kutasov-89,Gerganov-L-M-01}. As in the $N=2$ case, the model has a gapless (respectively fully gapped) spectrum when $ \lambda'  >0$ (respectively $ \lambda'  <0$). The Bethe-exact spectrum of the elementary excitations for $ \lambda'  >0$ consists of massless particles which transform in the fundamental representation of the  $\SU(N)$ group and are $\SU(N)$ spinons \cite{Johannesson-86}. The exact massless scattering amplitudes for the right and left moving $\SU(N)$ spinons have been given in Ref.~\cite{Fendley-01}. In the $ \lambda'  >0$ case, the  $\SU(N)$  spinons become massive excitations whose mass as a function of  $ \lambda'$  is known from the exact solution of the model as
well as its factorized S-matrix \cite{Berg-W-78,Koberle-K-S-79}.

The crucial point in concluding on the IR properties of the model is then the sign of $\lambda' $  generated by the RG flow starting from $\mathcal{V} (G)$.  The basic strategy in this paper is illustrated in Fig.~\ref{fig:strategy}.   
The idea in the above argument is to skip such higher-order computation by modifying $\mathcal{V} (G)$: We regularize $|O(x)|^2$ via the point-splitting procedure as 
\begin{equation}
    |O(x)|^2\Rightarrow \int \diff^2 r F_\varepsilon(|r|) O(x+r/2) O^*(x-r/2), 
    \label{eqn:pt-split-OO}
\end{equation}
where the minimal requirement of the weight function $F_\varepsilon$ is  
\begin{itemize}
    \item the positivity: $F_\varepsilon(|r|)\ge 0$.
    \item the locality: $F_\varepsilon(|r|)$ is sufficiently localized in the region $|r|\lesssim \varepsilon$. 
\end{itemize}
The positivity is required for the regularization procedure \eqref{eqn:pt-split-OO} not to change the sign of the potential $\mathcal{V}(G)$ which is crucial in the minimization. The locality is also necessary so that the regularized operator is also local. 
Now, we can establish the correspondence between $|O(x)|^2$ and the lowest dimensional operator [the current-current operator in the above example] just by using the short-distance OPE instead of solving higher-order RG flows. 

To get the meaningful result in this prescription, we also need to assume some regularity of $F_\varepsilon(|r|)$ at the origin. 
For this purpose, we have to look at the short-distance OPE of $O(x+r/2)O^*(x-r/2)$. 
If the lowest-dimensional operator that has the same symmetry of $|O(x)|^2$ is the current-current operator, the short-distance OPE takes the form of 
\begin{equation}
    O(x+r/2)O^*(x-r/2)\sim \frac{1}{|r|^{2\Delta_O}}\left(1+\# |r|^2 (\text{current-current})+\cdots\right) \; ,
\label{eqn:OO-to-JJ}
\end{equation}
where $\Delta_O$ is the scaling dimension of $O$. For the above integration being finite, $F_\varepsilon(|r|)$ should converge to $0$ sufficiently fast as $|r|\to 0$. The exact details of the weight function would not be important as we suspect that the long-wavelength physics is insensitive to such UV details, but as an example, one may take 
\begin{align}
    |O(x)|^2\Rightarrow \frac{1}{\varepsilon^{2}}\int \diff^2 r |r|^{2\Delta_O} \rme^{-|r|/\varepsilon} O(x+r/2) O^*(x-r/2), 
    \label{eq:regularization_scheme}
\end{align}
which satisfy all the above criterion. Then, the nontrivial leading contribution becomes the current-current operator, and the sign of its coefficient is identical to that of $\#$ in \eqref{eqn:OO-to-JJ}. 
We will apply this prescription to study the IR behaviors of several nonlinear $\sigma$ models, in particular, $\SU(N)/\U(1)^{N-1}$ and $\SU(N)/\SO(N)$ nonlinear $\sigma$ models at the specific $\theta$ angles in the remaining sections of this paper.  

\section{Continuous deformations between \texorpdfstring{$\SU(N)_1$}{SU(N) level-1} WZNW model and flag-manifold \texorpdfstring{$\sigma$}{sigma}  model}
\label{sec:flagsigmamodel}
The work by Bykov~\cite{Bykov-12,Bykov-13} and also by Wamer, Lajko, Mila, and Affleck~\cite{Lajko-W-M-A-17,Wamer-L-M-A-20}  discovered that the $\SU(N)$ analogue of the antiferromagnetic Heisenberg spin chain with the local spins belonging to the $p$-box symmetric representation 
($\underbrace{{\tiny \yng(2)}\cdots{\tiny \yng(1)}}_{p}$) 
can be also described by the relativistic nonlinear $\sigma$ model, whose target space is the complete flag manifold $\SU(N)/\U(1)^{N-1}$ 
(see, e.g., Ref.~\cite{Affleck-B-W-22} for a recent review). 
This model has $N-1$ independent topological angles which are specified as $\theta_a =\frac{2\pi p}{N} a$ with $a=1,\ldots, N-1$. 
The IR properties of the flag-manifold  nonlinear $\sigma$  model with the $\theta$ terms are determined in this section in close parallel to the recent investigation of Refs.~\cite{Tanizaki-S-18,Ohmori-S-S-19}.

\subsection{The \texorpdfstring{$\SU(N)/U(1)^{N-1}$}{SU(N)/U(1)N-1} flag-manifold nonlinear \texorpdfstring{$\sigma$}{sigma} model}
We start with the $\SU(N)_1$ WZNW model (\ref{WZW}), whose internal global symmetry is given by 
\begin{equation}
\left(\frac{\SU(N)_L \times \SU(N)_R}{\mathbb{Z}_N} \right) \rtimes \mathbb{Z}_2 \; .
\label{WZWsymCFT}
\end{equation}
The global chiral transformation $(\Omega_L, \Omega_R)\in \SU(N)_L\times \SU(N)_R$ acts as $G   \rightarrow \Omega_L G \,\Omega_R^{\dagger}$, and the $\mathbb{Z}_N$ quotient is taken because the diagonal center $(\rme^{2\pi \im/N}\bm{1}_N, \rme^{2\pi \im/N}\bm{1}_N)\in \SU(N)_L\times \SU(N)_R$  does not affect $G$. 
The action \eqref{WZW} is also invariant under the ${\mathbb{Z}}_2$ charge conjugation $G \rightarrow G^{*}$, which is an independent symmetry except for $N=2$ where it is already included in $[\SU(2)_L\times \SU(2)_R]/\mathbb{Z}_2$. 
The WZNW model is conformal with the central charge $c=N-1$ \cite{Knizhnik-Z-84}, and the above global chiral symmetry is a part of the larger one, 
$G(z, \bar z)   \rightarrow \Omega_L(z) G(z, \bar z)  \Omega_R^{\dagger} (\bar z)$, where $\Omega_L(z)$ and $\Omega_R^{\dagger} (\bar z)$ 
are arbitrary (anti-)holomorphic $\SU(N)$-valued functions, and $z = \tau + i x$ ($\tau$ being the imaginary time).  

Following the approach of Affleck and Haldane \cite{Affleck-Haldane}, we consider a deformation of
the $\SU(N)$$_1$ WZNW model and relate it with the $\SU(N)/\U(1)^{N-1}$ nonlinear $\sigma$ model. 
To this end, as a natural extension of \eqref{SU2wzwfieldmanifold}, the classical moduli space for the WZNW field $G$ is parametrized as~\cite{Tanizaki-S-18}
\begin{eqnarray}
G &=& U \Omega U^{\dagger} \nonumber \\
\Omega &=&  \omega^{- (N-1)/2} 
\begin{pmatrix}
1  & 0 & \cdots & 0 \\
0 & \omega   & \cdots & 0 \\
 \vdots & \cdots & \omega^{N-2} & 0\\
0 & \cdots & 0 & \omega^{N-1}
\end{pmatrix} , \label{wzwfieldmanifold}\\
\nonumber  
\end{eqnarray}
where $U$ is a ``locally'' $\U(N)$-valued field, and $\omega =e^{i 2 \pi/N}$.
As a specific deformation to realize this classical moduli, we can add multiple double-trace potentials~\cite{Ohmori-S-S-19}:\footnote{Here, we add the double-trace term of $\tr G^n$ following Ref.~\cite{Ohmori-S-S-19}. However, our proposal explained in Section~\ref{sec:proposal} uses the form of $|O(x)|^2$ with a primary operator $O(x)$, and they may look inconsistent at first. We here note that this is not problematic. A WZNW primary takes the form of $\tr G^2-(\tr G)^2$ instead of $\tr G^2$ itself, and our proposal naturally uses the potential of the form $\lambda'_1 |\tr G|^2 + \lambda'_2 |\tr G^2-(\tr G)^2|+\cdots$ with $\lambda'_i>0$. We can see that these two potentials have the same minima, so we do not need to discriminate them seriously: First, the minimization of the $\lambda'_1$ term sets $\tr G=0$, and thus setting $\tr G^2=0$ and $\tr G^2-(\tr G)^2=0$ are the same. It is straightforward to check this is true for $n=2,\ldots, [N/2]$.   }
\begin{eqnarray}
{\cal S} =  {\cal S}_{{\rm WZNW}_1} + \sum_{n=1}^{\left[N/2\right]} \int_{M_2} d^2 x  \; \lambda_n \left| {\rm Tr} \left[ G^n\right] \right|^2, 
\label{WZWint}
\end{eqnarray}
with $\lambda_n>0$. In the strong-coupling regime $\lambda_n \rightarrow + \infty$ for $n=1,\ldots, \left[N/2\right]$, the classical moduli of Eq. (\ref{WZWint}) selects an $\SU(N)$ matrix $G$ such that ${\rm Tr} \left[ G^n\right] = 0$ with $n=1,\ldots, \left[N/2\right]$, which can be parametrized as~\eqref{wzwfieldmanifold}. 
As noted in Ref.~\cite{Ohmori-S-S-19}, just adding $|\tr G|^2$ is not enough to select the $\SU(N)/\U(1)^{N-1}$ moduli space for $N\ge 4$, while it works for $N=2$ and $3$. 

The double-trace potential explicitly violates the continuous chiral symmetry, while it preserves the vector-like symmetry and the discrete chiral symmetry. The full global internal symmetry of model (\ref{WZWint}) is thus for $N>2$:
\begin{equation}
\left(\PSU(N)_V \times (\mathbb{Z}_N)_L \right) \rtimes \mathbb{Z}_2 , 
\label{WZWsympert}
\end{equation}
where $\PSU(N)_V=\SU(N)_V/\mathbb{Z}_N\ni [V]$ acts as $G\to V G V^{\dagger}$, and $(\mathbb{Z}_N)_L\subset \SU(N)_L$ as $G\to \rme^{2\pi \im/N}G$. 
This is the manifest symmetry that can be maintained at the $\SU(N)$ antiferromagnetic Heisenberg chain 
in the $p$-box symmetric representation 
($\underbrace{{\tiny \yng(2)}\cdots{\tiny \yng(1)}}_{p}$), where $\PSU(N)$ describes its spin rotational symmetry and $(\mathbb{Z}_N)_L$ is the low-energy realization of the one-unit lattice translation. 
There is the mixed 't~Hooft anomaly between $\PSU(N)$ and $(\mathbb{Z}_N)_L$: When we promote the global symmetry $\PSU(N)_V\times (\mathbb{Z}_N)_L$ to the local gauge redundancy, the system should be interpreted as the boundary of the $3$d SPT state described by~\cite{Tanizaki-S-18}
\begin{align}
    \calS_{\text{$3$d SPT}}=\frac{2\pi}{N}\int_{M_3} A_L\cup w_2(\PSU(N)), 
\end{align}
where $A_L$ is the background gauge field for $(\mathbb{Z}_N)_L$ and $w_2(\PSU(N))$ is the background two-form gauge field for $\PSU(N)$ describing the obstruction to lift the $\PSU(N)$ bundle to the $\SU(N)$ bundle. 
Then, the anomaly-inflow argument states that the low-energy effective theory has to reproduce the same 't~Hooft anomaly, which gives the LSM-type constraint \cite{Yao-H-O-19}.  
This constrains the possible scenarios of the low-energy physics as: 
\begin{itemize}
    \item The ground states are $N$-fold degenerate due to the spontaneous breaking of $(\mathbb{Z}_N)_L$. 
    \item The ground state preserves the symmetry (\ref{WZWsympert}) but has gapless excitations. 
\end{itemize}
The first option is realized by flipping the sign of $\lambda_n$ in the model~\eqref{WZWint}: When $\lambda_n\to -\infty$ for $n=1,2,\ldots, [N/2]$, the potential minima are given by 
\begin{equation}
    G=\rme^{\frac{2\pi \im}{N}k}\bm{1}_N,
    \label{eq:ClassicalVacuaNegativeDT}
\end{equation}
with $k=1,\ldots, N$, and we obtain $N$-fold gapped vacua that violate $(\mathbb{Z}_N)_L$. In the lattice model, they correspond to the valence-bond-solid  states that break spontaneously the one-unit lattice translation but preserve the $N$-unit translation. 

We now present the details for the case of $\lambda_n>0$, where the classical moduli is given by \eqref{wzwfieldmanifold}, $G=U\Omega U^\dagger$. As $\Omega$ is a diagonal matrix, this decomposition has the $\U(1)^N$ gauge redundancy, 
\begin{align}
    U(x)\sim U(x)\diag(\rme^{\im \alpha_1(x)}, \ldots, \rme^{\im \alpha_N(x)}), 
\end{align}
and the physical target space is given by the flag manifold, $\U(N)/\U(1)^{N-1} \simeq \SU(N)/\U(1)^{N-1}$. 
Let us introduce $N^2$ complex scalar fields $z_{i j}$ ($i,j = 1, \ldots, N$) 
such that 
\begin{eqnarray}
    U(x)=[z_{ij}(x)]_{i,j=1,\ldots,N}=[\vec{z}_1(x), \ldots, \vec{z}_N(x)],
\label{complexscalarfields}
\end{eqnarray}
with $(\vec{z}_j)_i=z_{ij}$. 
These fields are constrained to be orthonormal complex vectors, $\vec z^{\dagger}_i \cdot \vec z_j = \delta_{ij}$, to enforce the $\U(N)$ property, $ U^{\dagger} U = 1_N$. 
The original global symmetries of the action (\ref{WZWint}) have a direct interpretation on the complex fields $\vec z_i$ via the identification (\ref{wzwfieldmanifold}):
\begin{itemize}
    \item The $\PSU(N)_V$ spin symmetry: $\vec{z}_i\mapsto V\vec{z}_i$ with $V\in \SU(N)_V$. 
    \item The $(\mathbb{Z}_N)_L$ chiral symmetry: $\vec{z}_i\mapsto \vec{z}_{i-1}$, where the flavor index is understood in mod $N$. 
    \item The $\mathbb{Z}_2$ charge conjugation: $\vec{z}_i\mapsto \vec{z}^{\,*}_{N-i+1}$. 
\end{itemize}

The next step is to derive the low-energy effective Lagrangian for the complex fields $\vec z_i$ by substituting \eqref{wzwfieldmanifold} into the WZNW action~\eqref{WZW}. 
The calculation of the kinetic term is straightforward, whereas the evaluation of the WZNW term requires the extension of $G(x)=U(x) \Omega U(x)^\dagger$ to a three-dimensional manifold $M_3$ with $\partial M_3=M_2$. 
The derivation has been done in Refs.  \cite{Ohmori-S-S-19, Tanizaki-S-18}, which propose to take $M_3=(M_2\times [0,1])/(M_2\times \{1\})$ and $G(x,x_3)=U(x) \Omega(x_3) U(x)^\dagger$ with $\Omega(x_3)=\omega^{-(N-1)/2}\diag(\rme^{\im \theta_1(x_3)},\ldots, \rme^{\im \theta_N(x_3)})$. Here, we just require that $\theta_a(0)=\frac{2\pi}{N} (a-1)$ and $\theta_a(1)=\theta'$ ($a = 1, \ldots, N$), where $\theta'$ is arbitrary.  
The Lagrangian then takes the form of a nonlinear $\sigma$ model on the flag manifold  $\SU(N)$/U(1)$^{N-1}$ with topological $\theta$ terms: 
\begin{align}
    {\cal L} &= \frac{N}{4\pi} \sum_{a=1}^{N} \left( |\partial_{\mu}  {\vec z_{a}} |^2
    -  |{\vec z^{\,\dagger}}_{a} \cdot \partial_{\mu}  {\vec z_{a}} |^2  \right) + \sum_{1 \le a < b \le N}  \left(g_{ab} \delta^{\mu \nu}  +  b_{ab} \epsilon^{\mu \nu} \right) \left({\vec z^{\,\dagger}}_{a} \cdot \partial_{\mu}  {\vec z_{b}}  \right) \left( {\vec z^{\, \dagger}}_{b} \cdot \partial_{\nu}  {\vec z_{a}} \right) 
    \notag\\ 
    & \quad +  \sum_{a=1}^{N}  \frac{\theta_a}{2\pi} \epsilon^{\mu \nu} 
    \partial_{\mu}  {\vec z^{\,\dagger}}_{a} \cdot \partial_{\nu}  {\vec z_{a}} \; ,
    \label{flagsigmalagrangian} 
\end{align}
where the first two terms are nontopological couplings $g_{ab} =  \cos( 2\pi (a-b)/N)/2\pi$ and $b_{ab}  
=  \sin( 2\pi (a-b)/N)/2\pi$ and the topological angles appearing in the last term are defined by 
\begin{align}
	\theta_a = \theta_a(0)-\theta_a(1)= \frac{2\pi}{N} (a-1) - \theta^{\prime}  \quad (a = 1, \ldots, N) \; . 
\end{align}
These $N$ topological angles $\theta_a$ couple to a set of integer topological charges: 
\begin{equation}
q_{a} = \frac{i}{2\pi}   \int d^2 x  \epsilon^{\mu \nu} 
\partial_{\mu}  {\vec z^{\,\dagger}}_{a} \cdot \partial_{\nu}  {\vec z_{a}} \; .
\label{topocharges}
\end{equation}
However, the topological charges are not all independent due to the orthonormalization constraint, 
$\vec z^{\,\dagger}_i \cdot \vec z_j = \delta_{ij}$, and they satisfy $\sum_{a=1}^{N} q_{a} = 0$ since it can be shown that 
$\sum_{a=1}^{N} \vec z^{\, \dagger}_{a} \cdot \partial_{\mu}  {\vec z_{a}}  = 0$ \cite{Tanizaki-S-18,Lajko-W-M-A-17}. 
This implies that the model \eqref{flagsigmalagrangian} is left invariant by shifting all the topological angles by the same amount: $\theta_a \rightarrow \theta_a + \theta$, 
and there are thus only $N-1$ independent topological angles, which is in full agreement with the value of the second homotopy group for the flag manifold: $\pi_2(\SU(N)/\U(1)^{N-1})\simeq \mathbb{Z}^{N-1}$. This explains why the effective Lagrangian appears to contain an undetermined constant $\theta'$. In the following we fix its value as  
\begin{align}
	\theta_a = \frac{2\pi}{N} a \quad (a = 1, \ldots, N-1)
\label{topangleflag}
\end{align}
in the model (\ref{flagsigmalagrangian}). 

\subsection{\texorpdfstring{$\SU(N)_1$}{SU(N) level-1} CFT  approach}
In this section, we provide an argument suggesting that the $\SU(N)/\U(1)^{N-1}$ nonlinear $\sigma$ model becomes gapless at the $(\mathbb{Z}_N)_L$ symmetric point with $\theta_a=2\pi a/N$ ($a = 1, \ldots, N-1$). 
For $N=2$, the flag nonlinear $\sigma$ model becomes the famous $\mathbb{C}P^1$ nonlinear $\sigma$ model with a $\theta=\pi$  topological term with its well-known SU(2)$_1$ massless behavior as reviewed in Sec. \ref{sec:demo_SU2}. The flag  nonlinear $\sigma$ model (\ref{flagsigmalagrangian}) with topological angles (\ref{topangleflag})  and 
the $\SU(N)$$_1$ CFT  share the same mixed global P$\SU(N)$ $\times \mathbb{Z}_N$ anomaly \cite{Ohmori-S-S-19, Tanizaki-S-18,Yao-H-O-19}, and a massless RG flow between the two theories is consistent with the 't Hooft anomaly matching, generalizing the $N=2$ result to $N>2$. 
However, the anomaly matching argument provides only  the kinematical constraint, and a more detailed analysis of the dynamics is required to determine whether  the massless RG flow is actually realized. 

Before presenting our argument, let us note that there is evidence supporting the massless RG flow from the flag nonlinear $\sigma$ model to the $\SU(N)_1$ CFT. 
As we briefly mentioned, the flag nonlinear $\sigma$ model (\ref{flagsigmalagrangian}) with topological angles  $\theta_a = 2\pi p a/N$ governs the IR properties of  $\SU(N)$ Heisenberg spin chain in symmetric rank-$p$ tensor representation in the large $p$ limit 
\cite{Lajko-W-M-A-17,Wamer-L-M-A-20}.  A gapless phase in the $\SU(N)$$_1$ universality class has been predicted for model (\ref{flagsigmalagrangian}) when $p$ and $N$ are coprime while a spectral gap is formed in other situations \cite{Ohmori-S-S-19, Tanizaki-S-18,Wamer-A-20, Wamer-L-M-A-20}. In the particular $p=2$ and $N=3$ case, which 
shares the same topological angles as in Eq. (\ref{topangleflag}) for $N=3$, a large-scale  DMRG calculation has shown the existence of a gapless SU(3)$_1$ behavior with central charge $c=2$ \cite{Nataf-G-M-21}.  

To understand the possible RG flow, the list of relevant and marginal scalar operators for the $\SU(N)_1$ CFT is needed. For this purpose, let us give a brief reminder of the $\SU(N)_1$ current algebra. 
The $\SU(N)$$_1$ CFT has $J_{R,L}^A$ chiral currents, defined by the following OPE (we use the same conventions as in Refs. \cite{Affleck-NP86, Affleck-88,Gogolin-N-T-book,James-K-L-R-T-18}):
\begin{equation}
J^{A}_{L} (z)  J^{B}_{L}(0) \sim \frac{\delta^{AB}}{8 \pi^2 z^2} +   \frac{i f^{ABC}  J^{C}_{L}(0) }{2 \pi z}  \; ,
 \label{OPESUN1curr}
\end{equation}
with a similar definition for the right current. In Eq. (\ref{OPESUN1curr}), $f^{ABC}$ denotes the antisymmetric structure constants of the $\SU(N)$ group and $z = \tau + i x$ ($\tau$ being the imaginary time). 
This OPE defines an infinite-dimensional algebra called the Kac-Moody algebra, and the counterparts of the highest-weight representations are called the (WZNW) primary operators. 
As an example of the primary operator, the defining OPE of the $\SU(N)$$_1$ WZNW $G$ field reads as \cite{DiFrancesco-M-S-book}:
\begin{eqnarray}
J_{ L}^A\left(z\right)  G_{\alpha \beta} (0,0) &\sim& - \frac{1}{2 \pi z} \; T^{A}_{\alpha \gamma}
G_{\gamma \beta} (0,0) \nonumber \\
J_{L}^A\left(z\right)  G^{\dagger}_{\beta \alpha} (0,0) &\sim& \frac{1}{2 \pi z}  \;
 G^{\dagger}_{\beta  \gamma} (0,0) T^{A}_{\gamma \alpha} \; ,
\label{OPEWZW}
\end{eqnarray}
$ T^{A}$ being the generators which transform in the fundamental  $\bolN$-representation of the 
$\SU(N)$ group with the normalization:   $\text{Tr}(T^{A} T^{B})=\delta^{AB}/2$. 

The $\SU(N)_1$ CFT admits $N$ primary fields $\Phi^{(a)}$ which transform in the fully antisymmetric representations of the $\SU(N)$ group, $a=0,..,N-1$ being the number of boxes of the underlying Young tableau with a single column. 
They are generated by fusion from the WZNW field $G=\Phi^{(1)}$ which transforms in the  $\bolN$-representation. 
These primary fields are relevant operators with scaling dimension $a(N-a)/N$. 
Under the $({\mathbb{Z}}_N)_L$ center symmetry,  
they transform as $\Phi^{(a)}  \rightarrow e^{i 2 a \pi/N} \Phi^{(a)}$. Since the action  (\ref{WZWint}) is ${\mathbb{Z}}_N$ invariant by construction, these relevant contributions cannot be generated under the RG flow~\cite{Ohmori-S-S-19}. 
Based on this observation, Ref.~\cite{Ohmori-S-S-19} conjectured that the $\SU(N)/\U(1)^{N-1}$ model with $\theta_a = \frac{2\pi a}{N}$ flows to the $\SU(N)_1$ WZNW CFT. 

To complete this argument, we have to take care of the $\SU(N)_V$-invariant current-current term, 
\begin{equation}
    -\gamma\sum_{A=1}^{N^2-1} J_L^A J_R^A  \;  ,
\end{equation}
which is marginally irrelevant for $\gamma> 0$ but marginally relevant for $\gamma < 0$. 
In this respect, the key point of the analysis is to determine the precise sign of the coupling constant $\gamma$ of the  $\SU(N)$$_1$  current-current term from the double-trace deformations. 
To make the connection between them, we employ the idea presented in Sections~\ref{sec:demo_SU2_OPE} and~\ref{sec:proposal}: We take the point-splitting of double-trace operators as 
\begin{align}
	|\tr G^n(x)|^2 \Rightarrow \frac{1}{\varepsilon^{2}}\int \diff^2 r |r|^{2n(N-n)/N} \rme^{-|r|/\varepsilon} \tr G^n(x+r/2) \tr G^{\dagger n}(x-r/2) \; , 
\end{align} 
and compute the right-hand-side by OPE to extract the current-current term as the leading nontrivial contribution. 

The computation of OPE can be performed by using the exact duality between the $\SU(N)_1$ WZNW CFT and $N-1$ massless compact bosons $\Phi_m$ ($m=1,\ldots, N-1$) that we combine into a single vector field
$\vec{\Phi}$. As discussed in Appendix~\ref{sec:AppendixA}, its periodicity is described by the root
lattice which is spanned by the simple root vectors $\vec{\alpha}_m$ and reads as follows\footnote{For details on the periodicity of the compact boson, see discussions in Section~\ref{sec:SUSp_massiveRGflow} and, in particular, Appendix~\ref{app:FreeBosonWZW_Summary}. }
\begin{align}
    \vec{\Phi}\sim \vec{\Phi}+\sqrt{\pi}\vec{\alpha}_m \;  ,
    \label{eq:PhiPeriodicity}
\end{align}
and its kinetic term of the Lagrangian is given by the canonical one, $\frac{1}{2}(\partial_\mu \vec{\Phi})^2$. 
In the compact boson representation, the building block of the interaction 
of model (\ref{WZWint}) admits a simple free-field representation:
\begin{eqnarray}
 {\rm Tr} \; G  =   \frac{1}{\sqrt{N}} \sum_{\alpha=1}^{N} : \rme^{i \sqrt{4 \pi}   {\vec  e}_{\alpha} \cdot {\vec  \Phi}} :  \; , 
\label{trGboso}
\end{eqnarray}
where $ {\vec  e}_{\alpha}$ are the weight vectors of the $\bolN$-representation, which are $(N-1)$-component vectors defined by Eq. (\ref{eq:SpecificWeight}). 
As a result, we can find that the short-distance OPE is given by 
\begin{align}
	\tr G^n(x+r/2) \tr G^{\dagger n}(x-r/2)=\frac{1}{|r|^{2n(N-n)/N}}\left(\frac{n! N!}{N^n (N-n)!}-\frac{8\pi^2 n n! (N-2)!}{N^n (N-n-1)!}|r|^2 J_L^A J_R^A+\cdots \right), 
\end{align}
whose full derivation is presented in Appendix~\ref{sec:AppendixB} [see Eqs.~\eqref{TrGnGnApp} and \eqref{TrGnGnNfinApp}]. 
By substituting this expression into the point-splitting form of the potential, we obtain the effective Lagrangian as 
\begin{equation}
    \mathcal{L}_{\text{eff}} =\frac{1}{2}(\partial_\mu \vec{\Phi})^2 
    - \gamma_{\text{eff}} \left\{ \frac{1}{8\pi}(\partial_\mu \vec{\Phi})^2-\frac{1}{4\pi^2}\sum_{1 \le \alpha<\beta \le N}\cos\left(\sqrt{4\pi}(\vec{e}_\alpha-\vec{e}_\beta)\cdot \vec{\Phi}\right)\right\}   \; ,
    \label{eq:CompactBosonLagrangian_Flag}
\end{equation}
where the coupling constant $\gamma_{\rm eff}$ is defined as 
\begin{equation}
\gamma_{\rm eff} =96 \varepsilon^2 \pi^3 (N-2)! \sum_{n=1}^{\left[N/2\right]} \;  \frac{ \lambda_n n n!}{N^n (N-n-1)!} .
\label{couplingfinal}
\end{equation} 
Equivalently, the effective Hamiltonian for model (\ref{WZWint}) in the general $N$ case is given by the Sugawara energy-momentum tensor with the $\SU(N)$ current-current term:
\begin{equation}
 \mathcal{H}_{N} = \frac{2\pi }{N+1} \left( : J^A_{R} J^A_{R}: + : J^A_{L} J^A_{L}: 
\right)   - \gamma_{\rm eff} J_R^{A} J_L^{A} .
\label{HcurrentcurrentN}
\end{equation}
As seen, the stabilization of the flag manifold  $\SU(N)$/U(1)$^{N-1}$  from the action 
(\ref{WZWint}) requires that all the couplings $\lambda_n, n=1, \ldots \left[N/2\right]$ are positive\footnote{Here, it is important to note that each deformation gives the same sign for the current-current interaction. If some couplings were opposite, we had to discuss which deformation dominates the others. However, our prescription given in \eqref{eq:regularization_scheme} is UV-cutoff dependent, and thus it would be dangerous to compare the magnitude between different deformations. It is an interesting question if there is an extension of our prescription that can treat those general cases.} and we have 
$\gamma_{\rm eff}  > 0$ from Eq.~(\ref{couplingfinal}).  
Then, the current-current operator in (\ref{eq:CompactBosonLagrangian_Flag}, \ref{HcurrentcurrentN}) is marginally irrelevant, and a massless $\SU(N)$$_1$ behavior is obtained in the far IR  limit. We thus conclude that the IR properties of the flag nonlinear $\sigma$ model (\ref{flagsigmalagrangian}) with topological angles $\theta_a = 2\pi a/N$ are critical and belong to the $\SU(N)$$_1$ universality class as conjectured in previous studies~\cite{Ohmori-S-S-19, Tanizaki-S-18,Wamer-A-20, Wamer-L-M-A-20}.

Let us check the consistency by flipping the sign of $\lambda_n$, which makes $\gamma_{\mathrm{eff}}<0$. 
Model (\ref{HcurrentcurrentN}) is an integrable field theory with a non-perturbative mass gap when 
$\gamma_{\mathrm{eff}}<0$. The nature of the vacua can be determined by minimizing the $\cos$-term in the effective Lagrangian~\eqref{eq:CompactBosonLagrangian_Flag}. This gives $N$-fold gapped vacua:
\begin{equation}
    \langle \vec{\Phi} \rangle =\sqrt{\pi}k \vec{e}_1, 
\end{equation}
with $k=0,1,\ldots, N-1$. We note that $\langle \vec{\Phi} \rangle =\sqrt{\pi}k \vec{e}_i$ with any $i=1,\ldots, N$ also minimize the potential, but they are gauge equivalent to the above configuration due to $\vec{\Phi}\sim \vec{\Phi}+\sqrt{\pi} \vec{\alpha}_m$ ($\vec{\alpha}_m = \vec{e}_m - \vec{e}_{m+1}$), and thus there are only $N$ physically distinct minima. 
For the gauge-invariant operator $\tr(G)$, we find that 
\begin{align}
	\langle \tr G\rangle \approx \frac{1}{\sqrt{N}}\sum_{\alpha=1}^{N} \rme^{\im \sqrt{4\pi}\vec{e}_\alpha\cdot \langle \vec{\Phi}\rangle}
	=\sqrt{N} \rme^{-2\pi \im k/N},
\end{align}
as $\vec{e}_\alpha\cdot \vec{e}_\beta=\delta_{\alpha \beta}-\frac{1}{N}$ (see Eq. \ref{weightSUN}). 
Up to an overall magnitude, these vacua are consistent with the ones~\eqref{eq:ClassicalVacuaNegativeDT} obtained by the classical analysis for the double-trace terms with $\lambda_n<0$. 

\section{\texorpdfstring{$\SU(N)/\SO(N)$}{SU(N)/SO(N)} nonlinear \texorpdfstring{$\sigma$}{sigma}  model at \texorpdfstring{$\theta=\pi$}{theta=pi}}
\label{sec:SU/SOsigmamodel}
In this section, we investigate the IR properties of the $\SU(N)/\SO(N)$ nonlinear $\sigma$  model at $\theta= \pi$ by a similar approach. For $N=2$, this model reduces to the $\mathbb{C}P^1$ nonlinear $\sigma$ model at $\theta=\pi$ discussed in Sec.~\ref{sec:demo_SU2}. 
For $N\ge 3$, this symmetric space admits a $\mathbb{Z}_2$ topological term since $\pi_2(\SU(N)/ \SO(N) ) \simeq  \mathbb{Z}_2$. In this respect, the value of $\theta$ cannot be tuned at will and its allowed value $\theta=\pi$ is fixed.
The resulting model is expected to be a massless integrable field theory with the $\SU(N)_1$ criticality according to the analysis of Refs.~\cite{Fendley-01,Fendley-JHEP-01,Marino-M-R-23}, and we will examine it using our approach. A comment on the IR properties of the $\SU(N)/\USp(N)$ nonlinear $\sigma$ model for even $N=2k$ will also be given at the end of this section.

\subsection{Connecting the \texorpdfstring{$\SU(N)/\SO(N)$}{SU(N)/SO(N)} nonlinear \texorpdfstring{$\sigma$}{sigma} model at \texorpdfstring{$\theta=\pi$}{theta=pi} and the \texorpdfstring{$\SU(N)_1$}{SU(N) level-1} WZNW model}
The $\SU(N)/\SO(N)$ coset is described as\footnote{This expression follows from the general result of the Cartan embedding: Let $\sigma:G\to G$ be an automorphism and $K=\{g\in G\, |\, g=\sigma(g)\}$. Then, the map  $g\mapsto g\sigma(g^{-1})$ of $G$ to $G$ induces the embedding $G/K\to G$, which is called the Cartan embedding, and thus we can identify $G/K$ as the image of this map. In our case, we can take $\sigma$ as the complex conjugation for unitary matrices, and then $K=\SO(N)$. As $U\sigma(U^{-1})=UU^T$, we find this expression for the coset $\SU(N)/\SO(N)$. We thank the anonymous referee for pointing out this clear explanation. }  
\begin{align}
    \SU(N)/\SO(N)=\{G\in \SU(N) \, | \, \exists U\in \SU(N)\,\,\mathrm{s.t.}\,\, G=UU^{\text{T}}\},  
\end{align}
since $G$ is left invariant under the right $\SO(N)$ action on $U$, $U\mapsto U O$ with $O\in \SO(N)$. 
To obtain it as the classical moduli space, it is convenient to use the fact that $G= UU^{\text{T}}$ for some $U\in \SU(N)$ is equivalent 
to the condition $G=G^{\text{T}}$ \cite{Ohmori-S-S-19}. 
We thus consider the following $\SU(N)_1$ deformed action
to select the $\SU(N)/\SO(N)$ manifold:
\begin{equation}
{\cal S} =  {\cal S}_{{\rm WZNW}_1} + \lambda  \int_{M_2} d^2 x  \;  {\rm Tr} \left[ \left( G - G^{\text{T}}\right) 
\left( G - G^{\text{T} }\right)^{\dagger}\right]  .
\label{WZWintSU(N)/SO(N)}
\end{equation}
In  the strong-coupling regime $ \lambda \rightarrow +\infty$, the classical moduli for $G(x)$ is described as 
\begin{align}
    G(x)=U(x) U(x)^{\text{T}}, 
\label{moduliSO}
\end{align}
where $U(x)$ is ``locally'' an $\SU(N)$-valued scalar field but it is associated with the $\SO(N)$ gauge redundancy, $U(x)\sim U(x) O(x)$. 
This defines an $\SO(N)$ principal bundle on $M_2$. 

As expected, when $N=2$, this model reduces to the one in Sec.~\ref{sec:demo_SU2} by changing the variable. 
For $\SU(2)$ matrices, we have $U^{\text{T}}=\sigma_2 U^\dagger \sigma_2$, and thus the moduli space condition (\ref{moduliSO}) becomes $G=U\sigma_2 U^\dagger \sigma_2$. By changing $G\to G'(\im \sigma_2)$ and $U\to U'\frac{\bm{1}_2+\im \sigma_1}{\sqrt{2}}$, this relation becomes $G'=U' (\im \sigma_3) U'^{\dagger}$ 
which is Eq. (\ref{SU2wzwfieldmanifold}).
Then, we obtain the two-dimensional $\mathbb{C}P^1$ nonlinear $\sigma$ model at $\theta=\pi$ for $N=2$ when $\lambda \to \infty$. 
In the following, we focus on the case $N\ge 3$. 

We can now derive the effective Lagrangian for the $\SU(N)/\SO(N)$ nonlinear $\sigma$ model by plugging $G=UU^T$ into the WZNW action~\eqref{WZW}, and the computation of the kinetic term is straightforward, which gives the standard form obtained by the Callan-Coleman-Wess-Zumino procedure \cite{Coleman-W-Z-69,Callan-C-W-Z-69}. 
The nontrivial part is, again, to determine the topological term from the WZNW term, and the situation turns out to be rather complicated for $N\ge 3$. 
We first give the formal derivation using some mathematical machineries and then provide a more explicit derivation. 

Let us first give the list for the possible topological terms for the $\SU(N)/\SO(N)$ nonlinear $\sigma$ model, which are consistent with the locality and unitarity of relativistic quantum field theories (QFTs)~\cite{Freed:2004yc, Freed:2016rqq, Freed:2017rlk, Lee:2020ojw, Kobayashi:2021qfj}. 
The idea is that the topological term itself can be regarded as an invertible QFT, and one can employ techniques to classify SPT phases~\cite{Wen:2013oza, Kapustin:2014tfa, Freed:2016rqq, Yonekura:2018ufj}:
For bosonic QFTs, this can be classified by the $\SO$-bordism group, $\widetilde{\Omega}^{\SO}_{\bullet}(\SU(N)/\SO(N))$, which is identical to the integral homology (i.e. homology with the integer coefficient)\footnote{Instead of an integral homology, one may use the $\U(1)$-valued cohomology, which has the equivalent data due to the universal coefficient theorem. We shall freely interchange these notions depending on the context.}, $H_{\bullet}(\SU(N)/\SO(N);\mathbb{Z})$, at low degrees. For $N\ge 3$, we have (the table can be found in Ref.~\cite{Lee:2020ojw})
\begin{align}
    H_2(\SU(N)/\SO(N);\mathbb{Z})\simeq \pi_2(\SU(N)/\SO(N))\simeq  \mathbb{Z}_2,\quad 
    H_3(\SU(N)/\SO(N);\mathbb{Z})\simeq 0. 
\end{align}
Therefore, the unique possibility for the topological term of the $\SU(N)/\SO(N)$ nonlinear $\sigma$ model is the discrete $\theta$ term,
\begin{equation}
    \pi w_2(\SO(N)), 
    \label{eq:Z2_theta_term}
\end{equation}
where $w_2(\SO(N))$ is the 2nd Stiefel-Whitney class for the $\SO(N)$ principal bundle.\footnote{The 2nd Stiefel-Whitney class characterizes the obstruction of promoting the $\SO(N)$-bundle to an $\Spin(N)$-bundle:  Taking a good cover $\{U_\alpha\}_{\alpha}$ of the manifold, then we have the $\SO(N)$-valued transition function $g_{\alpha\beta}$ on  each double overlap $U_{\alpha\beta}=U_\alpha\cap U_\beta$, which satisfies $g_{\alpha\beta}g_{\beta\gamma}g_{\gamma\alpha}=1$ in $\SO(N)$ on the triple overlap $U_{\alpha\beta\gamma}=U_\alpha\cap U_\beta\cap U_\gamma$. Take a lift of $\SO(N)$ transition functions to $\Spin(N)$, which we denote as $\tilde{g}_{\alpha\beta}$, then $\tilde{g}_{\alpha\beta}\tilde{g}_{\beta\gamma}\tilde{g}_{\gamma\alpha}=(-1)^{n_{\alpha\beta\gamma}}$ in $\Spin(N)$. The collection of $\{n_{\alpha\beta\gamma}\}$ defines a $\mathbb{Z}_2$-valued $2$-cocycle, and its equivalence class gives $w_2(\SO(N))$.} 
For fermionic QFTs, the classification is given by the $\Spin$-bordism, which can be computed using the Atiyah-Hirzebruth spectral sequence~\cite{Lee:2020ojw}; $\widetilde{\Omega}^{\mathrm{Spin}}_2(\SU(N)/\SO(N))\simeq \mathbb{Z}_2$ and $\widetilde{\Omega}^{\mathrm{Spin}}_3(\SU(N)/\SO(N))$ can only have torsion elements, and thus the only possible two-dimensional topological term is again given by the second Stiefel-Whitney class of the $\SO(N)$ bundle. 

With this knowledge, the remaining task is to demonstrate that the $\SU(N)_1$ WZNW term evaluated on the $\SU(N)/\SO(N)$ moduli  (\ref{moduliSO}) gives the nontrivial element of the $\Spin$-bordism group $\widetilde{\Omega}^{\mathrm{Spin}}_2(\SU(N)/\SO(N))\simeq \mathbb{Z}_2$. Since we can embed the nontrivial configuration for $N=2$ into the larger $N$, it is almost evident that the WZNW term gives the nontrivial $\mathbb{Z}_2$ element~\eqref{eq:Z2_theta_term}. As a consequence, we find that the effective Lagrangian on the classical moduli becomes
\begin{equation}
    {\cal S}_{\mathrm{eff}}=(\text{kinetic term for the $\SU(N)/\SO(N)$ field})+\im\pi w_2(\SO(N)). 
\label{SeffSOtop}
\end{equation}

While the above discussion provides an abstract derivation, let us also present a simpler explanation of the form of the topological term. When we denote $G=UU^T$ to parametrize the $\SU(N)/\SO(N)$-valued field, we should note that $U$ is not necessarily globally defined as the $\SU(N)$-valued field. When we go from one patch to another, $U$ is affected by the $\SO(N)$ gauge transformation, which defines the $\SO(N)$ bundle.  
What we would like to show is that the $\SU(N)_1$ WZNW term on $G=UU^T$ is identical to $\pi w_2(\SO(N))$, and let us make this statement more concrete. 
The group $\SO(N)$ has the nontrivial first homotopy $\pi_1(\SO(N))\simeq \mathbb{Z}_2$, and its universal cover is given by the $\Spin(N)$ group. 
Thus, it is a nontrivial question if the $\SO(N)$ bundle can be lifted to (or regarded as) the $\Spin(N)$ bundle: On two-dimensional closed manifolds, this is possible if and only if the second Stiefel-Whitney class $w_2(\SO(N))$ vanishes. 
Therefore, the problem reduces to checking the following properties: 
\begin{itemize}
    \item[(i)] The WZNW term vanishes mod $2\pi$ when the $\SO(N)$ bundle can be lifted to $\Spin(N)$. 
    \item[(ii)] If the lift does not exist, the WZNW term gives $\pi$ mod $2\pi$. 
\end{itemize}

Let us begin with the statement (i). When the $\SO(N)$ bundle can be lifted to the $\Spin(N)$ bundle on two-dimensional manifolds, there exists a global section as the $\Spin(N)$-bundle on $2$-manifolds is a trivial bundle thanks to $\pi_1(\Spin(N))=0$: That is, we can take $U$ as the globally defined $\SU(N)$ valued field in a suitable gauge choice. 
Using the Polyakov-Wiegmann formula \cite{Polyakov-W-84}, we find 
\begin{equation}
\begin{split}
    {\rm WZNW}[UU^T] &=\underbrace{{\rm WZNW}[U]+{\rm WZNW}[U^T]}_{\text{cancel}}+\frac{1}{8\pi}\int_{M_2}\underbrace{\tr[(U^{-1}\diff U)\wedge (U^{-1}\diff U)^T]}_{=0}    \\
    &=0 \; . 
\end{split}
\end{equation}
Thus, the WZNW term vanishes when the $\SO(N)$ bundle has the lift to the $\Spin(N)$ bundle. 

Next, let us check the statement (ii). The above argument using the Polyakov-Wiegmann identity can be generalized to show that the WZNW term does not change under the small deformation of the fields with $G=UU^T$. 
Therefore, we just have to give an example of the field configuration $G=UU^T$, for which the WZNW term gives $\pi$. 
However, this is already known for the $N=2$ case as $\SU(2)/\SO(2)\simeq \mathbb{C}P^1$ and we can consider the monopole (or hedgehog) configuration, on which the $\SU(2)_1$ WZNW term gives $\pi$. Embedding this $\SU(2)$ configuration into $\SU(N)$ is straightforward as one can just put it in the first $2\times 2$ block for the identity matrix, and then the $\SU(N)_1$ WZNW term gives $\pi$ for that configuration. 
This completes the pedagogical derivation for the effective Lagrangian (\ref{SeffSOtop}). 

\subsection{Global symmetry and 't~Hooft anomaly}
Here, let us discuss the global symmetry of model (\ref{WZWintSU(N)/SO(N)}) and the 't~Hooft anomaly for the $\SU(N)/\SO(N)$ nonlinear $\sigma$ model at $\theta=\pi$. The structure of the global symmetry is distinct between even and odd $N$:
\begin{align}
    \text{Even $N$ : }& \quad \frac{\SU(N)_{\tilde{V}}\times (\mathbb{Z}_N)_L}{(\mathbb{Z}_2)_{\tilde{V}}\times \mathbb{Z}_{N/2}}\rtimes \mathbb{Z}_2, \label{SOeven} \\
    \text{Odd $N$ : }& \quad \SU(N)_{\tilde{V}}\rtimes \mathbb{Z}_2.  \label{SOodd}
\end{align}
Here, $\SU(N)_{\tilde{V}}$ is the anomaly-free subgroup of $\SU(N)_L\times \SU(N)_R$ defined by the embedding, $\SU(N)_{\tilde{V}}\ni V\hookrightarrow (V,V^{*})\in \SU(N)_L\times \SU(N)_R$, and it acts on the WZNW field as 
\begin{align}
    G(x)\mapsto V G(x) V^{\text{T}}.
\end{align}
It is important to notice that this is different from the vector-like $\SU(N)_V$ symmetry considered for the flag-manifold nonlinear $\sigma$ model. 
The deformation~\eqref{WZWintSU(N)/SO(N)} also preserves the discrete chiral symmetry $(\mathbb{Z}_N)_L$, $G\mapsto \rme^{2\pi \im/N}G$, but it has an overlap with $\SU(N)_{\tilde{V}}$: The center of $\SU(N)_{\tilde{V}}$ acts as $G\to \rme^{4\pi \im/N}G$, and thus $(\mathbb{Z}_N)_L$ is completely included when $N$ is odd while only $(\mathbb{Z}_{N/2})_L\subset (\mathbb{Z}_N)_L$ is included for even $N$. In this regard, for even $N$, it is convenient to define $\widetilde{\SU(N)}_{\tilde{V}}=[\SU(N)_{\tilde{V}}\times (\mathbb{Z}_N)_L]/\mathbb{Z}_{N/2} = \{U\in \U(N) \,|\, \det U=\pm 1\}$, and we can treat their actions in a simple way as $G\to V G V^T$ with $V\in \widetilde{\SU(N)}_{\tilde{V}}$. 
For even $N$, $-\bm{1}_N\in \SU(N)_{\tilde{V}}$ acts trivially on $G$, so we have to take another $\mathbb{Z}_2$ quotient ($(\mathbb{Z}_2)_{\tilde{V}}$). Combining these and by adding the 
$ \mathbb{Z}_2$ complex conjugation symmetry $G\mapsto G^*$, we obtain the above symmetry groups
(\ref{SOeven}, \ref{SOodd}) for even and odd $N$. 

After decomposing $G=UU^T$, we need to describe the symmetry transformation on the $U$ field. First, $V\in \SU(N)_{\tilde{V}}$ acts as 
\begin{equation}
    U(x)\mapsto V U(x). 
\end{equation}
When $N$ is even, the $\SO(N)$ gauge redundancy, $U(x)\sim U(x) O(x)$, can absorb $-1$, and we can understand the $\mathbb{Z}_2$ quotient. For odd $N$, $\SO(N)$ does not have a nontrivial center, so such quotient structure does not appear. What is somewhat nontrivial is the action of $\rme^{\frac{\pi \im}{N}}\bm{1}_N \in \widetilde{\SU(N)}_{\tilde{V}}$ (or $\rme^{2\pi \im/N}\in (\mathbb{Z}_N)_L$) for even $N$: 
\begin{equation}
    U(x)\mapsto \rme^{\frac{\pi \im}{N}}U(x) C, 
\end{equation}
where $C\in O(N)$ with $\det C=-1$ gives the outer automorphism on the $\SO(N)$ gauge transformation. 
The $\mathbb{Z}_2$ complex conjugation, $G\mapsto G^*$, is given by $U(x)\mapsto U(x)^*$. 

Next, we discuss the 't~Hooft anomaly of this symmetry. Let us first show that there is no 't~Hooft anomaly for odd $N$. For this purpose, we consider the symmetry-preserving local perturbations that makes the ground state unique and gapped. 
This can be achieved by regarding the $\SU(N)/\SO(N)$ nonlinear $\sigma$ model as the $\SO(N)$ gauge theory coupled to the $\SU(N)$ scalar field, and then the above discrete $\theta$ angle for the nonlinear $\sigma$ model becomes that of $2$d $\SO(N)$ gauge theory. 
Then, we can take the linear $\sigma$ model realization for the $\SU(N)$-valued scalar and deform its potential to the massive one instead of the wine-bottle-type potential, which ends up with the $2$d pure $\SO(N)$ gauge theory at $\theta=\pi$ as the low-energy effective theory of the deformed model. 
The Hilbert space of the $2$d pure gauge theory is solely given by the Lorentz-invariant state with the Wilson line (or test quark) at spatial infinities, whose energy density is specified the quadratic Casimir of the Wilson-line representation. 
Taking $\theta=\pi$ for $\SO(N)$ gauge theory is equivalent to restrict those quark representations to the projective ones, which are representations of $\Spin(N)$ without the promotion to $\SO(N)$ representations. 
Since $\SO(N)$ with odd $N$ has the unique spinor representation, we get the unique gapped vacuum, which shows that the 't~Hooft anomaly is absent. 

If we perform the same deformation for even $N$, we obtain two-fold gapped vacua because $\SO(N)$ with even $N$ has two inequivalent spinor representations, which are exchanged by the outer automorphism $C$, and thus we have the spontaneous breaking,
${\widetilde{\SU(N)}_{\tilde{V}}}/{\mathbb{Z}_2}\xrightarrow{\mathrm{SSB}} {\SU(N)_{\tilde{V}}}/{\mathbb{Z}_2}$. 
Indeed, we can show that there is the $\mathbb{Z}_2$ 't~Hooft anomaly for the global symmetry, $\widetilde{\SU(N)_{\tilde{V}}}/\mathbb{Z}_2$: Its $3$d SPT action can be formally written as 
\begin{align}
    \calS_{\text{$3$d SPT}}=\pi \int_{M_3} C\cup w_2(\SU(N)_{\tilde{V}}/\mathbb{Z}_2), 
\end{align}
where $C$ denotes the $\mathbb{Z}_N$ gauge field for the action $U\to \rme^{\pi \im/N}U C$ and $w_2(\SU(N)/\mathbb{Z}_2)$ is the two-form gauge field characterizing the obstruction of lifting the $\SU(N)_{\tilde{V}}/\mathbb{Z}_2$ bundle to the $\SU(N)_{\tilde{V}}$ bundle. 
Maybe, the simplest way to derive it is again to consider the pure $\SO(N)$ gauge theory limit of the above deformation, and the consideration of this limit is sufficient as the 't~Hooft anomaly is preserved under any local symmetric deformations of the theory. In this limit, gauging of $\SU(N)_{\tilde{V}}/\mathbb{Z}_2$ is equivalent to gauge the $\mathbb{Z}_2$ one-form symmetry of the $\SO(N)$ gauge theory, which introduces the $\mathbb{Z}_2$ two-form gauge field, $B=w_2(\SU(N)_{\tilde{V}}/\mathbb{Z}_2)$. 
This has the nontrivial interplay with the dynamical two-form gauge field $w_2(\SO(N))$, and its detail depends on whether $N=4k+2$ or $N=4k$ because the center of $\Spin(N)$ is $\mathbb{Z}_4$ for $N=4k+2$ but $\mathbb{Z}_2\times \mathbb{Z}_2$ for $N=4k$. In both cases, the outer automorphism $C:U\to \rme^{\pi\im/N}UC$ has the same action on $w_2(\SO(N))$ as 
\begin{align}
    C: w_2(\SO(N))\mapsto w_2(\SO(N))+B,
\end{align}
and thus it gives the $\mathbb{Z}_2$ anomaly at $\theta=\pi$ given by inflow of the above SPT action.

\subsection{Massless RG flow from the \texorpdfstring{$\SU(N)/\SO(N)$}{SU(N)/SO(N)} model at \texorpdfstring{$\theta=\pi$}{theta=pi} to the \texorpdfstring{$\SU(N)_1$}{SU(N) level-1} CFT}
\label{sec:SUSOsigma_masslessRGflow}
Here, we study the action (\ref{WZWintSU(N)/SO(N)}) in the vicinity of the $\SU(N)_1$ WZNW CFT and discuss the massless RG flow from the $\SU(N)/\SO(N)$ nonlinear $\sigma$ model at $\theta=\pi$. 
We can rewrite the action (\ref{WZWintSU(N)/SO(N)}) as
\begin{eqnarray}
{\cal S} =  {\cal S}_{{\rm WZNW}_1} - 2 \lambda  \int_{M_2} d^2 x  \;  {\rm Tr} \left[ G G^{*}\right]  , 
\label{WZWintSU(N)/SO(N)bis}
\end{eqnarray}
up to an unimportant constant. 
Next, we express the potential part in terms of the fields of the $\SU(N)$$_1$ CFT by computing the OPE after separating the points: 
\begin{align}
	\tr(GG^*(x))\Rightarrow \frac{1}{\varepsilon^2}\int \diff^2 r  |r|^{2(N-1)/N}\rme^{-|r|/\varepsilon}\tr[G(x-r/2)G^*(x+r/2)]. 
\end{align} 
As the potential preserves the  $(\mathbb{Z}_N)_L$ transformation, $G \rightarrow e^{i 2  \pi/N} G$, no relevant primary fields can be generated by this procedure. 
Therefore, up to irrelevant operators, one is left with $\SU(N)_{\tilde{V}}$-invariant  current-current contributions
\begin{align}
    \tr(J_L J_R^{\text{T}})=T^A_{\alpha \beta}T^B_{\alpha \beta}J^A_L J^B_R \; . 
\end{align}
As has been done in Sec.~\ref{sec:flagsigmamodel}, the precise identification of the perturbing field can be derived by means of the Abelian bosonization approach (\ref{freefieldrepWZWGfield}) of the $G$ field in terms of $N-1$ chiral bosonic fields ${\vec \Phi}_{R,L}$. In particular, in Appendix \ref{sec:AppendixB}, we explicitly show that the leading nontrivial contribution of OPE is given as
\begin{equation}
\nord{ {\rm Tr} (G G^{*} ) }  =  - \frac{16 \pi^2}{N} T^{A}_{\alpha\beta} T^{B}_{\alpha\beta} J_L^{A} J_R^{B} 
\label{perturbSUN/SON}
\end{equation}
[the symbol $\nord{\! AB}$ denotes the first non-constant term in the OPE $A(z, \bar z) B(0, 0)$; see Eq.~\eqref{TraceGGdagApp}].  
With this result in hands, we deduce the leading contribution of the Hamiltonian density for model (\ref{WZWintSU(N)/SO(N)bis})  in the general $N$ case: 
\begin{equation}
 \mathcal{H}_{N} = \frac{2\pi }{N+1} \left( : J^A_{R} J^A_{R}: + : J^A_{L} J^A_{L}: 
\right)   + 2 \gamma_{\rm eff}   \; T^{A}_{\alpha\beta} T^{B}_{\alpha\beta} J_L^{A} J_R^{B}  \;  ,
\label{HcurrentcurrentSUN/SON}
\end{equation}
with  $\gamma_{\rm eff}  =  \frac{192 \pi^3\lambda  \epsilon^2}{N}$.
The perturbation in Eq. (\ref{HcurrentcurrentSUN/SON}) is different from the one (\ref{HcurrentcurrentN}) obtained for the flag nonlinear $\sigma$ model 
in that we now contract the currents with the metric $T^{A}_{\alpha\beta} T^{B}_{\alpha\beta}$ instead of $\delta^{AB}$. Such a marginal deformation of 
the $\SU(N)_{1}$ WZNW model has already been discussed by two of us in the context of $\SU(N)$ self-dual sine-Gordon models \cite{Lecheminant-T-06-SDSG}. Its IR effect can be elucidated by a simple one-loop RG approach \cite{Lecheminant-T-06-SDSG,Itoi-K-97} or by a non-perturbative duality approach \cite{Boulat-A-L-09}. 

Let us first consider the one-loop RG approach. Defining $g =\gamma_{\rm eff} $, the one-loop beta function for the model (\ref{HcurrentcurrentSUN/SON}) is given by \cite{Lecheminant-T-06-SDSG}:
\begin{equation}
{\dot g} = - \frac{N}{4\pi} g^2, 
\label{1loopSUN/SON}
\end{equation}
so that $g(l) =\frac{4 \pi g(0)}{4 \pi+g(0)N l}$ and  $g(l)$
goes to zero in the IR regime ($l \rightarrow \infty$) when $g (0) >0$, i.e., $\lambda >0$, whereas it diverges when $g (0) <0$.  This leads us to conclude that the two-dimensional $\SU(N)/ \SO(N)$  nonlinear $\sigma$ model at $\theta= \pi$ enjoys the same IR massless $\SU(N)_1$ criticality as the flag nonlinear $\sigma$ model though the directions of the marginally irrelevant deformations are expected to be different for these two models 
when $N>2$ (Fig.~\ref{fig:NLsigma-vs-flow}). This may explain the fact that whereas the $\SU(N)/ \SO(N)$ nonlinear $\sigma$ model at $\theta= \pi$ is integrable \cite{Fendley-01,Fendley-JHEP-01}, the flag nonlinear $\sigma$ model (\ref{flagsigmalagrangian}) with topological angles $\theta_a = 2\pi a/N$ is apparently not integrable \cite{Komatsu-M-S-19}.  
 
One can go beyond the one-loop RG approach by considering the compact boson effective Lagrangian.
For this purpose, we have to introduce the $T$-dual field, whose periodicity is given by
\begin{align}
    \vec{\Theta}\sim \vec{\Theta}+\sqrt{4\pi}\vec{e}_m. 
    \label{eq:ThetaPeriodicity}
\end{align}
In terms of the chiral boson $\vec{\Phi}_{L,R}$, the original field is given by $\vec{\Phi}=\vec{\Phi}_L+\vec{\Phi}_R$ and the dual field is given by $\vec{\Theta}=\vec{\Phi}_L- \vec{\Phi}_R$. 
Then, the effective Lagrangian can be denoted as 
\begin{align}
    \mathcal{L}_{\mathrm{eff}} =\frac{1}{2}(\partial_\mu \vec{\Theta})^2 
    + \gamma_{\rm eff}  \left\{ 
    -\frac{1}{8\pi}(\partial_\mu \vec{\Theta})^2+\frac{1}{4\pi^2}\sum_{1 \le \alpha<\beta \le N}\cos\left(\sqrt{4 \pi}(\vec{e}_\alpha-\vec{e}_\beta)\cdot \vec{\Theta}\right)\right\}  \; ,
    \label{eq:SU/SOLagrangian_CompactBoson}
\end{align}
which takes  exactly the same form as the flag nonlinear $\sigma$ model~\eqref{eq:CompactBosonLagrangian_Flag} by replacing $\vec{\Theta}\Rightarrow\vec{\Phi}$. 
In terms of the current algebra, this formal equivalence is best seen by considering 
a duality transformation \cite{Boulat-A-L-09}:
\begin{eqnarray}
J_L^{A} &\Rightarrow&  {\tilde J}_L^{A}  = K^{AB} J_L^{B} , \nonumber \\
J_R^{A} &\Rightarrow&  {\tilde J}_R^{A}  = J_R^{A} ,
\label{currentrans}
\end{eqnarray}
where $K^{AB}=-2 T^{A}_{\alpha \beta} T^B_{\alpha \beta}$.  Written explicitly, $\tilde{J}_L^A$ is given by 
\begin{align}
    {\tilde J}_L^{A} = \left\{
    \begin{array}{cl}
       J_L^{A}  &  (\text{$A \in $ Anti-symmetric ($\SO(N)$) parts}),\\
      -J_L^{A}  &  (\text{$A \in$ Symmetric, Diagonal parts}) \; ,
    \end{array}\right.
\label{currentransdual}
\end{align}
where we have used the fact that the $\SU(N)$ generators $T^{A}$ can be decomposed into three distinct
classes of matrices: antisymmetric, symmetric, and diagonal.  
In terms of the new currents ${\tilde J}_L^{A}$ that satisfy the same $\SU(N)_1$ Kac-Moody algebra \eqref{OPESUN1curr} as the original ones, 
the model \eqref{HcurrentcurrentSUN/SON} now reads as follows:  
\begin{equation}
 \mathcal{H}_{N} = \frac{2\pi }{N+1} \left( : {\tilde J}^A_{R} {\tilde J}^A_{R}: + : {\tilde J}^A_{L} {\tilde J}^A_{L}: 
\right)   - \gamma_{\rm eff}  {\tilde J}_L^{A} {\tilde J}_R^{A} \; ,
\label{HcurrentcurrentSUN/SONdual}
\end{equation}
which takes the form of an integrable field theory in terms of the new  $\SU(N)_1$ currents.
We thus observe the equivalence of the ``local physics'' between the model \eqref{HcurrentcurrentSUN/SON} and the flag nonlinear $\sigma$ model \eqref{HcurrentcurrentN} after the duality transformation \eqref{currentrans}. Both models share the same IR fixed point, i.e., the $\SU(N)_1$ WZNW CFT, on the marginally irrelevant side. 
We note, however, that the duality transformation \eqref{currentrans} does not respect the periodicity of the fields, and we cannot naively apply it when discussing the global aspects of the physics. The correct treatment of the periodicity of the fields becomes important on the marginally {\em relevant} side as we will see in the next subsection.

To the best of our knowledge, unlike in the $\mathbb{C}P^1$ and $\SU(N)/\U(1)^{N-1}$ cases, no simple physical systems are known so far from which the $\theta=\pi$ $\SU(N)/ \SO(N)$ nonlinear $\sigma$ model is derived in the low-energy limit.  
Nevertheless, our approach may give a useful hint to look for such systems.  
In fact, as has been pointed out in Ref.~\cite{Lecheminant-T-06-SDSG}, the effective Hamiltonian \eqref{HcurrentcurrentSUN/SON} describes 
low-energy physics of several interesting lattice spin systems.  For instance, the criticality of one of the phase boundaries found in a certain frustrated spin-1 system \cite{Corboz-L-T-T-07} is described precisely by \eqref{HcurrentcurrentSUN/SON}.   Therefore, those systems may provide us with a good starting point 
to derive the $\SU(N)/ \SO(N)$ nonlinear $\sigma$ model from actual physical systems.  
\subsection{Gapped phases and \texorpdfstring{$\SU(N)/\USp(N)$}{SU(N)/USp(N)} nonlinear $\sigma$ model}
\label{sec:SUSp_massiveRGflow}
The effect of the marginal interaction $- \lambda \tr(J_L J_R^{\text{T}})$ is very different depending on the sign of $\lambda$; the system becomes gapless for $\lambda>0$ but gapped for $\lambda<0$. 
In the previous section, we have seen that the $\lambda>0$ side corresponds to a massless flow from the $\theta=\pi$ $\SU(N)/\SO(N)$ nonlinear $\sigma$ model.   
In this section, we consider the vacuum properties on the opposite side $\lambda<0$.  
To this end, the first important task for studying $\lambda<0$ is to correctly identify the number of gapped vacua and the discrete symmetry breaking. 

We begin by identifying the periodicity of the compact bosons more carefully. For this purpose, we note that the WZNW field $G$ is given by the combination of the original field $\vec{\Phi}$ and its $T$-dual field $\vec{\Theta}$ as
[see Eq.~\eqref{freefieldrepWZWGfield}]
\begin{align}
    G_{ji}\sim \exp\left\{ \im \sqrt{\pi} \left[(\vec{e}_i+\vec{e}_j)\cdot \vec{\Phi}+(\vec{e}_i-\vec{e}_j)\cdot \vec{\Theta}\right] \right\}  \; .
\end{align}
The gauge redundancy must maintain $G$ invariant, and thus the correct periodicity of the fields is identified as
\begin{align}
    \begin{pmatrix}
        \vec{\Phi}\\
        \vec{\Theta}
    \end{pmatrix}
    \sim \begin{pmatrix}
        \vec{\Phi}\\
        \vec{\Theta}
    \end{pmatrix}
    +\sqrt{\pi}\begin{pmatrix}
        \vec{\alpha}_k\\
        \vec{\alpha}_k
    \end{pmatrix}, \quad 
        \begin{pmatrix}
        \vec{\Phi}\\
        \vec{\Theta}
    \end{pmatrix}
    \sim \begin{pmatrix}
        \vec{\Phi}\\
        \vec{\Theta}
    \end{pmatrix}
    +\sqrt{4\pi}\begin{pmatrix}
        0\\
        \vec{e}_k
    \end{pmatrix}. 
\end{align}
The second one is nothing but \eqref{eq:ThetaPeriodicity}. However, the periodicity for the  $\vec{\Phi}$ field, \eqref{eq:PhiPeriodicity}, should also transform the $\vec{\Theta}$  field, and this plays the pivotal role to correctly find the physically distinct vacua when $\vec{\Theta}$  acquires the vacuum expectation value. 

\begin{figure}[t]
\begin{center}
\includegraphics[scale=0.7]{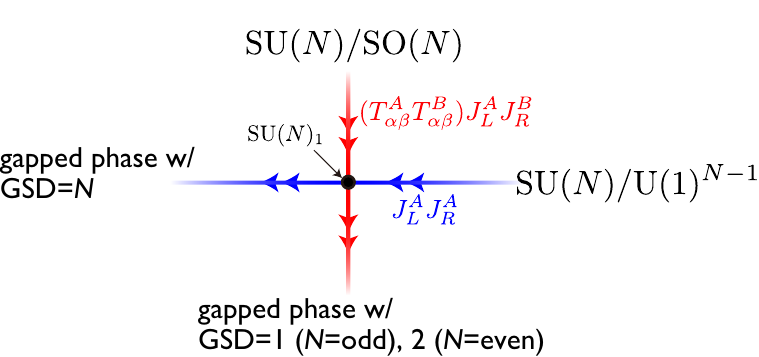}
\end{center}
\caption{Two directions of marginal deformations $J^A_{L} J^A_{L}$ and $T^A_{\alpha \beta}T^B_{\alpha \beta}J^A_L J^B_R$ 
around the $\SU(N)_{1}$ fixed point and the corresponding $\mathrm{G}/\mathrm{H}$ nonlinear $\sigma$ models [$\mathrm{G}=\SU(N)$].  
When the IR phases are gapped, the ground-state degeneracy (GSD) is also shown. 
\label{fig:NLsigma-vs-flow}}
\end{figure}

Now, let us analyze the effective Lagrangian \eqref{eq:SU/SOLagrangian_CompactBoson} for $\lambda<0$, i.e., 
$\gamma_{\rm eff} < 0$. 
To minimize the potential term, we find the vacua labeled by 
\begin{eqnarray}
    \langle \vec{\Theta}\rangle = \sqrt{\pi}\sum_{k}n_k \vec{e}_k. 
\end{eqnarray}
The periodicity by the root vector tells that physically distinct vacua must have different $N$-ality, and thus all the vacua with the same $\sum_k n_k$ (mod $N$) are the same. 
Moreover, we have to identify $\vec{\Theta}\sim \vec{\Theta}+2\vec{e}_1$, and thus the physically distinct vacua must have different $\sum_{k}n_k$ in mod $\gcd(N,2)$. 
As a result, we have for $\lambda<0$:
\begin{itemize}
    \item Odd $N$: The unique gapped vacuum, and no spontaneous symmetry breaking occurs. 
    \item Even $N$: Two-fold gapped vacua, and there is the spontaneous breaking, $\widetilde{\SU(N)}\xrightarrow{\mathrm{SSB}}\SU(N)$. 
\end{itemize}
We summarize the RG flows corresponding to two distinct current-current deformations of the $\SU(N)_1$ WZNW CFT in Fig.~\ref{fig:NLsigma-vs-flow}. 

We now investigate the IR properties of the $\SU(2k)/\USp(2k)$ nonlinear $\sigma$ model. To this end, we return to the WZNW description with the $\tr(GG^*)$ deformation \eqref{WZWintSU(N)/SO(N)bis},  the negative $\lambda$ amounts to considering the following action instead of \eqref{WZWintSU(N)/SO(N)}:
\begin{equation}
{\cal S} =  {\cal S}_{{\rm WZNW}_1} 
+ | \lambda |  \int_{M_2} d^2 x  \;  {\rm Tr} \left[ \left( G + G^{\text{T}}\right) \left( G + G^{\text{T} }\right)^{\dagger}\right]  \; , 
\end{equation}
which means that the classical moduli for even $N$ is given by\footnote{We note that this condition cannot be achieved for odd $N$. Taking the determinant of both sides, we get $\det(G^T)=\det(-G)=(-1)^N\det(G)$. Since $G\in \SU(N)$, this gives $(-1)^N=1$, so the classical moduli for odd $N$ cannot be expressed in this way. } 
\begin{align}
    G^{\text{T}}=-G \; . 
\end{align}
As $G$ on the classical moduli is an antisymmetric $\SU(N)$ matrix, we can define its Pfaffian and it becomes $\pm 1$. Therefore, the moduli space for even $N$ ($N=2k$) has two disconnected components:  
\begin{equation}
    \{U \mathcal{J}_+ U^{\text{T}}\,|\, U\in \SU(N) \} \sqcup \{U \mathcal{J}_- U^{\text{T}}\,|\, U\in \SU(N)\} \; ,
    \label{eq:ClassicalModuli_SU/USp}
\end{equation}
where $\mathcal{J}_+=\bm{1}_{k}\otimes \im\sigma_2=\begin{pmatrix}
    \bm{0} & \bm{1}_{k}\\
    -\bm{1}_{k} & \bm{0}
\end{pmatrix}$ is the metric of the $\USp(2k)$ group with the positive Pfaffian, and $\mathcal{J}_-=\mathrm{diag}(1,\ldots, 1,-1)\otimes \im\sigma_2$ is the counterpart with negative Pfaffian.  The gauge redundancy is given by the $\USp(2k)$ transformation, $U\to U V$ with $V\mathcal{J}V^T=\mathcal{J}$. Therefore, the low-energy effective theory is given by the double copy of the $\SU(2k)/\USp(2k)$ nonlinear $\sigma$ model. 
We note that the $\SU(2k)/\USp(2k)$ nonlinear $\sigma$ model does not have the $\theta$ parameter, and thus it is natural to suspect that it has the unique gapped vacuum.  This nonlinear  $\sigma$ model is known to be a massive integrable field theory whose S-matrix and thermodynamic Bethe ansatz equations have been obtained \cite{Babichenko-03,Babichenko-T-03}.
Since  each choice of the sign in \eqref{eq:ClassicalModuli_SU/USp} gives a gapped vacua, we have the two-fold gapped ground state predicted by the above anomaly discussion when $N=2k$.  We thus conclude that 
the two-dimensional  $\SU(2k)/\USp(2k)$ nonlinear $\sigma$ model is a massive field theory with a unique ground state.

\section{Grassmannian and \texorpdfstring{$\mathbb{C}P^{N-1}$}{CP(N-1)}  nonlinear \texorpdfstring{$\sigma$}{sigma} models at \texorpdfstring{$\theta= \pi$}{theta=pi}}
\label{sec:Grassmanniansigmamodel}
In this section, we examine the IR properties of the two-dimensional nonlinear $\sigma$ model at $\theta= \pi$ based on the complex Grassmannian target space  Gr($N$, $k$) =  $ \frac{\rm{U}(N)}{\rm{U}(k)  \times \rm{U}(N-k)}$.
Specifically, we focus on the cases of  U($2k$)/U($k$)$\times$ U($k$) and $\mathbb{C}P^{N-1}$   models since these field theories enjoy many interesting applications  in condensed matter physics and have been extensively studied over the years.

\subsection{Generalities}
Complex Grassmannian nonlinear $\sigma$  models have been introduced mainly as a generalization of the $\mathbb{C}P^{N-1}$  $=$ Gr($N$, 1) nonlinear  $\sigma$ model which is a field theory of $N$-component complex $z$ fields constrained to $z^{\dagger} z = 1$ \cite{DAdda-L-V-78,MacFarlane-79,Pisarski-79,Dubois-Violette-G-79,Brezin-I-Z-Z-79,Brezin-H-Z-80}. This class of nonlinear $\sigma$ models is based on the complex Grassmannian target space  Gr($N$, $k$) and are described by a $N\times N$ Hermitian projector matrix $P$ such that $P^{\dagger} = P = P^2$, and   ${\rm Tr} P = k$. The two-dimensional Euclidean action of this model reads as follows: 
\begin{equation}
 {\cal S}_{{\rm Gr}(N, k)} = \frac{1}{2 g^2} \int_{M_2} d^2 x \; {\rm Tr} \; \left( (\partial_{\mu} P )^2  \right) 
+ \frac{ \theta}{4\pi}   \int_{M_2} d^2 x \; \epsilon^{\mu \nu} 
 {\rm Tr} \; (P \partial_{\mu} P \partial_{\nu} P)  \; ,
\label{ComplexGrassmann}
\end{equation}
where a $\theta$ term has been added since $\pi_2 ({\rm Gr}(N, k)) =\mathbb{Z} $.   The model \eqref{ComplexGrassmann} is invariant under a global P$\SU(N)$ symmetry: $ P \rightarrow V P V^{\dagger}$, $V$ being an $\SU(N)$ matrix. At $\theta= 0, \pi$, it posseses an additional $\mathbb{Z}_2 $ charge conjugation symmetry C: $P \rightarrow P^{*}$. A local U$(k)$ gauge redundancy becomes manifest if we introduce a parametrization of $P = \Phi \Phi^{\dagger}$ in terms of  a  $N \times k$ complex matrix-valued complex fields $\Phi$ satisfying $ \Phi^{\dagger} \Phi = \bm{1}_k$. 

The action (\ref{ComplexGrassmann}) can be rewritten as a field theory of $k$ $N$-component complex scalar fields 
$ \Phi_{ a l} = (\vec \Phi_l)_a$, with $a=1, \ldots, N$ and $l=1, \ldots, k$ with the constraint ${\vec \Phi}^{*}_l \cdot {\vec \Phi}_{l^{'}} = \delta_{l l^{'}}$. This is a natural  generalization of the $\mathbb{C}P^{N-1}$ model with the U(1) gauge invariance to the one with a local U($k$) non-Abelian gauge invariance: ${\vec \Phi}_l \rightarrow   U_{m l}(x) {\vec \Phi}_m$ ($U$ is an U($k$) matrix so that $P$ is left invariant). The global P$\SU(N)$ symmetry $ P \rightarrow V P V^{\dagger}$ of the model \eqref{ComplexGrassmann} is now described by: $ (\vec \Phi_l)_a  \rightarrow   V_{a b}  (\vec \Phi_l)_b$  whereas
the $\mathbb{Z}_2 $ charge symmetry C becomes $\Phi_{ a l} \rightarrow \Phi_{ a l}^{*}$. The action (\ref{ComplexGrassmann})  reads then as follows:  
\begin{equation}
 {\cal S}_{{\rm Gr}(N, k)}  = \frac{1}{g^2} \int_{M_2} d^2 x \; \left[ \partial_{\mu} {\vec \Phi}^{*}_l \cdot 
  \partial_{\mu} {\vec \Phi}_l  
  +  \left({\vec \Phi}^{*}_l \cdot \partial_{\mu} {\vec \Phi}_{l^{'}} \right)
   \left({\vec \Phi}^{*}_{l^{'}} \cdot \partial_{\mu} {\vec \Phi}_{l} \right) \right] 
  - \frac{ \theta}{2\pi}   \int_{M_2} d^2 x \; \epsilon^{\mu \nu}  \partial_{\mu} {\vec \Phi}^{*}_l 
   \cdot  \partial_{\nu} {\vec \Phi}_l  \; .
\label{ComplexGrassmannfields}
\end{equation}

Grassmannian nonlinear $\sigma$  models (\ref{ComplexGrassmannfields}) with a $\theta$ term  turns out to have many interesting applications in condensed matter physics. As already discussed in the introduction, the Gr($2k$, $k$) nonlinear $\sigma$ model at $\theta= \pi$  when $k\rightarrow 0$ is expected to describe the integer quantum Hall plateau phase transition.  This field theory also captures the low-energy properties of an SU($2k$) Heisenberg spin chain with explicit dimerization  where the spin operators belong to self-conjugate fully antisymmetric representation of the SU($2k$) group \cite{Pruisken-D-S-22,Nguyen-S-23}.  Similarly, two-dimensional $\mathbb{C}P^{N-1}$  nonlinear $\sigma$  model with $\theta=\pi p$ can be realized by considering a $p$-leg $\SU(N)$ Heisenberg spin ladder with alternating Heisenberg spin chains in the $\bolN$-$\bar{\bolN}$ representations of the $\SU(N)$ group \cite{Beard-P-R-W-05,Laflamme-E-al-16,Laflamme-E-al-16,Evans-G-H-W-18}. The latter spin lattice system might be realized experimentally using alkaline-earth ultracold atoms \cite{Laflamme-E-al-16}. Other related realization 
is  $\SU(N)$ Heisenberg spin chain with alternating $\SU(N)$ irreps described by symmetric representation with $p$ boxes
on even sites and its conjugate representation on odd sites \cite{Affleck-85,Read-S-NP-89,Nguyen-M-T-U-23}.

The IR properties of  $\mathbb{C}P^{N-1}$  nonlinear $\sigma$  model with a $\theta=\pi$ term are rather well understood. For even $N$, the model exhibits a mixed anomaly between P$\SU(N)$ and the $\mathbb{Z}_2 $ charge conjugation C symmetry  \cite{Sulejmanpasic-T-18,Ohmori-S-S-19,Nguyen-M-T-U-23}. The IR physics  is then non-trivial as the result of this mixed anomaly: a trivial fully gapped phase with a unique ground state is not allowed. By contrast, the vacuum at $\theta=0$ is trivial for even $N$. As discussed in Sec. II, the $\mathbb{C}P^{1}$ nonlinear $\sigma$  model at $\theta= \pi$ is massless in the SU(2)$_1$ universality class.  However, this massless behavior at $\theta= \pi$ does not generalize to larger value of $N$.  In contrast, for even $N>2$, a massive phase is formed with a two-fold ground-state degeneracy which stems from the spontaneous symmetry breaking of  the $\mathbb{Z}_2 $ charge conjugation symmetry \cite{Seiberg-84,Beard-P-R-W-05}.

In the odd-$N$ case, the global symmety does not have a mixed anomaly but there is a global inconsistency between the $\theta=0$ and $\theta=\pi$ theories \cite{Kikuchi-T-17,Komargodski-S-T-19,Nguyen-M-T-U-23}: The gauged action cannot be modified by a local symmetric counterterm such that both $\theta=0, \pi$ are anomaly free. 
The global inconsistency condition tells (i) the physics at $\theta=\pi$ has a nontrivial IR physics as in the case of even $N$, or (ii) if the system at $\theta=\pi$ is trivially gapped, then it has to be in a different SPT state compared with the $\theta=0$ theory.
The various previous analysis predict that the $\mathbb{C}P^{N-1}$ nonlinear $\sigma$ model at $\theta=\pi$ has two-fold degenerate gound states that realize the scenario (i), and they include a large-$N$ analysis, a strong-coupling approach, and a numerical investigation for $N=3$ \cite{DAdda-L-V-78,Seiberg-84,Beard-P-R-W-05}. This observation is compatible with the fact that the $\SU(N)$ Heisenberg spin chain with alternating $\bolN$-$\bar{\bolN}$ representations, whose low-energy properties are governed by the $\mathbb{C}P^{N-1}$   nonlinear $\sigma$  model at $\theta=\pi$,  has a fully gapped phase with a two-fold ground-state degeneracy, a conclusion derived from exact results \cite{Affleck-SUN-90,Klumper-90}. As a function of $\theta$, the model displays a first-order phase transition at $\theta=\pi$ as  first predicted in Refs. \cite{DAdda-L-V-78,Seiberg-84}.

In the  general complex Grassmannian ${\rm Gr}(N, k)$ case,  there is a mixed anomaly between P$\SU(N)$ and 
the charge conjugation C symmetry only when $N$ is even, $k$ odd and $\theta= \pi$ \cite{Dunne-T-U-18}.  
A global inconsistency between the $\theta=0$ and $\theta=\pi$ theories is found in the remaining cases \cite{Dunne-T-U-18}.  When $N$ is even and $k=N/2$, a  fully gapped phase with a two-fold ground state degeneracy is expected on general grounds \cite{Affleck-88}. The latter model with $\theta=\pi$ arises as the long-distance effective field theory of the $\SU(N)$ Heisenberg spin chain where the spins belong to the fully antisymmetric self-conjugate representation. This lattice model is known to be dimerized with a two-fold ground state degeneracy which stems from the spontaneous breaking of the one-step translation symmetry \cite{Affleck-88,Read-S-NP-89,Nonne-L-C-R-B-11,Dufour-N-M-15}. In the $N=4$ case ($k=2$), the existence of this 
doubly-degenerate massive phase has been reported in various different numerical works \cite{Onufriev1999,Assaraf-A-B-C-L-04,Paramekanti-M-07,Dufour-N-M-15}.

\subsection{Connecting the \texorpdfstring{${\rm Gr}(2k, k)$}{Gr(2k,k)} nonlinear \texorpdfstring{$\sigma$}{sigma} model at \texorpdfstring{$\theta=\pi$}{theta=pi} and the \texorpdfstring{$\SU(2k)_1$}{SU(2k) level-1} WZNW model}
We propose to apply our general strategy to the ${\rm Gr}(2k, k)$ and $\mathbb{C}P^{N-1}$  
nonlinear $\sigma$  models at $\theta= \pi$ to reproduce all these results.
Let us first consider the Grassmaniann case by deforming the SU($2k$)$_1$ WZNW model by a suitable potential to select the ${\rm Gr}(2k, k)$ manifold
in the strong-coupling regime:
\begin{equation}
{\cal S}_{{\rm Gr}(2k, k)} =  {\cal S}_{{\rm WZNW}_1} + \int_{M_2} d^2 x  \;  \left\{ \lambda_1 \left[ {\rm Tr} \left( G^2\right) + \text{c.c.} \right] 
 + \lambda_2  | {\rm Tr} \left( G \right) |^2 \right\}  \; .
\label{WZWintGr(2k,k)}
\end{equation}
The potential is invariant under the PSU($2k$) symmetry: $ G(x) \rightarrow V G(x) V^{\dagger}$, $V$ being an SU($2k$) matrix, and under two independent $\mathbb{Z}_2 $ symmetries: $G \rightarrow - G$ and $ G \rightarrow G^{*}$. The strong-coupling regime with $ \lambda_{1,2} \rightarrow \infty $ stabilizes the ${\rm Gr}(2k, k)$ manifold:
\begin{equation}
G =  U \Omega U^{\dagger} 
 \label{Gr(2k,k)}\\
\end{equation}
with $ \Omega  = {\rm diag} (i \ldots i -i \dots -i)$ with $k$ eigenvalues $\pm i$  and 
$U$ is a general U($2k$) matrix.\footnote{As $\tr(G)=\tr(\Omega)=0$, this minimizes $|\tr(G)|^2=0$. 
Since $\tr(G^2)=\tr(\Omega^2)=\tr(-\bm{1})=-2k$, this also minimizes the $\lambda_1$ term. 
We can also check all the minima take this form by reversing the logic: To minimize $\tr(G^2)+\text{c.c.}$, we would like to set $G^2=-\bm{1}$. 
Then, the eigenvalues of $G$ have to be $\pm\im$, and we want to choose the same numbers of $\im$ and $-\im$ to make $\tr(G)=0$, which gives $G=U\Omega U^\dagger$. }  
There is a redundancy in the description since one can find U($2k$) matrices $V$ such that 
\begin{equation}
 V \Omega V^{\dagger}  = \Omega ,
 \label{redundancyGr(2k,k)}
\end{equation}
by choosing $V$ as:
\begin{equation}
V  =
\begin{pmatrix}
A &    0_k \\
  0_k  &  B \\
\end{pmatrix} 
\; , \label{U(k)xU(k)} 
\end{equation}
where $A$ and $B$ are independent U($k$) matrices. The ground-state manifold (\ref{Gr(2k,k)}) corresponds 
thus to the  ${\rm Gr}(2k, k)$ Grassmannian manifold. 

We then introduce $4 k^2$ complex scalar fields $\Phi_{a b}$ ($a,b = 1, \ldots, 2k$) such that 
$ U_{a b} (x) = \Phi_{a b} (x) = (\vec \Phi_b)_a(x)$ 
and $G_{a b} = \sum_c  \Phi^{*}_{b c}   \Omega_{cc}  \Phi_{a c}$ from Eq. (\ref{Gr(2k,k)}). These fields are constraint to be orthonormal complex vectors: $\vec \Phi^{*}_a \cdot \vec \Phi_b = \delta_{ab}$ to enforce the U($2k$) property: $ U^{\dagger} U = \bm{1}_{2k}$ and distinct scalar fields take value in the ${\rm Gr}(2k, k)$ Grassmannian manifold. The PSU($2k$) symmetry on the fields acts as  $ (\vec \Phi_b)_a  \rightarrow   V_{a c}  (\vec \Phi_b)_c$, whereas the $\mathbb{Z}_2 $ charge conjugation C symmetry 
$\Phi_{ a b} \rightarrow \Phi_{ a b}^{*}$ corresponds to $G\rightarrow - G^{*}$ from the identification (\ref{Gr(2k,k)}). The latter is obviously a symmetry of the potential (\ref{WZWintGr(2k,k)}). 
The symmetry $G\to G^*$ is realized by ${\vec \Phi}_{l} \rightarrow {\vec \Phi}_{k+l}^*, {\vec \Phi}_{k+l} \rightarrow {\vec \Phi}_{l}^*$, $l=1, \dots, k$.  

The derivation of the low-energy effective Lagrangian for the complex scalar fields ${\vec \Phi}_l $ is similar to
the one for the flag nonlinear $\sigma$ model.  We extend $G(x)=U(x) \Omega U(x)^\dagger$ to a three-dimensional manifold $M_3$ with $\partial M_3=M_2$ with $M_3=(M_2\times [0,1])/(M_2\times \{1\})$. In this respect, we consider $G(x,x_3)=U(x) \Omega(x_3) U(x)^\dagger$ with $\Omega(x_3)=\diag(\rme^{\im \theta_1(x_3)},\ldots, \rme^{\im \theta_N(x_3)})$. Here, we just require that $\theta_a(0)= \pi/2$, $\theta_{a+k} (0)=- \pi/2$ ($a=1, \ldots, k$) and $\theta_a(1)=0$ ($a=1, \ldots, 2k$). The action then 
takes the form of that of the ${\rm Gr}(2k, k)$ Grassmannian nonlinear $\sigma$  model (\ref{ComplexGrassmannfields}) at $\theta=\pi$. This identification has already been obtained by Affleck in Ref. \cite{Affleck-88} using a different potential.
 
We now turn to the direct analysis of action (\ref{WZWintGr(2k,k)}) by exploiting the SU($2k$)$_1$ CFT.  The main difference with actions (\ref{WZWint}) and (\ref{WZWintSU(N)/SO(N)}) stem from the fact that action (\ref{WZWintGr(2k,k)}) is not $\mathbb{Z}_N $ symmetric  but only invariant under a $\mathbb{Z}_2 $: $G \rightarrow - G$.
It means that some relevant operators are now allowed and one expects on general grounds no $\SU(N)$$_1$ massless behavior in this problem. We can use the free-field representation of the SU($2k$)$_1$ CFT to derive this result. First,
using  the result of Appendix \ref{sec:AppendixB} (see Eq. (\ref{TrGGdagcurcur})),  the $\lambda_2$ term in Eq. (\ref{WZWintGr(2k,k)}) is a marginal irrelevant current-current interaction: 
$\lambda_2   \nord{ {\rm Tr} \; G  \; {\rm Tr} \; G^{\dagger} }  = -   \frac{4\pi^2 \lambda_2 }{k} J_R^{A} J_L^{A}$. 
One can thus safely neglect this contribution to analyse the IR properties of the action (\ref{WZWintGr(2k,k)}). Second,
in stark contrast, the $\lambda_1$ term in Eq. (\ref{WZWintGr(2k,k)}) is a strongly relevant perturbation. Using the free-field representation (\ref{TrG2App}), the Hamiltonian density associated to the action (\ref{WZWintGr(2k,k)}) takes then the form of a generalized sine-Gordon model \cite{Boyanovsky-H-91}:
\begin{equation}
 \mathcal{H} _{\rm{Gr}(2k,k)} \simeq      \left(\partial_x  {\vec  \Phi}_{R} \right)^2 + 
\left(\partial_x  {\vec  \Phi}_{L} \right)^2   
 - \frac{2 \lambda_1}{k} \sum_{1 \le \alpha < \gamma \le 2k} 
 : \cos \left[ \sqrt{4 \pi} \left({\vec  e}_{\gamma} + {\vec  e}_{\alpha} \right) \cdot {\vec  \Phi}\right]:   \; .
\label{HSGGr(2k,k)}
\end{equation}
The perturbation has a scaling dimension $\Delta = \left({\vec  e}_{\gamma} + {\vec  e}_{\alpha} \right)^2 = 2(k-1)/k$ 
as  the SU($2k$)$_1$ primary field which belongs to the fully antisymmetric irrep of SU($2k$) with two boxes. It is a strongly relevant perturbation which opens a mass gap for either sign of $\lambda_1$. In the vacuum, the bosonic fields
are pinned into minima of the sine-Gordon potential (\ref{HSGGr(2k,k)}). 
Let us work on the case $\lambda_1>0$, then we can find two physically inequivalent minima up to the gauge redundancy $\vec{\Phi}\sim \vec{\Phi}+\sqrt{\pi}\vec{\alpha}_m$ as
\begin{equation}
\langle {\vec  \Phi} \rangle = {\vec  0} , \; \; \langle {\vec  \Phi} \rangle = \frac{\sqrt{\pi}}{2} \sum_{n=1}^{2k-1}  n{\vec {\alpha}}_n.
 \label{GSGr(2k,k)}
\end{equation}
In these ground states, one has $\langle G  \rangle \simeq \pm 1_{2k}$ from Eq. (\ref{freefieldrepWZWGfield}) and the $\mathbb{Z}_2 $ charge conjugation C  ($G\rightarrow - G^{*}$, i.e., $\Phi_{ a b} \rightarrow \Phi_{ a b}^{*}$) is spontaneously broken. We thus find that ${\rm Gr}(2k, k)$ nonlinear $\sigma$ model with $\theta= \pi$ is a fully gapped phase with a two-fold ground-state degeneracy which stems from the spontaneous breaking of the C symmetry. This result is in full agreement with the known result on the lattice regularization of the model in terms of an SU($2k$) Heisenberg spin chain in self-conjugate fully antisymmetric representation of the SU($2k$) group \cite{Nonne-L-C-R-B-11,Dufour-N-M-15,Onufriev1999,Assaraf-A-B-C-L-04,Paramekanti-M-07}.

\subsection{\texorpdfstring{$\mathbb{C}P^{N-1}$}{CP(N-1)}  nonlinear \texorpdfstring{$\sigma$}{sigma} model at \texorpdfstring{$\theta=\pi$}{theta=pi}}
We now turn to the $\mathbb{C}P^{N-1}$ case, where the analysis should be divided based on the parity of $N$. 
In this respect, we consider the following two deformed actions with positive $\lambda$:
\begin{equation}
{\cal S}_{\mathbb{C}P^{N-1}} =  
\begin{cases}
{\cal S}_{{\rm WZNW}_1} + \lambda  \int_{M_2} d^2 x  \,  \left| {\rm Tr} \left( G\right)   + (N -2)  \right|^2  & \text{when $N$ odd} \\
{\cal S}_{{\rm WZNW}_1} + \lambda  \int_{M_2} d^2 x  \,  \left|  \rme^{-i \frac{\pi}{N}} {\rm Tr} \left( G\right)   + (N -2)  \right|^2  & \text{when $N$ even}  \; .
\end{cases}
\label{WZWintGr(N-k even)}
\end{equation}

The potential is invariant under the P$\SU(N)$ symmetry: $ G \rightarrow V G V^{\dagger}$, with $V$ being an $\SU(N)$ matrix, and under a $\mathbb{Z}_2 $ conjugation symmetry: $ G \rightarrow G^{*}$ (respectively $\rme^{- i \frac{\pi}{N}} G \rightarrow  \rme^{ i \frac{\pi}{N}} G^{*}$) for odd (respectively even) $N$. As it can be checked by a numerical minimization,\footnote{One can easily check that the following configuration minimizes the potential term. A nontrivial task is to confirm if any minima can be expressed in this form. We confirm this by performing the numerical global minimization of the potential in two ways: One is the random search, i.e. the local minimization starting from the random configuration, and the other is the stimulated annealing. If one wants to evade this numerics, one may add an extra term, such as $-\lambda'[\tr(G^2)+\text{c.c.}]$ for odd $N$. The minimization of this extra term is achieved by requiring $G^2=\bm{1}$, so the eigenvalues of $G$ are $\pm1$. Then, to minimize the $\lambda$ term, we need to choose $-1$ as much as possible, and the $\mathbb{C}P^{N-1}$ moduli is obtained. We can give the similar discussion for even $N$ by adding $-\lambda'[ \rme^{-2\pi\im/N}\tr(G^2)+\text{c.c.}]$. For $\mathbb{C}P^{N-1}$, however, we believe that this addition term is just a mathematical trick that makes the proof easier, and we do not introduce them in the discussion of the main text.} 
the strong-coupling regime with $ \lambda \rightarrow \infty $ gives
\begin{equation}
G =   U \Omega U^{\dagger} ,
 \label{Gr(N,k)even}\\
\end{equation}
$U$ being a general U($N$) matrix and 
\begin{eqnarray}
\Omega &=&  {\rm diag} (1, -1, \dots, -1)  \; \text{when $N$ odd} \\
\Omega &=& \rme^{i \frac{\pi}{N}}  {\rm diag} (1, -1, \dots, -1)   \; \text{when $N$ even} 
\label{CPmanifold)}
\end{eqnarray}
with $N-1$ eigenvalues $-1$. For both $N$, $G$ is an $\SU(N)$ matrix. The description \eqref{Gr(N,k)even} is redundant with respect to the local transformation $U\to UV$ by U($N$) matrices V that satisfy 
\begin{equation}
 V \Omega V^{\dagger}  = \Omega \; .
 \label{redundancyGr(N,k)even}
\end{equation}
By choosing $V$ as:
\begin{equation}
V =
\begin{pmatrix}
e^{i \theta} &    0_{1,N-1} \\
  0_{N-1,1}  &  B \\
\end{pmatrix} , \label{U(1)xU(N-1)}
\end{equation}
with  a phase $\theta$ and an U($N-1$) matrix $B$, one observes that $V$ satisfies Eq. (\ref{redundancyGr(N,k)even}). The ground-state manifold (\ref{Gr(N,k)even}) corresponds thus to 
the  ${\rm Gr}(N, 1)$ Grassmannian manifold, i.e., $\mathbb{C}P^{N-1}$.  We then introduce $N$ complex scalar fields $(\vec{z})_i=U_{i1}$ ($i = 1, \ldots, N$) constrained to $\vec z^{\dagger} \cdot \vec z = 1$ associated with 
the $\mathbb{C}P^{N-1}$ manifold. The original global symmetries of action (\ref{WZWintGr(N-k even)}) have a direct interpretation on the complex fields $\vec z$ via the identification (\ref{Gr(N,k)even}):
\begin{itemize}
    \item The $\PSU(N)$ spin symmetry: $\vec{z} \mapsto V\vec{z} $ with $V\in \SU(N)$. 
    \item The $\mathbb{Z}_2$ charge conjugation: $\vec{z} \mapsto \vec{z}^{\,*}$ .
\end{itemize}

The derivation of the low-energy effective Lagrangian for the complex scalar fields $\vec z$  is similar to
the one for the flag and Grassmannian nonlinear $\sigma$ models.  We extend $G(x)=U(x) \Omega U(x)^\dagger$ to a three-dimensional manifold $M_3$ with $\partial M_3=M_2$ with $M_3=(M_2\times [0,1])/(M_2\times \{1\})$. The matrix $G$ is parametrized  as $G(x,x_3)=U(x) \Omega(x_3) U(x)^\dagger$ with $\Omega(x_3)=\diag(\rme^{\im \theta_1(x_3)},\ldots, \rme^{\im \theta_N(x_3)})$.  We choose $\theta_1(0)= \alpha \pi/N$, $\theta_{a} (0)= \pi + \alpha \pi/N$ ($a=2, \ldots, N$)  with
$\alpha = 0,1$ for  odd and even $N$ respectively, and $\theta_a(1)=0$ ($a=1, \ldots, N$). Pluging Eq. (\ref{Gr(N,k)even}) into the WZNW action (\ref{WZW}) leads to the $\mathbb{C}P^{N-1}$ nonlinear $\sigma$  model at $\theta=\pi$ 
\cite{Ohmori-S-S-19}.

The leading contribution which governs the IR properties of the action (\ref{WZWintGr(N-k even)}) is 
\begin{equation}
\begin{split}
& {\cal S}_{\mathbb{C}P^{N-1}} \simeq  {\cal S}_{{\rm WZNW}_1} + \lambda (N - 2) \int_{M_2} d^2 x  \; 
 \left [ \rme^{-i \alpha \frac{\pi}{N}} {\rm Tr} \; G + \text{c.c.} \right]  \quad (\lambda >0 )  \\
 & \alpha = 0 \;(N=\text{odd}), \; 1\; (N=\text{even}) \; ,
 \end{split}
\label{WZWintGr(N-k even)leading}
\end{equation}
where we have neglected 
the marginal irrelevant perturbation $ \nord{ {\rm Tr} \; G  \; {\rm Tr} \; G^{\dagger} }  = -   \frac{8\pi^2 }{N} J_R^{A} J_L^{A}$ 
(see Eq. (\ref{TrGGdagcurcur})). 

The perturbation in Eq. (\ref{WZWintGr(N-k even)leading}) is  strongly relevant with scaling dimension $(N-1)/N<2$.  It opens a mass gap and the nature of the ground state for positive $\lambda$ depends on the parity of $N$. When $N$ is odd ($\alpha=0$),  the phase is two-fold degenerate with $ G_{\pm} = - e^{\pm i \pi/N} 1_N$. The  $\mathbb{Z}_2 $ charge conjugation symmetry  $ G \rightarrow G^{*}$ ($\vec{z} \mapsto \vec{z}^{\,*}$) is spontaneously broken in this phase. In the even $N$ case ($\alpha=1$), we find also a two-fold degenerate fully gapped phase for $\lambda > 0$ with 
$ G = - 1_N, G = - \rme^{i 2\pi/N} 1_N$ that breaks spontaneously the  $\mathbb{Z}_2 $ charge conjugation symmetry  $ G \rightarrow \rme^{ i 2 \pi/N}  G^{*}$.  Our results lead to the conclusion that the $\mathbb{C}P^{N-1}$  nonlinear $\sigma$  model with a $\theta=\pi$ term describes a massive phase when $N>2$ which is two-fold degenerate as the result of the spontaneous symmetry breaking of the charge conjugation symmetry. 
All these are fully consistent with what is known for the lattice $\SU(N)$ spin models \cite{Read-S-NP-89,
Affleck-SUN-90,Klumper-90, Beard-P-R-W-05,Laflamme-E-al-16,Laflamme-E-al-16,Evans-G-H-W-18}.

\section{Concluding remarks}
\label{sec:conclusion}
In this paper, we explored the IR properties of two-dimensional nonlinear $\sigma$ models at nonzero topological angles.
The target spaces considered in our study include (i) $\SU(N)/\U(1)^{N-1}$, (ii) $\SU(N)/\SO(N)$, and  some complex Grassmannian manifolds  (iii) $\mathrm{Gr}(2k,k)$ and (iv) $\mathbb{C}P^{N-1}$. 
After studying the structure of the global symmetries and its 't~Hooft anomalies, we examined the detailed dynamics to determine how the IR limit of the nonlinear $\sigma$ model realizes the symmetry constraint. Specifically, we addressed the question: Does the system become gapless, or does it acquire a mass gap with degenerate ground states?  

To obtain the reasonable scenario for the dynamics, we established an explicit connection between those nonlinear 
$\sigma$ models and the $\SU(N)_1$ WZNW CFT with specific potentials, following the approach pioneered by
Affleck and Haldane~\cite{Affleck-Haldane} for the $\mathbb{C}P^1$ nonlinear $\sigma$ model.  
We then exploited the fact that the $\SU(N)_1$ WZNW CFT admits a free-field representation in terms of 
$N-1$ compactified bosons to relate the potential term properly UV-regularized by the point-splitting with the leading relevant/marginal interactions 
for the WZNW CFT.  
For the $\SU(N)/\U(1)^{N-1}$ nonlinear $\sigma$ model at $\theta_a=\frac{2\pi}{N}a$ ($a=1, \ldots, N-1$) and the $\SU(N)/\SO(N)$ nonlinear $\sigma$ model at $\theta=\pi$, this procedure relates the perturbing potentials to two different marginal current-current interactions having the specific signs for which they become marginally irrelevant. As a result, we found that the RG flows of those nonlinear $\sigma$ models logarithmically approach back to the $\SU(N)_1$ WZNW CFT. 

As a consistency check, we also analyzed the IR physics by flipping the sign of the coefficient for the potentials in those models. The classical analysis of the sign-flipped potentials predicts gapped vacua with specific degeneracy. A similar point-splitting UV regularization procedure combined with the short-distance OPE now yields a marginally-relevant current-current interaction. Using the free compact boson description for the $\SU(N)_1$ WZNW CFT,  the number of gapped vacua generated by this marginally relevant current-current interaction can be determined, reproducing the result of the classical analysis. 

For the $\SU(N)/\SO(N)$ nonlinear $\sigma$ model at $\theta=\pi$, there is already a strong evidence for its massless RG flow to the $\SU(N)_1$ WZNW CFT, based on the integrability~\cite{Fendley-01,Fendley-JHEP-01,Marino-M-R-23}.
It is worth noting here that integrability requires the sufficiently large symmetry to constrain the RG flow, and the case of the complete flag manifold $\SU(N)/\U(1)^{N-1}$ most likely is not integrable.  However, our approach, which combines the Affleck-Haldane argument with the OPE and point-splitting regularization of the potential, does not require integrability and has much wider applicability.  
Our results not only reproduce the known massless flow in the $\theta=\pi$ $\SU(N)/\SO(N)$ nonlinear $\sigma$ model from a different perspective, 
but also strongly suggest that the non-integrable $\SU(N)/\U(1)^{N-1}$ nonlinear $\sigma$ model at $\theta_a=\frac{2\pi}{N}a$ undergoes a similar massless RG flow  to $\SU(N)_1$, supporting the conjecture by Refs.~\cite{Lajko-W-M-A-17, Tanizaki-S-18, Ohmori-S-S-19}.  

For the Grassmannian cases [i.e., ${\rm Gr}(2k, k)$ and $\mathbb{C}P^{N-1}$] at $\theta=\pi$, the deformations connecting the  $\SU(N)_1$  WZNW CFT to these nonlinear $\sigma$ models are strongly relevant from the beginning, leading to the formation of a mass gap and two-fold ground-state degeneracy. This degeneracy is associated to the spontaneous breaking of the $\mathbb{Z}_2 $ charge conjugation symmetry.  The resulting IR behavior is fully consistent with the results well-established for the corresponding $\SU(N)$ lattice spin models. 

We expect that the approach developed in this paper provides a systematic method to determine whether two-dimensional nonlinear $\sigma$ models exhibit a massless RG flow. For example, let us consider a G/H nonlinear $\sigma$ model with nontrivial $\theta$ angles which arises as a semiclassical approximation of a quantum spin chain with global continuous symmetry G. A key question is whether it flows to the $\mathrm{G}_1$ WZNW CFT  assuming that the analysis of symmetry and anomaly is consistent with such a massless flow. If the model preserves a sufficiently large discrete chiral symmetry [like $(\mathbb{Z}_N)_L$ in Sec.~\ref{sec:flagsigmamodel}] as the low-energy realization of the one-step lattice translation symmetry, this chiral symmetry  forbids relevant terms by scalar primary operators~\cite{Affleck-Haldane,Affleck-88,Ohmori-S-S-19}. In such cases, the leading operator to the WZNW CFT is the marginal current-current interaction, but then the fate of the RG flow crucially depends on the sign of the marginal coupling, which determines whether the operator is marginal relevant or irrelevant. Our claim is that this sign can be determined unambiguously by combining the point-splitting regularization of the potential with the short-distance OPE obtained from a free-field representation of the $\mathrm{G}_1$ CFT.  While we do not have a rigorous proof that this prescription always yields the correct answer, it produces a consistent picture across all the examples examined in this paper. Further applications to two-dimensional nonlinear $\sigma$ models with real Grassmannian manifolds and $\Sp(N)$ series at $\theta=\pi$  will be explored elsewhere.

\section*{Acknowledgements}
The authors wish to dedicate this paper to the memory of Ian Affleck, whose groundbreaking contributions 
to our field and his lectures profoundly influenced our careers. 
YT also appreciates the YITP-RIKEN iTHEMS conference ``Generalized symmetries in QFT 2024'' (YITP-W-24-15) and the YITP long-term
workshop ``Hadrons and Hadron Interactions in QCD 2024'' (YITP-T-24-02) for providing
the opportunities of useful discussions. 


\paragraph{Funding information}
This work is supported by the IRP project ``Exotic Quantum Matter in Multicomponent Systems (EXQMS)'' from 
CNRS.  It is also supported in part by Japan Society for the Promotion of Science (JSPS) KAKENHI Grant No. 21K03401 (KT) and 23K22489 (YT), and also by Center for Gravitational Physics and Quantum Information at Yukawa Institute for Theoretical Physics. 


\begin{appendix}
\numberwithin{equation}{section}
\section{Abelian-bosonization description of the $\SU(N)$$_1$ CFT}
\label{sec:AppendixA}
In this Appendix, we  review a simple free-field description of the $\SU(N)$$_1$ with central charge $c=N-1$ in terms of $N-1$ bosonic fields. This identification will be important for the investigation of the IR properties of perturbed $\SU(N)$$_1$ CFT.
\subsection{Review of the multi-component Abelian bosonization for the Hubbard model}
The starting point is  the one-dimensional U($N$) Hubbard chain in the large repulsive $U$ limit for a $1/N$-filling with Fermi-wavector $k_{F} = \pi/(Na_0)$:
\begin{eqnarray}
{\cal H}_{\text{Hubbard}} &=& - t \sum_i \sum_{\alpha=1}^{N} \left( c^{\dagger}_{\alpha,i+1} c_{\alpha,i} + \text{H.c.} \right)
+ \frac{U}{2} \sum_{i, \alpha,\beta} n_{\alpha,i}   n_{\beta,i} \left( 1 - \delta_{\alpha\beta} \right),
\label{hubbardSUN}
\end{eqnarray}
where $c^{\dagger}_{\alpha,i}$ creates a spinless fermion in the band $\alpha=1, \ldots N$ of the site $i$ and $
n_{\alpha,i} = c^{\dagger}_{\alpha,i} c_{\alpha,i}$ is the occupation number. 
When the Hubbard interaction $U$ is sufficiently large, the system becomes a Mott insulator and 
the charge degrees of freedom gets decoupled from the low-energy physics.  The physical properties of this
Mott-insulating phase are described by the $\SU(N)$ Sutherland model \cite{Affleck-NP86,Affleck-88,James-K-L-R-T-18,Itoi-K-97,Assaraf-A-C-L-99,Manmana-H-C-F-R-11}:
\begin{equation}
{\cal H}_{\text{Sutherland}} = J   \sum_i \sum_{A=1}^{N^2-1} S^{A}_{i+1} S^{A}_i \; ,
\label{Sutherland-75app}
\end{equation}
where $J= 4 t^2/U$ is the spin-exchange and $S^{A}_i = c^{\dagger}_{\alpha,i}  T^{A}_{\alpha \beta}   c_{\beta,i} $
is the $\SU(N)$ spin operator on site $i$. A summation over repeated indices is implied in the following and the $\SU(N)$
generators are normalized such that: $\mbox{Tr} (T^A T^B) = \delta^{AB}/2$.
The model (\ref{Sutherland-75app}) is integrable and has $N-1$ gapless modes that are described by the $\SU(N)$$_1$ CFT with central charge $c=N-1$. 

A continuum description of the $\SU(N)$ spin operator can be derived from the continuum
limit of the fermion in terms of $N$ Dirac fermions: $c_{\alpha,n}/\sqrt{a_0} \rightarrow \rme^{-i k_F x } L_{\alpha} (x)
+ \rme^{i k_F x }  R_{\alpha}(x) $ with $x=na_0$. One thus obtains the identification:
\begin{eqnarray}
 & S^{A}_n/ a_0  \rightarrow J^{A}_{R} + J^{A}_{L} + \rme^{i 2k_F x }  L^{\dagger}_{\alpha}T^{A}_{\alpha \beta}  R_{\beta} 
+ \rme^{-i 2k_F x }  R^{\dagger}_{\alpha}T^{A}_{\alpha \beta}  L_{\beta} \nonumber \\
& = J^{A}_{R} + J^{A}_{L} + \rme^{i 2k_F x } N^{A} +  \rme^{-i 2k_F x } N^{A \dagger} ,
\label{spinopapp}
\end{eqnarray}
where $J^{A}_{L} = L^{\dagger}_{\alpha}T^{A}_{\alpha \beta}  L_{\beta}$ is the left $\SU(N)$$_1$ current with a similar definition for the right one. The 2$k_F$ $\SU(N)$ spin density $N^{A}$ of Eq. (\ref{spinopapp}) is thus: 
$N^{A} = \langle  L^{\dagger}_{\alpha}T^{A}_{\alpha \beta}  R_{\beta} \rangle_c$, the average being over the fully gapped
charge mode. We then introduce $N$ left-right moving bosons $\varphi_{\alpha L,R}$  to bosonize the Dirac fermions \cite{James-K-L-R-T-18,Gogolin-N-T-book} :
\begin{eqnarray}
 L_{\alpha} = \frac{\kappa_{\alpha}}{\sqrt{2 \pi a_0}} \; \rme^{ - i \sqrt{4 \pi}\varphi_{\alpha L}}, \quad 
 R_{\alpha} = \frac{\kappa_{\alpha}}{\sqrt{2 \pi a_0}} \; \rme^{ i \sqrt{4 \pi}\varphi_{\alpha R}}  ,
\label{bosoabeleq}
\end{eqnarray}
where $[\varphi_{\alpha R}, \varphi_{\beta L} ] = i \delta_{\alpha\beta}/4$ and $\kappa_{\alpha}$ are the Klein factors that insure 
the anticommutation of the fermions with different colors: $\{\kappa_{\alpha}, \kappa_{\beta} \} = 2   \delta_{\alpha\beta} $,
$\kappa^{\dagger}_{\alpha} = \kappa_{\alpha}$. Our normalization conventions of these chiral bosons are:
 \begin{equation}
 \begin{split}
&  \langle \varphi_{\alpha L} (z) \varphi_{\beta L} (0)\rangle = - \frac{\delta_{\alpha\beta}}{4 \pi} \ln z   \\
 & \langle \varphi_{\alpha R} (\bar z) \varphi_{\beta R} (0)\rangle = - \frac{\delta_{\alpha\beta}}{4 \pi} \ln \bar z \; .
 \end{split}
 \label{bosonormalization}
\end{equation}
The periodicity of compact bosons will be discussed later after clarifying the fundamental fields for the WZNW CFT. The 2$k_F$ $\SU(N)$ spin density  can thus be expressed in terms
of these bosonic fields:
\begin{equation}
N^{A} =  \frac{\kappa_{\alpha} \kappa_{\beta} i^{\delta_{\alpha\beta}}}{2 \pi a_0}  T^{A}_{\alpha \beta}    
\langle \rme^{  i \sqrt{4 \pi}\varphi_{\alpha L} + i \sqrt{4 \pi}\varphi_{\beta R} } \rangle_c .
\label{BoseNA}
\end{equation}
The next step of the approach is to switch to a new basis where the charge degrees of freedom are single out
to perform the average in Eq. (\ref{BoseNA}). In this respect, we introduce a new basis with one charge bosonic field $\Phi_{cR,L}$ and $N-1$ spin fields $\Phi_{mR,L}$ ($ m=1, \ldots N-1$) through \cite{Assaraf-A-C-L-99}:
\begin{eqnarray}
&& \Phi_{cR,L} = \frac{1}{\sqrt{N}} \sum_{\alpha =1}^{N} \varphi_{\alpha R,L } \label{SUNbasis} \\
&& \Phi_{m R,L} =  \frac{1}{\sqrt{m(m+1)}} \left( \sum_{p=1}^{m} \varphi_{pR,L} - m
\varphi_{m+1 R,L} \right)  \nonumber,
\end{eqnarray}
the inverse transformation being:
\begin{eqnarray}
 \varphi_{\alpha R,L }  &=& \frac{\Phi_{cR,L}}{\sqrt{N}} + \sum_{m=1}^{N-1} e^{m}_{\alpha} \Phi_{m R,L} 
 \nonumber \\
 &=&   \frac{\Phi_{cR,L}}{\sqrt{N}} + {\vec  e}_{\alpha} \cdot {\vec  \Phi}_{R,L}  ,
\label{invSUN}
\end{eqnarray}
where ${\vec  e}_{\alpha}$ ($\alpha=1, \ldots, N$)  are $N-1$-dimensional vectors  which satisfy:
\begin{subequations}
\begin{align}
& \sum_{\alpha=1}^N \vec{e}_\alpha = {\vec 0}, \\
& \sum_{\alpha=1}^N [\vec{e}_\alpha]^m [\vec{e}_\alpha]^{m'} = \delta_{mm'}, \\
& \vec{e}_\alpha {\cdot} \vec{e}_\beta
= \delta_{\alpha \beta} -\frac{1}{N},
\end{align}
\label{weightSUN}
\end{subequations}
where $m=1, \ldots, N -1$ describes the components of the ${\vec  e}_{\alpha}$ vectors. An explicit choice is:
\begin{align} \label{eq:SpecificWeight}
[\vec{e}_\alpha]^m = \begin{cases} \frac{1}{\sqrt{m(m+1)}} & (m \geq \alpha) \\ -\sqrt{\frac{m}{m+1}} & (m=\alpha-1) \\ 0 & (m<\alpha-1) \end{cases}. 
\end{align}
That is, $\vec{e}_{\alpha}$ are weight vectors for the defining representation of $\SU(N)$. 

With these definitions, we obtain the non-Abelian formulation of the 2$k_F$-part of the $\SU(N)$ spin density 
$N^{A}  =   i  \lambda   \; {\rm Tr} ( G  T^A)$ defined by:
\begin{subequations}
\begin{align}
L^{\dagger}_{ \alpha} R_{\beta} & = i \lambda \; G_{\beta \alpha} \\
\lambda &= \frac{\sqrt{N}}{2 \pi a_0^{1/N}} \left\langle  \rme^{  i \sqrt{4\pi/N} \Phi_c} \right\rangle_c \label{lambda} \\
G_{\beta \alpha} &= \frac{\kappa_{\alpha} \kappa_{\beta} i^{\delta_{\alpha\beta}-1}}{\sqrt{N}}
: \rme^{  i \sqrt{4 \pi} {\vec  e}_{\alpha} \cdot {\vec  \Phi}_{L}
+ i \sqrt{4 \pi}  {\vec  e}_{\beta} \cdot {\vec  \Phi}_{R}  }:  ,
\end{align}
\label{freefieldrepchargefield}
\end{subequations}
where  $ \Phi_c =  \Phi_{cR} +  \Phi_{cL} $ is the total charge boson field and we take average over the charge degrees of freedom in the large $U$ Mott-insulating phase.  
We now consider the $-2 k_F$ part $N^{A \dagger}$:
\begin{subequations}
\begin{align}
R^{\dagger}_{ \alpha} L_{\beta} & = - i \lambda^{*}  \left(G^{\dagger}\right)_{\beta \alpha} \\
\left(G^{\dagger}\right)_{\alpha \beta} \equiv G^{\dagger}_{\alpha \beta} &= \frac{\kappa_{\beta} \kappa_{\alpha}  (-i)^{\delta_{\alpha\beta}-1}}{\sqrt{N}}
: \rme^{  - i \sqrt{4 \pi} {\vec  e}_{\alpha} \cdot {\vec  \Phi}_{L}
- i \sqrt{4 \pi}  {\vec  e}_{\beta} \cdot {\vec  \Phi}_{R}  }: .
\end{align}
\label{freefieldrepchargefieldag}
\end{subequations}
We now check that $G$ has the correct two-point function since it is a primary field
with scaling dimension $1-1/N$:
\begin{equation}
\begin{split}
& \langle G^{\dagger}_{\alpha \gamma} (z, \bar z)  G_{\gamma \beta } (w, \bar w)\rangle = 
\frac{1}{N}
 \sum_{\gamma} (-i)^{\delta_{\alpha\gamma}} i^{\delta_{\beta\gamma}} 
 \kappa_{\gamma} \kappa_{\alpha} \kappa_{\beta}\kappa_{\gamma}   \\
 & \phantom{ \langle G^{\dagger}_{\alpha \gamma} (z, \bar z)  G_{\gamma \beta } (w, \bar w)\rangle =  }
 \times \rme^{i \pi/2( {\vec  e}_{\beta} \cdot
  {\vec  e}_{\gamma} -  {\vec  e}_{\alpha} \cdot
  {\vec  e}_{\gamma}) } 
\left \langle  : \rme^{ -i \sqrt{4 \pi}  {\vec  e}_{\gamma} \cdot {\vec  \Phi}_{R} (\bar z) }: 
 : \rme^{ i \sqrt{4 \pi}  {\vec  e}_{\gamma} \cdot {\vec  \Phi}_{R} (\bar w) }:  \right\rangle \; \times
 \\
&  \left\langle  : \rme^{ -i \sqrt{4 \pi}  {\vec  e}_{\alpha} \cdot {\vec  \Phi}_{L} ( z) }: 
 : \rme^{ i \sqrt{4 \pi}  {\vec  e}_{\beta} \cdot {\vec  \Phi}_{L} ( w) }:   \right\rangle = 
  \frac{\delta_{\alpha \beta}}{|z-w|^{2(1-1/N)}}  \; ,
  \end{split}
\label{defWZWGfield}
\end{equation}
as it should be. 

\subsection{\texorpdfstring{$\SU(N)_1$}{SU(N) level-1} WZNW CFT }
\label{app:FreeBosonWZW_Summary}
Let us now summarize the outcome of the Abelian bosonization in the above discussion. 
The $\SU(N)$$_1$ WZNW primary field $G$ has the following free-field representation in terms of $N-1$ bosonic fields: 
\begin{equation}
\begin{split}
& G_{\beta \alpha} = \frac{\kappa_{\alpha} \kappa_{\beta} i^{\delta_{\alpha\beta}-1}}{\sqrt{N}}
: \rme^{  i \sqrt{4 \pi} {\vec  e}_{\alpha} \cdot {\vec  \Phi}_{L}
+ i \sqrt{4 \pi}  {\vec  e}_{\beta} \cdot {\vec  \Phi}_{R}  }:   \\
& \left(G^{\dagger}\right)_{\alpha \beta} \equiv G^{\dagger}_{\alpha \beta}
= \frac{\kappa_{\beta} \kappa_{\alpha}  (-i)^{\delta_{\alpha\beta}-1}}{\sqrt{N}}
: \rme^{  - i \sqrt{4 \pi} {\vec  e}_{\alpha} \cdot {\vec  \Phi}_{L}
- i \sqrt{4 \pi}  {\vec  e}_{\beta} \cdot {\vec  \Phi}_{R}  }: .
\end{split}
\label{freefieldrepWZWGfield}
\end{equation}
A more rigorous free-field representation of the $G$ WZNW  field can be found, for instance, 
in Ref. \cite{Fuji-L-17} (see Appendix B)  or in the book \cite{DiFrancesco-M-S-book}, where the Klein factors are constructed out of the zero mode operators of the bosonic fields.  

Let us now identify the periodicity of compact bosons. For this purpose, we combine the left- and right-movers in two ways: The first one is the standard compact boson
\begin{align}
    \vec{\Phi}=\vec{\Phi}_L+\vec{\Phi}_R, 
\end{align}
and the another is its dual field, 
\begin{align}
    \vec{\Theta}=\vec{\Phi}_L - \vec{\Phi}_R. 
\end{align}
The periodicity of $\vec{\Phi}$ can be understood by looking at the diagonal components of $G$:
\begin{align}
    G_{\alpha\alpha}=\frac{1}{\sqrt{N}}:\rme^{\im \sqrt{4\pi}\vec{e}_\alpha\cdot \vec{\Phi}}:. 
\end{align}
We introduce the $\SU(N)$ simple root vectors as 
\begin{align}
    \vec{\alpha}_i =\vec{e}_i-\vec{e}_{i+1} .
\end{align}
Then, the minimal periodicity for $\vec{\Phi}$ that leaves $G_{\alpha\alpha}$ invariant is given by 
\begin{align}
    \vec{\Phi}\sim \vec{\Phi}+\sqrt{\pi} \vec{\alpha}_i, 
\end{align}
and we impose it as the gauge redundancy. However, in terms of $\vec{\Phi}$ and $\vec{\Theta}$, the primary field $G$ takes the form of 
\begin{align}
    G_{\beta\alpha}\sim \exp\left[\im \sqrt{\pi}\left((\vec{e}_\alpha+\vec{e}_\beta)\cdot \vec{\Phi} + (\vec{e}_\alpha-\vec{e}_\beta)\cdot \vec{\Theta}\right)\right], 
\end{align}
and the above periodicity of $\vec{\Phi}$ flips the sign for the off-diagonal component of $G$. 
Thus, the correct gauge redundancy requires the diagonal shift of the original and dual compact bosons as\footnote{One may wonder why the gauge redundancy has such a nontrivial action. Recall that the basic idea of the Abelian boson description of the $\SU(N)_1$ WZNW CFT relies on the two facts: (1)~The $N$-flavor massless Dirac fermion can be dualized to the $\U(N)_1$ WZNW model via the non-Abelian bosonization, but (2)~the same system can be dualized by applying the Abelian bosonization to each flavor. Then, gapping out the flavor-singlet component, we obtain the equivalence between $N-1$ free bosons and $\SU(N)_1$ WZNW CFT. 
However, bosonization is gauging of the fermion parity symmetry. When we apply the Abelian bosonization to each flavor naively, fermion parity of each flavor is gauged, and the bosonic operator corresponding to $G_{ij}$ with $i\not=j$ is dropped from the local operator spectrum. To circumvent it, we have to be careful so that flavored fermion parity is not gauged while we need to gauge the diagonal fermion parity, which secretly performs the topological operations, such as adding invertible topological phases and gauging of discrete symmetry. Now, we can interpret that the effect of the topological operation ends up with correcting the gauge redundancy via the Witten effect. 
} 
\begin{align}
    \begin{pmatrix}
        \vec{\Phi}\\
        \vec{\Theta}
    \end{pmatrix} \sim 
    \begin{pmatrix}
        \vec{\Phi}\\
        \vec{\Theta}
    \end{pmatrix}
    +\sqrt{\pi}
    \begin{pmatrix}
        \vec{\alpha}_i\\
        \vec{\alpha}_i
    \end{pmatrix}. 
\end{align}
We also have another gauge redundancy that only shifts the dual field $\vec{\Theta}$:
\begin{align}
    \begin{pmatrix}
        \vec{\Phi}\\
        \vec{\Theta}
    \end{pmatrix} \sim 
    \begin{pmatrix}
        \vec{\Phi}\\
        \vec{\Theta}
    \end{pmatrix}
    +\sqrt{4\pi}
    \begin{pmatrix}
        0\\
        \vec{e}_i
    \end{pmatrix}. 
\end{align}
The field configurations related by these two redundancies should be physically identified.\footnote{Here, the kinetic term of the free boson is taken as the canonical one, $\frac{1}{2}(\partial_\mu \vec{\Phi})^2$. There is the other standard convention, which takes $\vec{\Phi}\sim \vec{\Phi}+2\pi \vec{\alpha}_m$, with the kinetic term $\frac{R^2}{4\pi}(\partial\vec{\Phi})^2$ with $R=\frac{1}{\sqrt{2}}$. In this case, the $T$-duality maps this theory to the another compact boson, $\vec{\Theta}\sim \vec{\Theta}+2\pi \vec{e}_m$, with the kinetic term $\frac{1/R^2}{4\pi}(\partial\Theta)^2$. 
In this convention, $G_{\beta\alpha}\sim \exp\left[\im\left(\frac{1}{2}(\vec{e}_\alpha+\vec{e}_{\beta})\cdot \vec{\Phi}+ (\vec{e}_\alpha-\vec{e}_\beta)\cdot \vec{\Theta}\right)\right]$, and the correct periodicity becomes $(\vec{\Phi},\vec{\Theta})\sim (\vec{\Phi},\vec{\Theta})+(2\pi \vec{\alpha}_i,\pi\vec{\alpha}_i)$ and $(\vec{\Phi},\vec{\Theta})\sim (\vec{\Phi},\vec{\Theta})+(0,2\pi \vec{e}_i)$. 
These two conventions are translated via $\vec{\Phi}\Leftrightarrow \frac{\sqrt{2\pi}}{R}\vec{\Phi}$ and $\vec{\Theta}\Leftrightarrow \sqrt{2\pi} R \vec{\Theta}$. }  
 
\section{OPE computations using free-field representations}
\label{sec:AppendixB}

Here, we calculate the OPEs of the $\SU(N)_1$ WZNW CFT using its free-field representation introduced in Appendix~\ref{sec:AppendixA}.

\subsection{OPEs for the \texorpdfstring{$SU(N)/U(1)^{N-1}$}{SU(N)/U(1){N-1}} flag-manifold \texorpdfstring{$\sigma$}{sigma} model}

We first consider the potential part of model (\ref{WZWint}) for the flag nonlinear $\sigma$ model and express it in terms of the $\SU(N)$$_1$ fields. 
From the symmetry argument, we know that the short-distance OPE takes the form of 
\begin{align}
    \tr G^n(z,\bar{z}) \tr G^{\dagger n}(0,0)\sim \frac{1}{|z|^{2n(N-n)/N}}\left(c_0 + c_1 |z|^2 \tr( J_R J_L) + \cdots\right), 
\end{align}
where $z=x_1+\im x_2$ is the complex coordinate, 
and we would like to determine the coefficient $c_1$ for $n=1,2,\ldots, [N/2]$. 
Since the computation for general $n$ is cumbersome, let us first demonstrate the computation for the $n=1$ case in detail to give an idea how the calculation goes. We then explain how we extend its computation to  the $n=2$ case and, lastly, derive the formula for general $n$. 

To compute the OPE of $\tr G(z,\bar{z}) \tr G^\dagger(0,0)$, we use its free-field representation and apply the Wick theorem of the normal-ordered product: 
\begin{equation}
\begin{split}
&    \tr G(z,\bar{z}) \tr G^\dagger (0,0)  \\
    &= \frac{1}{N} \sum_{\alpha=1}^{N} : \rme^{i \sqrt{4 \pi}   {\vec  e}_{\alpha} \cdot {\vec  \Phi}(z,\bar{z})}:\,  : \rme^{-i \sqrt{4 \pi}   {\vec  e}_{\alpha} \cdot {\vec  \Phi}(0,0)}:
     +  \frac{1}{N} \sum_{\alpha \ne \beta} : \rme^{i \sqrt{4 \pi}   {\vec  e}_{\alpha} \cdot {\vec  \Phi}(z,\bar{z})}:\,  : \rme^{-i \sqrt{4 \pi}   {\vec  e}_{\beta} \cdot {\vec  \Phi}(0,0)}:   \\
     &= \frac{1}{N} \sum_{\alpha=1}^{N} \frac{1}{|z|^{2 \vec{e_\alpha}^2}} : \rme^{i \sqrt{4 \pi}   {\vec  e}_{\alpha} \cdot ({\vec  \Phi}(z,\bar{z})-\vec{\Phi}(0,0))}:
     +  \frac{1}{N} \sum_{\alpha \ne \beta} \frac{1}{|z|^{2\vec{e}_\alpha\cdot \vec{e}_\beta}} : \rme^{i \sqrt{4 \pi}  ( {\vec  e}_{\alpha} \cdot {\vec  \Phi}(z,\bar{z})- {\vec  e}_{\beta} \cdot {\vec  \Phi}(0,0) )}: \, .
\end{split}
\end{equation}
Since $\vec{e}_\alpha\cdot \vec{e}_\beta=\delta_{\alpha \beta}-\frac{1}{N}$, we may simply set $z=0$ inside the normal-ordered product of the second term. 
For the first term, we have to expand as $\vec{\Phi}(z,\bar{z})-\vec{\Phi}(0,0)= z \partial \vec{\Phi}(0,0)+\bar{z}\bar{\partial}\vec{\Phi}(0)+\cdots$ with the holomorphic derivative $\partial =\partial_z=\frac{1}{2}(\partial_1-\im \partial_2)$. 
Stopping the expansion at the second order, we get 
\begin{equation}
\begin{split}
 &   : \rme^{i \sqrt{4 \pi}   {\vec  e}_{\alpha} \cdot ({\vec  \Phi}(z,\bar{z})-\vec{\Phi}(0,0))} \! :   \\
&  = 1+ \im \sqrt{4\pi} \vec{e}_\alpha\cdot (z\partial \vec{\Phi}+\bar{z}\bar{\partial}\vec{\Phi} + 
    \frac{z^2}{2} \partial^2 \vec{\Phi} +  \frac{\bar{z}^2}{2} \bar{\partial}^2\vec{\Phi})-\frac{4\pi}{2}(\vec{e}_\alpha\cdot (z\partial \vec{\Phi}+\bar{z}\bar{\partial}\vec{\Phi}))^2+\cdots \; .
\end{split} 
\end{equation}
Here, we note that we perform an integration over the relative coordinate $(z,\bar{z})$ with a rotation-symmetric weight. Thus, we only need to maintain the term with $|z|^2$ for our purpose: 
\begin{equation}
\begin{split}
&    : \rme^{i \sqrt{4 \pi}   {\vec  e}_{\alpha} \cdot ({\vec  \Phi}(z,\bar{z})-\vec{\Phi}(0,0))} \! :   \\
 & = 1- 4\pi |z|^2 (\vec{e}_\alpha\cdot \partial\vec{\Phi})(\vec{e}_\alpha\cdot \bar{\partial}\vec{\Phi})+(\text{non-rotation-symmetric or higher-order terms}) \; .
\end{split} 
\label{opebosondebapp}
\end{equation}
As a result, we find that 
\begin{align}
    \tr G(z,\bar{z}) \tr G^\dagger (0,0) = \frac{1}{|z|^{2(1-1/N)}}\left[1-\frac{|z|^2}{N}\left(4\pi \partial \vec{\Phi}\cdot \bar{\partial}\vec{\Phi}-\sum_{\alpha\not=\beta}  : \rme^{i \sqrt{4 \pi}  ( {\vec  e}_{\alpha} - {\vec  e}_{\beta}) \cdot {\vec  \Phi}(0,0)}: \right)+\cdots\right].
    \label{trGOPE1App}
\end{align}

Here, let us introduce the new symbol $\nord{ AB }$ to denote the first 
non-constant term in the OPE $A(z, \bar z) B(0, 0)$ (which is not necessarily non-vanishing in the limit $|z| \to 0$).
In the case of $\tr G \tr G^\dagger$, it extracts the following marginal operator as the leading non-constant contribution, 
\begin{equation}
\begin{split}
	\nord{ {\rm Tr} \; G  \; {\rm Tr} \; G^{\dagger} } \; 
	&= - \frac{4\pi}{N} \; \partial {\vec  \Phi}  \cdot {\bar \partial} {\vec  \Phi}  + \frac{2}{N} \sum_{1 \le \alpha < \beta \le N}  : \cos \left[ {\sqrt{4 \pi} \left( {\vec  e}_{\alpha} -  {\vec  e}_{\beta} \right) {\cdot} {\vec  \Phi}} \right] :   \\
	&= - \frac{4\pi}{N} \; \partial {\vec  \Phi}_L  \cdot {\bar \partial} {\vec  \Phi}_R  + \frac{2}{N} \sum_{1 \le \alpha < \beta \le N}  : \cos \left[ {\sqrt{4 \pi} \left( {\vec  e}_{\alpha} -  {\vec  e}_{\beta} \right) {\cdot} {\vec  \Phi}} \right] :   \\
	 &= - \frac{4\pi}{N} \; \partial_x {\vec  \Phi}_L  \cdot \partial_x {\vec  \Phi}_R  +  \frac{2}{N} \sum_{1 \le \alpha < \beta \le N}   : \cos \left[ {\sqrt{4 \pi} \left( {\vec  e}_{\alpha} -  {\vec  e}_{\beta} \right) {\cdot} {\vec  \Phi}} \right] :   \; 
\end{split}
\label{TraceGGdagApp}
\end{equation}
where we have used $\partial \vec{\Phi} =\partial \vec{\Phi}_L$, $\bar{\partial} \vec{\Phi}=\bar{\partial}\vec{\Phi}_R$, and $\partial_x = i ( \partial  - {\bar \partial})$ with $\bar{\partial} \vec{\Phi}_L=\partial \vec{\Phi}_R=0$.  
Now we show that the marginal operators \eqref{TraceGGdagApp} can be written in the form of an $\SU(N)$$_1$ current-current interaction $\tr(J_{ L} J_{ R})$. Using the fermionic representation \eqref{spinopapp}, the $\SU(N)$$_1$ currents reads as follows:
\begin{eqnarray}
J_{ L}^A &=&  L^{\dagger}_{\alpha} T^{A}_{\alpha\beta} L_{\beta} \nonumber \\
J_{ R}^A &=&  R^{\dagger}_{\alpha} T^{A}_{\alpha\beta} R_{\beta}  \;  .
\label{suNcurrentferApp}
\end{eqnarray}
We deduce then
\begin{equation}
 J_R^{A} J_L^{A}   =   \frac{1}{2} \;  R^{\dagger}_{\alpha} R_{\beta} L^{\dagger}_{\beta} L_{\alpha} \nonumber \\
- \frac{1 }{2N} : R^{\dagger}_{\alpha} R_{\alpha}: \, : L^{\dagger}_{\beta} L_{\beta}:  \;  ,
\label{currentcurrentferApp}
\end{equation}
where we have used the identity: 
\begin{eqnarray}
T^{A}_{\alpha\beta} T^{A}_{\gamma \delta} = \frac{1}{2} \left( \delta_{\alpha \delta} \delta_{ \beta \gamma} - \frac{1}{N} \delta_{\alpha \beta} \delta_{ \ \delta\gamma} \right)  \; .
\label{algebraidenApp}
\end{eqnarray}
Using the Abelian bosonization rules \eqref{bosoabeleq}, we find
\begin{equation}
 \tr(J_R J_L) =\frac{1}{2}J_R^{A} J_L^{A}   = \frac{1}{4\pi} \left\{ \partial_x {\vec  \Phi}_{R} \cdot  \partial_x  {\vec  \Phi}_{L}  
-  \frac{1 }{2\pi} \sum_{1 \le \alpha < \beta \le N} :\cos \left[ \sqrt{4\pi} \left( {\vec  e}_{\alpha} - {\vec  e}_{\beta}\right) \cdot {\vec  \Phi}\right] : \right\} \; ,
\label{currentcurrentbosoApp}
\end{equation}
from which  we get the following identification:
\begin{equation}
 \nord{  {\rm Tr} \; G  \; {\rm Tr} \; G^{\dagger}  }    \, 
= \,  - \frac{8\pi^2}{N} J_R^{A} J_L^{A}  \; .
\label{TrGGdagcurcur}
\end{equation}

Next, let us move on to the analysis for $n=2$, i.e. $\tr G^2(z,\bar{z}) \tr G^{\dagger 2}(0,0)$.  Although the basic strategy is the same, the $n\not=1$ case requires an extra step as $\tr G^n$ itself is not a chiral primary and we have to extract the primary part out of $\tr G^n$. 
To this end, we need to evaluate the following OPE for $\tr G^2$ (no implicit summation over $\gamma$ here):\footnote{Here, we carefully compute the overall phase of $\nord{\tr G^2}$ appearing from the Klein factor and the commutation of zero modes for completeness of the discussion. However, it anyway cancels when computing the product $\nord{\tr G^2 (z,\bar{z})} \nord{\tr G^{\dagger 2}(0,0)}$, so one may neglect the details on the phase factor. }  
\begin{align}
	& G_{\alpha \gamma} (z, \bar z)   G_{\gamma \beta} (0, 0) 
	=  
	- \frac{\kappa_{\gamma} \kappa_{\alpha} \kappa_{\beta}\kappa_{\gamma}  i^{\delta_{\alpha\gamma}}i^{\delta_{\gamma\beta}}}{N} 
	: \rme^{  i \sqrt{4 \pi} {\vec  e}_{\gamma} \cdot {\vec  \Phi}_{L}(z) + i \sqrt{4 \pi}  {\vec  e}_{\alpha} \cdot {\vec  \Phi}_{R}(\bar{z}) }: \, 
	: \rme^{  i \sqrt{4 \pi} {\vec  e}_{\beta} \cdot {\vec  \Phi}_{L} (0) + i \sqrt{4 \pi}  {\vec  e}_{\gamma} \cdot {\vec  \Phi}_{R} (0) }:  \notag \\
	&=  -  \frac{\kappa_{\gamma}  \kappa_{\alpha} \kappa_{\beta} \kappa_{\gamma}   i^{\delta_{\alpha\gamma}} i^{\delta_{\gamma\beta}}}{N} \rme^{ -i \frac{\pi}{2} ( {\vec  e}_{\alpha} \cdot {\vec  e}_{\gamma} +  {\vec  e}_{\beta} \cdot {\vec  e}_{\gamma})} \notag\\
	&\qquad \times : \rme^{  i \sqrt{4 \pi} {\vec  e}_{\gamma} \cdot {\vec  \Phi}_{L}(z)}: \, : \rme^{  i \sqrt{4 \pi} {\vec  e}_{\alpha} \cdot {\vec  \Phi}_{R}(\bar z) }: \, 
	: \rme^{  i \sqrt{4 \pi} {\vec  e}_{\beta} \cdot {\vec  \Phi}_{L}(0)}: \,  : \rme^{  i \sqrt{4 \pi} {\vec  e}_{\gamma} \cdot {\vec  \Phi}_{R}(0)}:  \notag \\
	&=  -  \frac{\kappa_{\gamma} \kappa_{\alpha} \kappa_{\beta}  \kappa_{\gamma} i^{\delta_{\alpha\gamma}} i^{\delta_{\gamma\beta}}}{N} \rme^{ -i \frac{\pi}{2} ( {\vec  e}_{\alpha} \cdot {\vec  e}_{\gamma} +  {\vec  e}_{\beta} \cdot {\vec  e}_{\gamma})} \rme^{ - i \pi {\vec  e}_{\alpha} \cdot {\vec  e}_{\beta}} \notag \\
	&\qquad \times : \rme^{  i \sqrt{4 \pi} {\vec  e}_{\gamma} \cdot {\vec  \Phi}_{L} (z)}: \, : \rme^{  i \sqrt{4 \pi} {\vec  e}_{\beta} \cdot {\vec  \Phi}_{L}(0) }: \,
	: \rme^{  i \sqrt{4 \pi} {\vec  e}_{\alpha} \cdot {\vec  \Phi}_{R}(\bar z)}: \,  : \rme^{  i \sqrt{4 \pi} {\vec  e}_{\gamma} \cdot {\vec  \Phi}_{R} (0) }: \notag \\
	& \sim -  \frac{\kappa_{\gamma} \kappa_{\alpha} \kappa_{\beta}  \kappa_{\gamma}  i^{\delta_{\alpha\gamma}} i^{\delta_{\gamma\beta}}}{N} \frac{z ^{\delta_{\gamma\beta}} {\bar z}^{\delta_{\gamma\alpha}}}{|z|^{2/N}}  \rme^{ -i \frac{\pi}{2} ( {\vec  e}_{\alpha} \cdot {\vec  e}_{\gamma} +  {\vec  e}_{\beta} \cdot {\vec  e}_{\gamma})} \rme^{ - i \pi {\vec  e}_{\alpha} \cdot {\vec  e}_{\beta}}
	: \rme^{  i \sqrt{4 \pi} \left( {\vec  e}_{\gamma} + {\vec  e}_{\beta} \right) \cdot {\vec  \Phi}_{L}}	\rme^{  i \sqrt{4 \pi} \left( {\vec  e}_{\gamma} + {\vec  e}_{\alpha} \right) \cdot {\vec  \Phi}_{R}} : (0,0) \notag \\  
	& \sim 
	-  \frac{i \kappa_{\gamma} \kappa_{\alpha} \kappa_{\beta} \kappa_{\gamma}  i^{\delta_{\alpha\gamma}}  (-i)^{\delta_{\alpha\beta}} i^{\delta_{\gamma\beta}}}{N} \frac{z ^{\delta_{\gamma\beta}} {\bar z}^{\delta_{\gamma\alpha}}}{|z|^{2/N}}  : \rme^{  i \sqrt{4 \pi} \left({\vec  e}_{\gamma} \cdot {\vec  \Phi} + {\vec  e}_{\beta}\cdot {\vec  \Phi}_{L}	+ {\vec  e}_{\alpha}\cdot {\vec  \Phi}_{R} \right) }: (0,0) + \cdots  \; .
	\label{G2App}
\end{align}
The leading term in the OPE (\ref{G2App}) for $\alpha= \beta$ is obtained when $\gamma \ne \alpha$ so that
\begin{eqnarray}
 \nord{  {\rm Tr} \; G^2  } =    - \frac{1}{N} \sum_{\gamma \ne \alpha =1}^{N}  
 : \rme^{  i \sqrt{4 \pi} \left({\vec  e}_{\gamma} + {\vec  e}_{\alpha} \right) \cdot {\vec  \Phi}}: ,
\label{TrG2App}
\end{eqnarray}
which has scaling dimension $\Delta = \left({\vec  e}_{\gamma} + {\vec  e}_{\alpha} \right)^2 = 2(N-2)/N$, and this is the $\SU(N)$$_1$ primary field corresponding to the $2$-box fully antisymmetric representation. 
Since we are interested in the leading nontrivial contribution, we may replace $\tr G^2 (z,\bar{z}) \tr G^{\dagger 2}(0,0)$ by $\nord{\tr G^2 (z,\bar{z})} \nord{\tr G^{\dagger 2}(0,0)}$.

The next step of the approach is to compute the following OPE:
\begin{align}
& \nord{{\rm Tr} \; G^2 (z, \bar z)} \nord{ {\rm Tr} \;  {G^2}^{\dagger}(0, 0)}  \notag\\
	&  =   \frac{1}{N^2} \sum_{\gamma \ne \alpha =1}^{N}  \sum_{\delta \ne \beta =1}^{N}  
	: \rme^{  i \sqrt{4 \pi} \left({\vec  e}_{\gamma} + {\vec  e}_{\alpha} \right) \cdot {\vec  \Phi}(z, \bar z)}:\, : \rme^{ -  i \sqrt{4 \pi} \left({\vec  e}_{\delta} + {\vec  e}_{\beta} \right) \cdot {\vec  \Phi}(0, 0)}:  \notag\\
	&  \sim \frac{1}{N^2} \sum_{\gamma \ne \alpha, \delta \ne \beta} |z|^{-2\left({\vec  e}_{\gamma} + {\vec  e}_{\alpha} \right) \cdot \left({\vec  e}_{\delta} + {\vec  e}_{\beta} \right)}  : \rme^{  i \sqrt{4 \pi} \left({\vec  e}_{\gamma} + {\vec  e}_{\alpha} - {\vec  e}_{\delta} -  {\vec  e}_{\beta} \right) \cdot {\vec  \Phi} (0, 0)} \rme^{  i \sqrt{4 \pi}  \left({\vec  e}_{\gamma} + {\vec  e}_{\alpha}\right) \cdot  \left(z \partial {\vec  \Phi}_L (0)  + {\bar z} {\bar \partial} {\vec  \Phi}_R (0)\right)} :  + \cdots \notag \\
	 & \sim \frac{2!}{N^2 |z|^{4(N-2)/N}} \sum_{\gamma \ne \alpha} \left[1 -  \frac{1}{2} 4\pi  \left( \left({\vec  e}_{\gamma} + {\vec  e}_{\alpha}\right) \cdot  \left(z \partial {\vec  \Phi}_L (0)   + {\bar z} {\bar \partial} {\vec  \Phi}_R (0)\right) \right)^2 \right] \notag \\
 & \phantom{=} \; 
 +  \frac{4}{N^2}
 \sum_{\gamma \ne \alpha,  \beta \ne \alpha, \gamma \ne \beta } |z|^{-2\left({\vec  e}_{\gamma} + {\vec  e}_{\alpha} \right) \cdot \left({\vec  e}_{\alpha} + {\vec  e}_{\beta} \right)}
 : \rme^{  i \sqrt{4 \pi} \left({\vec  e}_{\gamma}  - {\vec  e}_{\beta}\right)
  \cdot {\vec  \Phi} (0, 0)}: + \cdots  \notag \\ 
	& \sim \frac{1}{|z|^{4(N-2)/N}}\left[\frac{2(N-1)}{N} 
	- \frac{N-2}{N^2} |z|^2 \left(16\pi  \partial {\vec  \Phi}_L \cdot {\bar \partial} {\vec  \Phi}_R 
	-4 \sum_{\gamma \ne \beta}  : \rme^{  i \sqrt{4 \pi} \left({\vec  e}_{\gamma}  - {\vec  e}_{\beta}\right) \cdot {\vec  \Phi} (0, 0)}: \right)+ \cdots\right]
 \label{TrG2G2App}
\end{align}
since we have from Eqs. (\ref{weightSUN}):
\begin{equation}
\begin{split}
\sum_{\gamma \ne \alpha} \left(
  {\vec  e}_{\gamma} + {\vec  e}_{\alpha} \right) &=  {\vec  0}   \\
  \sum_{\gamma \ne \alpha}   [{\vec  e}_{\gamma} + {\vec  e}_{\alpha}]^{m}  [{\vec  e}_{\gamma} + {\vec  e}_{\alpha}]^{n} 
   &= 2 (N-2) \delta_{mn}  \\
\left( {\vec  e}_{\gamma} + {\vec  e}_{\alpha} \right) \cdot \left({\vec  e}_{\alpha} + {\vec  e}_{\beta} \right) &= 1 - \frac{4}{N} 
\quad (\gamma \ne \alpha \ne \beta )  \; .
\end{split}
\end{equation}
In Eq. (\ref{TrG2G2App}) the non-rotation-symmetric terms have been neglected as in Eq. (\ref{opebosondebapp}) since they average to zero after the integration over the relative coordinate.

We deduce that the first regular term in the OPE (\ref{TrG2G2App}) is the marginal current-current term:
\begin{equation}
\begin{split}
&  \nord{  (\nord{{\rm Tr} \; G^2})  (\nord{{\rm Tr} \;  {G^{\dagger 2} }})  }   \\
  & =  - \frac{32 \pi^2(N-2)}{N^2} \left\{ \frac{1}{2\pi}  \partial_x {\vec  \Phi}_{R} \cdot  \partial_x  {\vec  \Phi}_{L}  -  
 \frac{1}{4\pi^2} \sum_{1 \le \alpha < \beta \le N} :\cos \left[ \sqrt{4\pi} \left( {\vec  e}_{\alpha} 
 - {\vec  e}_{\beta}\right) \cdot {\vec  \Phi}\right]: \right\}   \\
 &=   - \frac{32 \pi^2(N-2)}{N^2} J_R^{A} J_L^{A} \; ,  
 \end{split}
\label{TrG2G2NfinApp}
\end{equation}
where we have used Eq. (\ref{currentcurrentbosoApp}).

We now consider the general $n$ case. The leading term in ${\rm Tr} \; G^n$ is 
\begin{eqnarray}
 \nord{{\rm Tr} \; G^n } =  \frac{i^n}{N^{n/2}} \sum_{i_1 \ne i_2 \ldots \ne i_n} 
 : \rme^{  i \sqrt{4 \pi} \left({\vec  e}_{i_1} + {\vec  e}_{i_2}+ \ldots  + {\vec  e}_{i_n} \right) \cdot {\vec  \Phi}}:  \; ,
\label{TrGnApp}
\end{eqnarray}
which has scaling dimension $\Delta_n = \left({\vec  e}_{i_1} + {\vec  e}_{i_2}+ \ldots  {\vec  e}_{i_n} \right)^2
= n (1 - 1/N) - n(n-1)/N = n(N-n)/N$, and this is the $\SU(N)$$_1$ primary field corresponding to the $n$-box antisymmetric representation.
We then need to evaluate the following OPE:
\begin{align}
	& \nord{{\rm Tr} \; G^n (z, \bar z)}\nord{{\rm Tr} \;  {G^n}^{\dagger}(0, 0) }   \notag\\
	&=   \frac{1}{N^n}  \sum_{i_1 \ne i_2 \ldots \ne i_n} \sum_{j_1 \ne j_2 \ldots \ne j_n}  
	 : \rme^{  i \sqrt{4 \pi} \left({\vec  e}_{i_1} + {\vec  e}_{i_2}+ \ldots  + {\vec  e}_{i_n} \right) \cdot {\vec  \Phi}(z, \bar z)  }:\,
	 : \rme^{ - i \sqrt{4 \pi} \left({\vec  e}_{j_1} + {\vec  e}_{j_2} +  \ldots  + {\vec  e}_{j_n} \right) \cdot {\vec  \Phi}(0, 0)}:  
 	 \notag \\
 	 & = \frac{1}{N^n} \sum_{i_1 \ne i_2 \ldots \ne i_n} \sum_{j_1 \ne j_2 \ldots \ne j_n}  
 	  |z|^{-2\left({\vec  e}_{i_1} + {\vec  e}_{i_2}+ \ldots  + {\vec  e}_{i_n} \right) \cdot \left({\vec  e}_{j_1} + {\vec  e}_{j_2}+ \ldots  + {\vec  e}_{j_n} \right)} 
 	   : \rme^{  i \sqrt{4 \pi} \left( \left({\vec  e}_{i_1} + {\vec  e}_{i_2}+ \ldots  + {\vec  e}_{i_n} \right) \cdot {\vec  \Phi} (z, \bar z)  
 	   - \left({\vec  e}_{j_1} + {\vec  e}_{j_2} + \ldots  + {\vec  e}_{j_n} \right) \cdot {\vec  \Phi} (0, 0) \right)}: 
 	   \notag \\
	 & \sim \frac{n !}{N^n |z|^{2n(N-n)/N}}  \sum_{i_1 \ne i_2 \ldots \ne i_n}
	 \left(1 -  \frac{1}{2}4\pi \left\{ \left({\vec  e}_{i_1} + {\vec  e}_{i_2}+ \ldots  + {\vec  e}_{i_n} \right) \cdot \left(z \partial {\vec  \Phi}_L (0)  + {\bar z} {\bar \partial} {\vec  \Phi}_R (0)\right) \right\}^2\right)
	  \notag \\
	  & \quad +  \frac{ n n!}{N^n} \sum_{\beta \ne \gamma \ne i_2 \ldots \ne i_n}  |z|^{-2\left({\vec  e}_{\gamma} 
	  + {\vec  e}_{i_2}+ \ldots  + {\vec  e}_{i_n} \right) \cdot \left({\vec  e}_{i_2}+ \ldots  + {\vec  e}_{i_n} + {\vec  e}_{\beta} \right)}  
	   : \rme^{  i \sqrt{4 \pi} \left({\vec  e}_{\gamma}  - {\vec  e}_{\beta}\right) \cdot {\vec  \Phi} (0, 0)}: + \cdots 
	\nonumber \\
	& \sim \frac{n ! }{N^n |z|^{2n(N-n)/N}}\left[ \frac{N!}{(N-n)!} 
	-  |z|^{2} \frac{4 \pi n (N-2)!}{ (N -n-1)! } \left(\partial {\vec  \Phi}_L \cdot {\bar \partial}{\vec  \Phi}_R 
	- \frac{1}{4\pi}
    \sum_{\gamma \ne \beta}  : \rme^{  i \sqrt{4 \pi} \left({\vec  e}_{\gamma}  - {\vec  e}_{\beta}\right) \cdot {\vec  \Phi} (0, 0)}:\right) + \dots\right],
 \label{TrGnGnApp}
\end{align}
where as before the non-rotation-symmetric terms have been neglected and we have used the following identities with help of Eqs. (\ref{weightSUN}):
\begin{equation}
\begin{split}
& \sum_{i_1 \ne i_2 \ldots \ne i_n}  1 = N!/(N-n)!   \\
& \left({\vec  e}_{\gamma} 
 + {\vec  e}_{i_2}+ \ldots  + {\vec  e}_{i_n} \right) \cdot \left({\vec  e}_{i_2}+ \ldots  + {\vec  e}_{i_n} + {\vec  e}_{\beta} \right)
=  n - 1- \frac{n^2}{N}    \quad  \left(  \gamma \ne i_2 \ne \ldots \ne i_n \ne \beta  \right)     \\
& \sum_{i_1 \ne i_2 \ne \ldots \ne i_n} \left[ {\vec  e}_{i_1}+ \ldots  + {\vec  e}_{i_n} \right]^{m} \left[{\vec  e}_{i_1}+ \ldots  + {\vec  e}_{i_n}  \right]^{p}
=  n \frac{(N-2)!}{(N-n-1)!} \delta_{mp}  \;  , \\
& \sum_{ i_2\ne \ldots \ne i_n (\ne \gamma \ne  \beta) } 1 = \frac{(N-2)!}{(N-n-1)!} \;  .  
\end{split}
\end{equation}
The leading non-trivial term in the OPE (\ref{TrGnGnApp}) is for $n=2, \ldots ,\left[N/2\right]$:
\begin{equation}
\begin{split}
&  \nord{ (\nord{{\rm Tr} \; G^n})(\nord{{\rm Tr} \;  {G^{\dagger n} } }) } \\
& =  - \frac{8 \pi^2  n n! (N-2)!}{N^n (N-n-1)!} \left[ \frac{1}{2\pi}  \partial_x {\vec  \Phi}_{R} \cdot  \partial_x  {\vec  \Phi}_{L} 
  -  \frac{1}{4\pi^2} \sum_{1 \le \alpha < \beta \le N} :\cos \left( \sqrt{4\pi} \left( {\vec  e}_{\alpha} 
 - {\vec  e}_{\beta}\right) \cdot {\vec  \Phi}\right): \right]   \\
& =    - \frac{8 \pi^2 n n! (N-2)!}{N^n (N-n-1)!}  J_R^{A} J_L^{A} \;  ,
 \end{split}
\label{TrGnGnNfinApp}
\end{equation}
where Eq. (\ref{currentcurrentbosoApp}) has been used.

\subsection{OPE for the \texorpdfstring{$\SU(N)/\SO(N)$}{SU(N)/SO(N)} nonlinear \texorpdfstring{$\sigma$}{sigma} model with \texorpdfstring{$\theta = \pi$}{theta=pi}}
We now investigate the free-field representation of the interacting term in the action (\ref{WZWintSU(N)/SO(N)bis}) which accounts for the physical properties of the $\SU(N)$/SO($N$) nonlinear $\sigma$ model with $\theta = \pi$. In this respect, we need to calculate the OPE $G_{\alpha \beta}(z, \bar z) (G^{\dagger})_{\alpha \beta}(0, 0)$. Using the results (\ref{freefieldrepWZWGfield}), we have\footnote{Here, the computation of the phase factor associated with the Klein factor and commutation of zero modes is the most subtle part, and we need to do it with great cares. In this case, however, there is a trick to evade its computation and it provides the useful consistency check. For the computation of the diagonal contribution, i.e. $\alpha=\beta$, the Klein factor becomes trivial in an obvious way. Moreover, no subtle phase associated with the chiral zero modes is also absent because the compact boson can be combined as $\vec{\Phi}=\vec{\Phi}_L+\vec{\Phi}_R$. This part is already computed when calculating the OPE of $\tr G(z,\bar{z}) \tr G^\dagger(0,0)$ in \eqref{trGOPE1App}. Thus, the subtle phase factor can appear only in the off-diagonal contribution, but the off-diagonal one is severely constrained by the symmetry as the marginal term should be combined as $\tr(J_L J_R^T)$, and one can skip its computation from this viewpoint. }
\begin{equation}
\begin{split}
&  \sum_{\alpha,\beta} G_{\alpha \beta}(z, \bar z) (G^{\dagger})_{\alpha \beta}(0, 0)   \\
&=   \frac{1}{N} \sum_{\alpha, \beta} 
 \kappa_{\beta}  \kappa_{\alpha} \kappa_{\beta} \kappa_{\alpha} 
: \rme^{  i \sqrt{4 \pi} {\vec  e}_{\beta} \cdot {\vec  \Phi}_{L} (z)
+ i \sqrt{4 \pi}  {\vec  e}_{\alpha} \cdot {\vec  \Phi}_{R} (\bar{z}) }: \,  
: \rme^{  - i \sqrt{4 \pi} {\vec  e}_{\alpha} \cdot {\vec  \Phi}_{L} (0)
- i \sqrt{4 \pi}  {\vec  e}_{\beta} \cdot {\vec  \Phi}_{R} (0) }:    \\
& = \frac{1}{N} \sum_{\alpha, \beta} 
 \kappa_{\beta}  \kappa_{\alpha} \kappa_{\beta} \kappa_{\alpha} \;
\rme^{- i\pi  {\vec  e}_{\alpha} \cdot {\vec  e}_{\beta}}
: \rme^{  i \sqrt{4 \pi} {\vec  e}_{\beta} \cdot {\vec  \Phi}_{L}(z)}: \, 
: \rme^{  i \sqrt{4 \pi} {\vec  e}_{\alpha} \cdot {\vec  \Phi}_{R}(\bar z)}: \, 
  : \rme^{ - i \sqrt{4 \pi} {\vec  e}_{\alpha} \cdot {\vec  \Phi}_{L}(0) }: \, 
: \rme^{ - i \sqrt{4 \pi} {\vec  e}_{\beta} \cdot {\vec  \Phi}_{R}(0)}:   \\
& 
= \frac{1}{N} \sum_{\alpha, \beta} 
 \kappa_{\beta}  \kappa_{\alpha} \kappa_{\beta} \kappa_{\alpha}  \;
 \rme^{ - i \pi {\vec  e}_{\alpha} \cdot {\vec  e}_{\beta} + i \pi  {\vec  e}_{\alpha} \cdot {\vec  e}_{\alpha} }
: \rme^{  i \sqrt{4 \pi} {\vec  e}_{\beta} \cdot {\vec  \Phi}_{L}(z)}: \,
 : \rme^{ - i \sqrt{4 \pi} {\vec  e}_{\alpha} \cdot {\vec  \Phi}_{L}(0)}: \, 
: \rme^{  i \sqrt{4 \pi} {\vec  e}_{\alpha} \cdot {\vec  \Phi}_{R}(\bar z)}: \,  
: \rme^{ - i \sqrt{4 \pi} {\vec  e}_{\beta} \cdot {\vec  \Phi}_{R}(0)}:    \\
& \sim - \frac{1}{N} \sum_{\alpha, \beta} 
 \kappa_{\beta}  \kappa_{\alpha} \kappa_{\beta} \kappa_{\alpha}  (-1)^{\delta_{\alpha \beta}} \;
 \frac{1}{|z|^{2 {\vec  e}_{\alpha} \cdot {\vec  e}_{\beta}}}
: \rme^{  i \sqrt{4 \pi} {\vec  e}_{\beta} \cdot {\vec  \Phi}_{L}  (z) 
 - i \sqrt{4 \pi} {\vec  e}_{\alpha} \cdot {\vec  \Phi}_{L} (0) } : \;
: \rme^{  i \sqrt{4 \pi} {\vec  e}_{\alpha} \cdot {\vec  \Phi}_{R} (\bar z)  
-  i \sqrt{4 \pi} {\vec  e}_{\beta} \cdot {\vec  \Phi}_{R} (0)}  :    \\
& \sim \frac{1}{|z|^{2(1-1/N)}} \left[ 1 - \frac{4 \pi}{N} |z|^{2} 
\left\{
\partial_x {\vec  \Phi}_{R} \cdot  \partial_x  {\vec  \Phi}_{L} +  \frac{1 }{2\pi} \sum_{1 \le \alpha < \beta \le N} :\cos \left( \sqrt{4\pi} \left( {\vec  e}_{\alpha} - {\vec  e}_{\beta}\right) \cdot {\vec  \Theta}\right): 
\right\} (0, 0) + \cdots\right] \; ,
\end{split}
\label{OPESUN/SONApp}
\end{equation}
where ${\vec  \Theta} = {\vec  \Phi}_{L} -  {\vec  \Phi}_{R}$ is the dual field to ${\vec  \Phi}$ and 
where we have used $\partial_x = i ( \partial  - {\bar \partial})$.  

The leading regular term in the OPE (\ref{OPESUN/SONApp}) is a marginal contribution and we need to express it as a current-current perturbation, $\tr(J_L J_R^T)$. 
We can use the fermionic description (\ref{spinopapp}) of the $\SU(N)$$_1$ currents so that
\begin{equation}
\begin{split}
\tr(J_L J_R^T) &= T^{A}_{\alpha\beta} T^{B}_{\alpha\beta} J_L^{A} J_R^{B}  \\
&= T^{A}_{\alpha\beta} T^{B}_{\alpha\beta} L^{\dagger}_{\gamma}T^{A}_{\gamma \delta}  L_{\delta}
R^{\dagger}_{\gamma_1}T^{B}_{\gamma_1 \delta_1}  R_{\delta_1}  \\
&=  \frac{1}{4} \left(R^{\dagger}_{\alpha} R_{\beta} L^{\dagger}_{\alpha} L_{\beta} 
- \frac{1}{N} :R^{\dagger}_{\alpha} R_{\alpha}:\,  :L^{\dagger}_{\beta} L_{\beta}:  \right)  \; ,
\end{split}
\label{currcurrdualApp}
\end{equation}
where we have used the identity (\ref{algebraidenApp}) on the $\SU(N)$ generators. The next step of the approach is to apply the Abelian bosonization of the Dirac fermions (\ref{bosoabeleq}) and the use of the basis (\ref{invSUN}), and we find 
\begin{equation}
\begin{split}
\tr(J_L J_R^T) &= \frac{1}{4 \pi}
\left[  
\partial_x {\vec  \Phi}_{R} \cdot  \partial_x  {\vec  \Phi}_{L} +  \frac{1 }{2\pi} \sum_{1 \le \alpha < \beta \le N} :\cos \left( \sqrt{4\pi} \left( {\vec  e}_{\alpha} - {\vec  e}_{\beta}\right) \cdot {\vec  \Theta}\right): 
\right]  \\
&=\frac{1}{4\pi}\left[ -\partial \vec{\Theta}\cdot \bar{\partial}\vec{\Theta} +  \frac{1 }{2\pi} \sum_{1 \le \alpha < \beta \le N} :\cos \left( \sqrt{4\pi} \left( {\vec  e}_{\alpha} - {\vec  e}_{\beta}\right) \cdot {\vec  \Theta}\right): 
\right] \; .
\end{split}
\label{currcurrdualend}
\end{equation}
We thus deduce the identification: 
\begin{equation}
\nord{  {\rm Tr} (G G^{*} ) } =   - \frac{16 \pi^2}{N} \tr(J_L J_R^T). \; 
\label{perturbSUN/SONApp}
\end{equation}

\end{appendix}





\end{document}